\documentclass[a4paper, 11pt]{article}

\usepackage{amsmath}
\usepackage{amsfonts}
\usepackage{graphics}
\usepackage{epsfig}
\usepackage{amssymb}
\usepackage{amsthm}
\usepackage{amscd}
\usepackage[all]{xy}
\usepackage{latexsym}
\usepackage{multirow}
\usepackage{geometry}
\usepackage{color}
\usepackage{bm}
\usepackage{mathrsfs} 
\usepackage{titlesec}
\usepackage{hyperref}
\usepackage{graphicx}
\usepackage{tikz}
\usetikzlibrary{arrows.meta,decorations.pathmorphing}
\usepackage{subcaption}

\newtheorem{thm}{Theorem}[section]

\theoremstyle{definition}

\newtheorem{rem}[thm]{Remark}

\numberwithin{equation}{section}
\numberwithin{figure}{section}
\numberwithin{table}{section}

\newcommand{\Res}{\operatorname*{Res}}
\newcommand{\dd}{\mathrm d}

\newcommand{\thc}[2]{\vartheta\!\begin{bmatrix}#1\\#2\end{bmatrix}}

\title{Resurgence and Modularity\\in the First Painlev\'e Equation}

\author{\large Kohei Iwaki\footnote{Graduate School of Mathematical
Sciences, The University of Tokyo.}
\qquad Noam Tamarin\footnote{CAMGSD, Departamento de Matem\'atica,
Instituto Superior T\'ecnico, Universidade de Lisboa,
1049-001 Lisboa, Portugal.
Email: \href{mailto:noam.tamarin@tecnico.ulisboa.pt}
{\texttt{noam.tamarin@tecnico.ulisboa.pt}}.}}

\date{}

\newcommand{\paperfigure}[2][]{\includegraphics[#1]{#2}}

\begin{document}

\maketitle

\begin{abstract}
For the first
Painlev\'e equation, we relate three nonlinear Stokes maps associated with
a full turn around infinity to modular transformations
of the two-parameter tau function through the identity
$\boldsymbol S_\pi\circ\boldsymbol S_0\circ
\boldsymbol S_{-\pi}=\mathsf S\circ\mathsf T$.
Here the subscripts label Borel directions, while
$\mathsf S$ and $\mathsf T$ denote the modular generators.
Working under the analytic assumptions of the
topological recursion and exact WKB constructions, we
compare Stokes multipliers of the associated isomonodromy
system while tracking the homology basis of its elliptic
spectral curve under rotations of the independent variable.
For the sectors considered, we identify the known parameter
transformation induced by an $8\pi/5$ rotation with the
modular $T$-action and derive a new explicit parameter
relation realizing the $S$-action for a $2\pi/5$ rotation.
We also determine the multiplicative factors in the
corresponding tau-function relations, including their
normalization constants.
Together, these results give an explicit realization of
single-valuedness after one full turn around infinity
for both the Painlev\'e transcendent and its tau function,
with the tau-function prefactors cancelling by Abel's
five-term identity for the Rogers dilogarithm.
\end{abstract}

\tableofcontents

\section{Introduction}
\label{sec:introduction}

The first Painlev\'e equation provides a setting in which
resurgence, isomonodromic deformations, and topological recursion
can be studied together.
Its solutions admit transseries representations whose parameters
transform nonlinearly across Stokes directions.
They can also be characterized by the Stokes multipliers of an
associated linear differential equation.
A further description is provided by topological recursion on an
elliptic spectral curve, from which a two-parameter tau function
is constructed by a discrete Fourier transform, also called
a Zak transform \cite{Iwaki2019,IILZ2025}.
The purpose of this paper is to investigate the relation between
nonlinear Stokes phenomena and modular transformations arising
from changes of the homology basis of this spectral curve.

The connection problems and Stokes phenomena of Painlev\'e
equations have been studied from several complementary
perspectives.
Kapaev, Kitaev, and others investigated connection problems and
asymptotic behavior using the Riemann--Hilbert method
\cite{KapaevKitaev1993,Kapaev2004}, while
Takei and others studied nonlinear Stokes phenomena through
exact WKB analysis \cite{Takei1995, Takei2000}.
The resurgent structure of Painlev\'e I, including
the large-order asymptotics of its instanton sectors
and the associated Stokes constants, was investigated
by Garoufalidis--Its--Kapaev--Mari\~no
\cite{GIKM12} using both Riemann--Hilbert
and resurgent methods.
Further developments by Aniceto--Schiappa--Vonk \cite{AnicetoSchiappaVonk2012}, 
Baldino--Schiappa--Schwick--Vega \cite{BSSV2022}, 
and van Spaendonck--Vonk \cite{vanSpaendonckVonk2022} provided detailed descriptions
of the two-parameter transseries and their
nonlinear Stokes transformations.
In this description, the change of lateral Borel sums across a
Stokes direction is compensated by a nonlinear transformation
of the transseries parameters, relating different asymptotic
representations of the same solution.

Eynard--Orantin's topological recursion \cite{EynardOrantin2007} provides another
approach to these questions.
The first-named author \cite{Iwaki2019} constructed a two-parameter tau function
for the first Painlev\'e equation as a Zak transform of the
partition function associated with an elliptic spectral curve.
The corresponding nonperturbative wave functions, obtained by
the Zak transform of the wave functions constructed by topological
recursion, solve the associated isomonodromy linear system.
This construction also provides a method for computing its
Stokes multipliers using exact WKB analysis.
Building on this framework, the first-named author and Mari\~no
\cite{IwakiMarino2024} derived a formula for the Stokes
automorphism from a mutation of the linear Stokes graph,
subject to suitable analytic assumptions.
The formula combines a logarithmic shift of a parameter with
a dilogarithmic prefactor in the transformation of the tau
function.
This formula is consistent with the Stokes
transformations of the two-parameter transseries
obtained from resurgence, as reviewed in Section~\ref{sec:stokes-review}.
The relation between the Fourier representation, topological
recursion, and the holomorphic anomaly approach was further
studied in \cite{IILZ2025}.

To understand the analytic continuation of the
tau function around infinity, it is natural to
consider not only Stokes transformations but also
the behavior of the underlying elliptic spectral
curve as the independent variable $z$ is rotated.
In particular, the choice of a Torelli marking,
namely a symplectic basis of the first homology
of the spectral curve, plays an essential role
in the construction by topological recursion.
A change of Torelli marking is described by an
action of the modular group, and the resulting
transformation properties of the topological
recursion free energies were studied by
Eynard--Orantin \cite{EynardOrantin2007}.

Furthermore, Eynard--Mari\~no \cite{EM11} constructed a non-perturbative partition function and established its modular transformation properties under changes of the Torelli marking.
We note that their non-perturbative partition function 
is closely related to the two-parameter tau function constructed in \cite{Iwaki2019}, both being obtained from the topological recursion partition function through the Zak transform.
We denote the two parameters of the tau function by $\nu$ and $\rho$,
where $\nu$ is the filling fraction associated with the chosen
$A$-cycle and $\rho$ is its Fourier-conjugate parameter.
These parameters play a role analogous to that of the theta characteristics in \cite{EM11}, rather than transforming simply as a pair of classical periods.
We show that, in suitable sectors, certain rotations of $z$ in Painlev\'e~I induce modular transformations on $(\nu,\rho)$, and clarify their relation to nonlinear Stokes phenomena.
It would be interesting to clarify the precise relation between our modular transformation formulas and those of \cite{EM11}, particularly at the level of normalization factors.

Our approach is to follow both the linear Stokes graph and the
Torelli marking as the independent variable $z$ is rotated.
Using the rotational symmetry of the first Painlev\'e equation,
we compare the Stokes multipliers before and after the rotation,
taking account of the induced relabeling of the Stokes sectors.
For the sectorial descriptions considered below, this identifies
rotations through $8\pi/5$ and $2\pi/5$ with the modular
$T$- and $S$-actions on $(\nu,\rho)$, respectively.
The transformation associated with the $8\pi/5$ rotation was
already known in the original transseries parameters
\cite[Section~7.3]{BSSV2022};
see also \cite[Section~4.2]{vanSpaendonckVonk2022}, though its modular nature was not identified there.
Here we give its geometric derivation and identify its modular
meaning.
For the $2\pi/5$ rotation, both the explicit relation between
the sectorial transseries parameters and its interpretation
as the modular $S$-action are new.

These modular transformations provide a second description
of analytic continuation around infinity. 
One description is obtained by composing the nonlinear Stokes
transformations encountered along the continuation path.
The other follows from comparing the Torelli markings at the
endpoints and applying the corresponding modular transformations.
We show that the two descriptions induce the same parameter
transformation, obtaining an explicit identity between a product
of Stokes maps and a composition of modular transformations.

The single-valuedness of Painlev\'e I transcendents is a
consequence of the Painlev\'e property.
Its realization in a transseries representation is nevertheless
nontrivial, since fractional powers and logarithms make this
representation apparently multivalued.
Previous work by Tamarin and collaborators
\cite{KagerEtAl2026Global,KagerEtAl2026Hierarchy,Tamarin2026Thesis}
addressed this through Stokes transformations, using analytic
continuation over several turns in the $z$-plane.
The modular description allows us to realize this closure
explicitly after a single turn in the $z$-plane. Our results suggest that the single-valuedness of Painlev\'e transcendents is realized through the compatibility of nontrivial nonlinear Stokes phenomena and modular transformation laws.

At the level of the tau function, the modular transformations
also carry exponential factors quadratic in the parameters,
together with normalization constants.
We determine the constant associated with $T$ from the
large-$z$ asymptotic expansion.
For the constant associated with $S$, we use a rotationally
symmetric solution that is regular at the origin, following
the strategy of Lisovyy--Roussillon
\cite{LisovyyRoussillon2017}.
Combining a Stokes transformation with a modular transformation
then yields a five-step recurrence for the parameters.
Abel's five-term identity for the Rogers dilogarithm implies
that the product of the corresponding tau-function prefactors
is equal to one.
Thus, the tau function also returns to its original value
after one full turn.\footnote{
A closely analogous prefactor calculation was already carried
out in \cite[Section~3.2]{Tamarin2026Thesis}; see also
\cite{KagerEtAl2026Hierarchy}.
The two-turn formulation of that calculation uses a different
cyclic operator, whereas the construction here closes after
one full turn.
}

Cyclic relations for Painlev\'e I connection coefficients and
their relation to dilogarithm identities were already established
by Lisovyy--Roussillon \cite{LisovyyRoussillon2017}.
Our emphasis here is on relating this structure to modular
transformations of the tau function constructed by topological
recursion and to Stokes automorphisms expressed in the same
parameters.

Our discussion relies on the results and analytic assumptions
of the preceding works.
In particular, the construction and Stokes analysis of the
tau function by topological recursion involve assumptions
concerning Borel summability, the discrete Fourier transform,
and the applicability of exact WKB connection formulas
\cite{Iwaki2019,IwakiMarino2024}.
We work under the same assumptions and do not attempt to
prove them here.
The Iwaki--Mari\~no formula is also consistent with predictions
obtained from the nonperturbative analysis of the holomorphic
anomaly equations in topological string theory
\cite{GuMarino2023,IwakiMarino2024}.
The underlying resurgent structure has substantial numerical
support in concrete examples of topological string theory,
including high-precision tests using Pad\'e approximations
of Borel transforms \cite{GuMarino2023}.

The paper is organized as follows.
Section~2 reviews the Stokes structure of the first Painlev\'e
equation from the perspectives of resurgence and topological
recursion.
This section contains no essentially new results; the detailed
review prepares the comparison of Stokes multipliers used
later.
Section~3 studies the rotational symmetry and derives the
transformations associated with $T$ and $S$.
Section~4 relates these modular transformations to Stokes
automorphisms, determines the normalization constants, and
discusses analytic continuation around infinity  
and the
associated dilogarithm identity. Section~\ref{sec:conclusions} discusses possible extensions
to other Painlev\'e equations and to higher members of
the Painlev\'e~I hierarchy.
Appendix~A collects the relevant theta-function formulas,
and Appendix~B discusses the transseries normalization and
the composition of Stokes transformations.

\section*{Acknowledgements}

We would like to thank 
Jasper Kager, 
Oleg Lisovyy, 
Jo\~ao Rodrigues, 
Shinji Sasaki, 
Ricardo Schiappa, 
Max Schwick, 
Yoshitsugu Takei, 
and 
Yuji Terashima 
for helpful discussions and valuable comments.
K.I. was supported by JSPS KAKENHI Grant Numbers 
24K00525, 23K17654, 22H00094.
N.T. thanks CAMGSD for its institutional support and acknowledges financial support from CAMGSD and the University of Tokyo for his research visit to Tokyo, during which part of this work was carried out.

\section{Review of Stokes Structure of Painlev\'e I}
\label{sec:stokes-review}

The results of this paper concern the first Painlev\'e equation
in the normalization
\begin{equation}
  u(z)^2-\frac13u''(z)=z.
  \label{eq:PI-unit-coupling}
\end{equation}
We review its Stokes structure from two complementary points of view.
In Section~\ref{sec:bssv}, we work directly with the nonlinear
equation, recalling its two-parameter transseries solution and the
Stokes automorphisms obtained from resurgence
\cite{BSSV2022,Tamarin2026Thesis}.
In Section~\ref{sec:iwaki-marino}, we review the approach of
Iwaki--Mari\~no \cite{IwakiMarino2024}, which uses topological
recursion and the associated linear isomonodromy system.
In this description, the nonlinear Stokes transformation is
determined by comparing the Stokes multipliers of the linear
system. Expressing the two descriptions in common tau-function
parameters prepares the comparison between resurgence and
modularity in Section~\ref{sec:resurgence-modularity}.

For the second approach, it is useful to introduce the string
coupling $g_s$ into the Painlev\'e equation and its associated
linear system. For fixed $g_s>0$, this formulation is equivalent
to \eqref{eq:PI-unit-coupling} by a simultaneous rescaling of $z$
and $u$; the precise rescaling and parameter conventions are
given in Section~\ref{sec:iwaki-marino}.
Keeping $g_s$ explicit is convenient for the topological recursion expansion and the exact WKB analysis of the linear system.
The associated elliptic spectral curve also provides the
geometric setting for the modular transformations studied in
Section~\ref{sec:modular-transformation}.
We therefore retain $g_s$ in that construction and the modular
derivation, returning to the $g_s=1$ convention when combining
the two descriptions in Section~\ref{sec:resurgence-modularity}.

\subsection{Stokes automorphisms from resurgence}
\label{sec:bssv}

We now recall the two-parameter transseries solution of
\eqref{eq:PI-unit-coupling} and its Stokes automorphisms,
following
\cite{AnicetoSchiappaVonk2012,BSSV2022,Tamarin2026Thesis,KagerEtAl2026Hierarchy}.
Our aim is to express the Stokes transformations of the
nonlinear equation in variables that can be compared with
the construction from the associated linear system in
Section~\ref{sec:iwaki-marino}.

The perturbative background and the elementary instanton action are
fixed by
\begin{equation}
 u_0(z)=\sqrt z,
 \qquad A'(z)^2=6u_0(z),
 \qquad A(z)=\widetilde A z^{5/4},
 \qquad \widetilde A=\frac{4\sqrt6}{5}.
 \label{eq:resurgent-background}
\end{equation}
There are four initial choices, from the two signs of $u_0$ and the two
signs of $A'$. We use \eqref{eq:resurgent-background} throughout, keeping
track of the argument of $z$ when continuing to other sheets.
In the conventions of \cite[Section~3]{Tamarin2026Thesis}, the
transseries is
\begin{equation}
\begin{aligned}
 u(z;\sigma_1,\sigma_2)
 &=\sqrt z\sum_{n,m\geq0}\sigma_1^n\sigma_2^m
 e^{-(n-m)\widetilde A z^{5/4}}
 z^{-\frac54\frac\alpha2(n-m)\mu}
 \sum_{g\geq0}u_g^{(n,m)}z^{-\frac54(g+\beta_{n,m})},\\
 \mu&=\sigma_1\sigma_2,
 \qquad \alpha=-2\sqrt{\frac23},
 \qquad
 \beta_{n,m}=\frac{n+m}{2}
 -\left\lfloor\frac{\min(n,m)-n\delta_{n,m}}2\right\rfloor .
\end{aligned}
\label{eq:resurgent-PI-transseries}
\end{equation}
Substitution in Painlev\'e~I determines the sector coefficients
recursively. The parameters multiply the elementary exponentials
$e^{-\widetilde A z^{5/4}}$ and $e^{+\widetilde A z^{5/4}}$, respectively.
We take the leading coefficients of these two sectors to be
$u_0^{(1,0)}=u_0^{(0,1)}=1$; this fixes the parameter normalization
used in the formulas below.\footnote{The value
$\alpha=-2\sqrt2/3$ printed in \cite[(3.46)]{Tamarin2026Thesis}
appears to be a typographical error. With unit leading coefficients
in the two elementary sectors, substitution in
$u^2-\frac13u''=z$ fixes the first logarithmic resonance by
$\sqrt6\,\alpha+4=0$, hence $\alpha=-2\sqrt{2/3}$.
We use this value throughout.}

Let us examine the structure of this solution. Its exponential weight
depends only on $n-m$, so that, for example, the sectors $(1,0)$ and
$(2,1)$ have the same exponential factor. The two elementary actions
are opposite, and their sum vanishes.\footnote{The matrix-model
interpretation of the sectors with both signs of the action is
developed in \cite{MarinoSchiappaSchwick2022}; their minimal-string
interpretation is discussed in \cite{SchiappaSchwickTamarin2026}.}
This is the resonance responsible
for the logarithmic sectors in the original transseries
\cite{AnicetoSchiappaVonk2012,BSSV2022}. In
\eqref{eq:resurgent-PI-transseries} these logarithms have already been
summed into a power of $z$: expanding
$z^{-\frac54\frac\alpha2(n-m)\mu}$ in $\mu$ reproduces them.
Thus the fractional powers and the logarithms both have to be followed
when the argument of $z$ changes.

\begin{figure}[htbp]
    \centering

    \begin{subfigure}[t]{0.48\textwidth}
        \centering
        \resizebox{\linewidth}{!}{%
        \begin{tikzpicture}[scale=0.75]

        \node[black] at (6.2,6.4) {\small sheet $-1$};
        
            \draw[gray, line width=1pt, ->] (-7,0) -- (7,0);
            \draw[gray, line width=1pt, ->] (0,-7) -- (0,7);

            \draw[black, line width=1pt, decorate,
                  decoration={zigzag, segment length=7pt, amplitude=1.6pt}]
                  (-6.8,0) -- (-0.25,0);
                  
            \node[gray] at (-5.8,6.2) {\small $z$-plane};

            \draw[red, dotted, line width=2pt] (0,0) -- (7,0);

            \draw[red, dotted,line width=2pt] (0,0) -- ({7*cos(144)},{7*sin(144)});

            \draw[red, dotted, line width=2pt] (0,0) -- ({7*cos(216)},{7*sin(216)});
            
            \draw[blue, line width=3pt] (0,0) -- ({6.5*cos(72)},{6.5*sin(72)});
            \node[blue] at ({6.5*cos(65)}, {6.5*sin(65)}) {$\mathfrak{S}_{2\pi}$};

            \draw[orange, line width=3pt] (0,0) -- ({6.5*cos(-72)},{6.5*sin(-72)});
            \node[orange] at ({6.5*cos(-65)}, {6.5*sin(-65)}) {$\mathfrak{S}_{3\pi}$};
        \end{tikzpicture}%
        }
        \caption{Sheet $-1$.}
        \label{fig:P1stokesAutomorphismsPrincipalMINUS2PISheet}
    \end{subfigure}
    \hfill
    \begin{subfigure}[t]{0.48\textwidth}
        \centering
        \resizebox{\linewidth}{!}{%
        \begin{tikzpicture}[scale=0.75]

        \node[black] at (6.2,6.4) {\small sheet $0$};
        
            \draw[gray, line width=1pt, ->] (-7,0) -- (7,0);
            \draw[gray, line width=1pt, ->] (0,-7) -- (0,7);

            \draw[black, line width=1pt, decorate,
                  decoration={zigzag, segment length=7pt, amplitude=1.6pt}]
                  (-6.8,0) -- (-0.25,0);
                  
            \node[gray] at (-5.8,6.2) {\small $z$-plane};

            \draw[blue, line width=3pt] (0,0) -- (7,0);
            \node[blue] at (5.5, 0.5) {$\mathfrak{S}_{0}$};
            
            \draw[orange, line width=3pt] (0,0) -- ({7*cos(144)},{7*sin(144)});
            \node[orange] at ({7.5*cos(151.2)}, {7.5*sin(151.2)}) {$\mathfrak{S}_{-\pi}$};
            
            \draw[orange, line width=3pt] (0,0) -- ({7*cos(216)},{7*sin(216)});
            \node[orange] at ({7.5*cos(208.8)}, {7.5*sin(208.8)}) {$\mathfrak{S}_{\pi}$};
            
            \draw[red,dotted, line width=2pt] (0,0) -- ({6.5*cos(72)},{6.5*sin(72)});
            \draw[red,dotted, line width=2pt] (0,0) -- ({6.5*cos(-72)},{6.5*sin(-72)});
            
        \end{tikzpicture}%
        }
        \caption{Sheet $0$.}
        \label{fig:P1stokesAutomorphismsPrincipalSheet}
    \end{subfigure}

    \caption{Stokes automorphisms on sheet $-1$ (left) and the principal sheet (right), for the background \eqref{eq:resurgent-background}. The subscripts are the continuously followed angles of $\widetilde x=z^{-5/4}$. The displayed automorphisms act for clockwise crossings; counterclockwise crossings use their inverses. Blue and orange denote forward and backward Stokes directions, and red dotted rays are anti-Stokes directions. Crossing the branch cut clockwise takes sheet $0$ to sheet $-1$ and interchanges the projected Stokes and anti-Stokes directions.}
    \label{fig:P1stokesAutomorphismsSheets}
\end{figure}
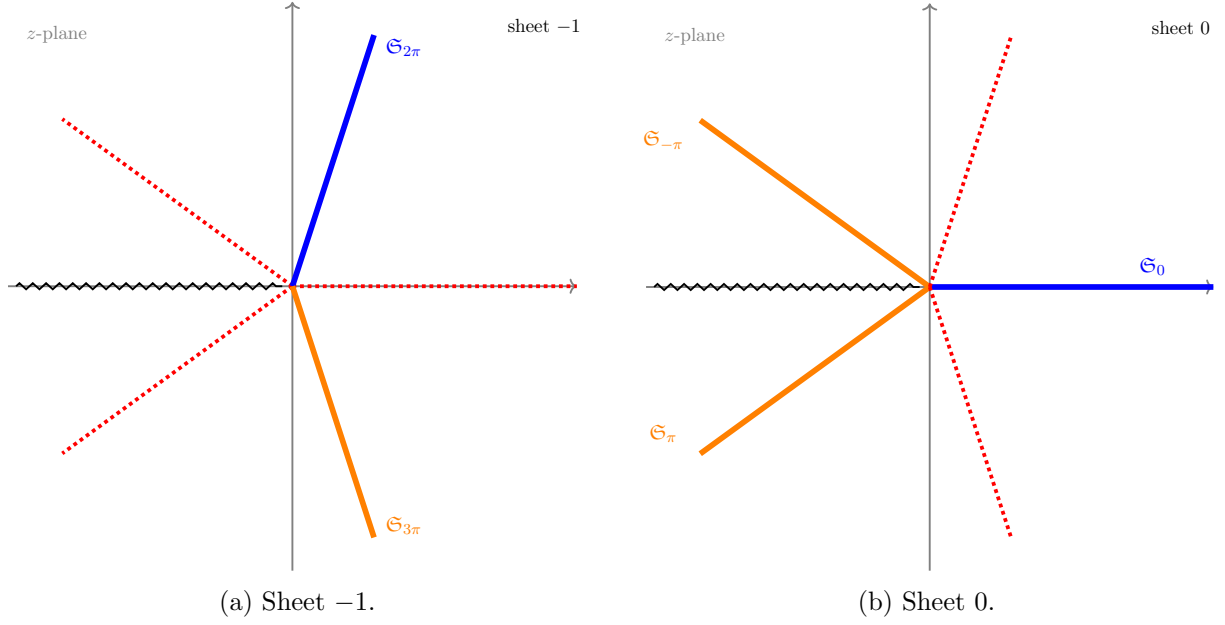

The formal power series in each sector have factorially growing
coefficients. Their large-order relations determine the Borel
singularities and the Stokes data relating the sectors
\cite{GIKM12, BSSV2022}. 
For the variable $\widetilde x=z^{-5/4}$, the nonzero
singular actions lie at integer multiples of $\widetilde A$. The
singular Borel directions are therefore $\arg\widetilde x=0,\pi$
modulo $2\pi$. On the principal sheet this gives
\begin{equation}
\begin{aligned}
 \text{Stokes lines:}\quad
 &\operatorname{Im}(\widetilde A z^{5/4})=0,
 &&\arg z=0,\ \pm\frac{4\pi}{5},\\
 \text{anti-Stokes lines:}\quad
 &\operatorname{Re}(\widetilde A z^{5/4})=0,
 &&\arg z=\pm\frac{2\pi}{5}.
\end{aligned}
\label{eq:resurgent-Stokes-rays}
\end{equation}
Across a Stokes line, the two lateral Borel sums of a fixed transseries
differ. Analytic continuation of the same solution is obtained by
compensating for this difference with a change of its transseries
parameters. An anti-Stokes line instead marks equality in magnitude
of the elementary exponentials. The positions in
\eqref{eq:resurgent-Stokes-rays} refer to the chosen sheet: a full turn
changes $z^{5/4}$ by a factor $i$, and interchanges the projected Stokes
and anti-Stokes directions. This is illustrated in
Figure~\ref{fig:P1stokesAutomorphismsSheets}.

We denote a Stokes automorphism by $\mathfrak S_\theta$ and its induced
parameter map by $\boldsymbol S_\theta$. The subscript is the
continuously followed Borel angle, $\theta=-\frac54\arg z$.
Our $\boldsymbol S_\theta$ implements a clockwise crossing in the
$z$-plane; a counterclockwise crossing uses its inverse, as in
\cite{Tamarin2026Thesis}. This convention also fixes the labels of the
two backward directions: $\theta=\pi$ lies at $\arg z=-4\pi/5$, whereas
$\theta=-\pi$ lies at $\arg z=4\pi/5$.

For comparison with the next subsection, it is useful to
express these transformations in Fourier variables.
The double transseries for $u$ in
\eqref{eq:resurgent-PI-transseries} can be integrated using
the Hamiltonian definition of $\tau$, recalled in
Section~\ref{sec:iwaki-marino}.
The resulting tau transseries can then be reorganized as a
discrete Fourier, or Zak, transform
\cite{KagerEtAl2026Local,KagerEtAl2026Global,Tamarin2026Thesis}.
We choose the Fourier variables so that the Stokes crossing
in the $\pi$ direction leaves $\rho$ fixed and shifts only $\nu$.
Their relation to the original transseries parameters is
\cite{Tamarin2026Thesis,KagerEtAl2026Global,KagerEtAl2026Hierarchy}
\begin{equation}
 \nu=\frac\alpha2\mu,
 \qquad
 e^{2\pi i\rho}=\rho_\pi
 =-i\sqrt\pi\sqrt{-\alpha}\,\sigma_1
 \frac{(48\sqrt6)^\nu e^{\pi i\nu}}{\Gamma(1+\nu)}.
 \label{eq:resurgent-Fourier-dictionary}
\end{equation}
The use of the same symbols $(\nu,\rho)$ anticipates their
identification with the filling fraction and Fourier parameter
of Section~\ref{sec:iwaki-marino}, after the rescaling described
in Remark~\ref{rem:unit-coupling-parameters}.
Here they are introduced through
\eqref{eq:resurgent-Fourier-dictionary}; identifying the two
descriptions also involves the asymptotic expansions, the sectors
in which they are resummed, and the normalization of the tau-function.
We return to these points in Section~\ref{sec:resurgence-modularity},
where the Stokes and modular transformations are used together.

With the parameter-dependent integration constant fixed accordingly,
the Fourier representation is
\begin{equation}
 \tau(z;\nu,\rho)=\sum_{n\in\mathbb Z}e^{2\pi in\rho}
 \mathcal Z(z,\nu+n),
 \label{eq:pinched-Fourier}
\end{equation}
where
\begin{equation}
\begin{aligned}
 \mathcal Z(z,\nu)
 ={}&z^{-1/48}e^{-2z^{5/2}/15}
 (2\pi)^{-\nu/2}(48\sqrt6)^{-\nu^2/2}
 e^{-\pi i\nu^2/2}G_2(1+\nu)\\
 &\hspace{8mm}\times e^{-\widetilde A\nu z^{5/4}}
 z^{-5\nu^2/8}
 \sum_{k\geq0}\widetilde D_k(\nu)z^{-5k/4}.
\end{aligned}
\label{eq:pinched-block}
\end{equation}
Here $G_2$ is the Barnes $G$-function, normalized by $G_2(1)=1$ and
$G_2(w+1)=\Gamma(w)G_2(w)$, and the $\widetilde D_k$ are polynomials
with $\widetilde D_0=1$ and
$\widetilde D_k(-\nu)=(-1)^k\widetilde D_k(\nu)$.
This form of the Fourier representation of the Painlev\'e~I
$\tau$-function was first proposed in \cite{BLMST2016}.
The reorganization and its large-$z$ expansion are discussed in
\cite[Sections~3.1--3.3]{KagerEtAl2026Local}, and its Stokes continuation
in \cite{KagerEtAl2026Global,KagerEtAl2026Hierarchy,Tamarin2026Thesis}.
The corresponding
construction for Painlev\'e~II appears in
\cite[Section~5]{IorgovLisovyyTamarinZhuravlov2026}.

In these coordinates the backward Stokes map takes the simple form
\begin{equation}
 \boldsymbol S_\pi(\nu,\rho)
 =\left(\nu-\frac{\log(1-e^{2\pi i\rho})}{2\pi i},\rho\right),
 \qquad
 \boldsymbol S_\pi^{-1}(\nu,\rho)
 =\left(\nu+\frac{\log(1-e^{2\pi i\rho})}{2\pi i},\rho\right).
 \label{eq:resurgent-Spi}
\end{equation}
Thus $\rho$ is unchanged, while $\nu$ undergoes a logarithmic shift.
The other Stokes maps are obtained by rotating the transseries
\cite{Tamarin2026Thesis,KagerEtAl2026Hierarchy}; the required relations are collected
in Appendix~\ref{app:parameter-dictionary}.
The logarithms are taken on a common local branch; changes of integer
representative are treated as in
Remark~\ref{rem:monodromy-parameter-branches}.
We will recover \eqref{eq:resurgent-Spi} in the next subsection by
comparing the linear Stokes multipliers. At the level of the tau
function, that comparison also supplies the dilogarithm prefactor
needed for the monodromy calculation.

\subsection{Stokes automorphism from isomonodromy}
\label{sec:iwaki-marino}

In this subsection, we consider the $g_s$-dependent Painlev\'e I equation, 
which is given by
\begin{equation}
  u(z)^2-\frac{g_s^2}{3}u''(z)=z,
  \label{eq:PI-gs}
\end{equation}
where \(g_s\) denotes the string coupling constant.
We take $\arg g_s=0$, so that $g_s>0$.

The relation to the equation \eqref{eq:PI-unit-coupling} considered
in the previous section is given by the rescaling
\begin{equation}
 \widehat z=g_s^{-4/5}z,
 \qquad \widehat u(\widehat z)=g_s^{-2/5}u(z).
\label{eq:resurgent-gs-scaling}
\end{equation}
Under this rescaling, \eqref{eq:PI-gs} becomes 
$\widehat u^2-\frac13\partial_{\widehat z}^2\widehat u=\widehat z$,
which is the equation \eqref{eq:PI-unit-coupling}.

In what follows, we recall the arguments in \cite{IwakiMarino2024},
and derive the Stokes jump formulas reviewed 
in the previous section from the perspectives of 
WKB analysis and topological recursion.
We also briefly discuss the isomonodromy system 
which we use for the derivation.

\subsubsection{Isomonodromy system}

The scalar isomonodromy system associated with Painlev\'e I in the
present normalization is
\begin{equation}\label{eq:PI-isomonodromy-system}
  \begin{aligned}
  \left[
    g_s^2\partial_x^2
    -\frac{g_s}{x-u}
       \bigl(g_s\partial_x-v\bigr)
    -\left(
       \frac{x^3}{18}
       -\frac{zx}{6}
       -\frac{\mathcal H}{3}
     \right)
  \right]\psi&=0,\\
  \left[
    g_s\partial_z
    +\frac{3}{x-u}
       \bigl(g_s\partial_x-v\bigr)
  \right]\psi&=0,
  \end{aligned}
\end{equation}
where
\begin{equation}
  \mathcal H(z,u,v)
  =-3v^2+\frac{u^3}{6}-\frac{zu}{2}.
\end{equation}
The compatibility condition of
\eqref{eq:PI-isomonodromy-system} gives
\begin{equation} \label{eq:PI-Hamiltonian-system}
  g_s\frac{\dd u}{\dd z}=-6v,
  \qquad
  g_s\frac{\dd v}{\dd z}=\frac{z-u^2}{2},
\end{equation}
and hence reproduces the Painlev\'e I equation \eqref{eq:PI-gs}.
In the $g_s$-dependent setting, the tau-function is defined in terms of the Hamiltonian function \({\mathcal H}\) by
\begin{equation}
  g_s^2\frac{d}{dz}\log \tau(z) = {\mathcal H}(z).
\end{equation}
In particular, one has
\begin{equation}
  u(z)
  =
  -2g_s^2\frac{d^2}{dz^2}\log \tau(z).
\end{equation}

As is well known, if $u$ satisfies the Painlev\'e I,
the Stokes multipliers at $x=\infty$ of the solutions of 
\eqref{eq:PI-isomonodromy-system} are independent of $z$ 
and hence define conserved quantities.
The existence of these conserved quantities reflects the
integrability of the Painlev\'e equation and forms the basis
of our discussion of Stokes phenomena and modularity below.

\subsubsection{Topological recursion construction of the tau-function}
Fix a Torelli marking, namely a symplectic basis $(A,B)$ with
$A\cdot B=1$, on the elliptic curve
\begin{equation}\label{eq:PI-spectral-curve}
 \Sigma_{z,\nu}:\qquad y^2=\frac{x^3}{18}-\frac{zx}{6}-\frac{h(z,\nu)}3,
\end{equation}
which can be regarded as the classical limit of 
the first equation in \eqref{eq:PI-isomonodromy-system}. 
Away from the discriminant locus, $h$ is determined locally by
\begin{equation}\label{eq:A-period-condition}
 \oint_A y\,\dd x=2\pi i\nu.
\end{equation}
The marking is transported continuously within the parameter domain.

We apply topological recursion to the marked curve
\eqref{eq:PI-spectral-curve} with the period condition
\eqref{eq:A-period-condition}.
Let $r_1,r_2,r_3$ be the finite ramification points of $x$, and
$\sigma(x,y)=(x,-y)$. The initial data are
\begin{equation}
 \omega_{0,1}=y\,\dd x,\qquad \omega_{0,2}(p,q)=B(p,q),
 \label{eq:TR-initial}
\end{equation}
where $B$ is the symmetric bidifferential with diagonal
principal part $\dd\xi(p)\dd\xi(q)/(\xi(p)-\xi(q))^2$ and vanishing
$A$-periods. With $p_I=(p_i)_{i\in I}$, the recursion reads
\begin{align}
 \omega_{g,n}(p_1,p_I)
 &=\sum_{j=1}^3\Res_{p=r_j}K(p_1,p)
 \Bigl[\omega_{g-1,n+1}(p,\sigma(p),p_I)\notag\\
 &\hspace{12mm}+\sum_{\substack{g_1+g_2=g\\J\sqcup J'=I}}'
 \omega_{g_1,|J|+1}(p,p_J)
 \omega_{g_2,|J'|+1}(\sigma(p),p_{J'})\Bigr],
 \label{eq:TR-recursion}\\
 K(p_1,p)&=\frac{\displaystyle\int_{\sigma(p)}^pB(p_1,q)}
 {2\bigl(y(p)-y(\sigma(p))\bigr)\dd x(p)}.
 \label{eq:TR-kernel}
\end{align}
Here $I=\{2,\ldots,n\}$, $2g-2+n>0$, negative-genus terms vanish,
and the prime excludes factors $\omega_{0,1}$; the integral is local.

In this normalization, convenient free energies are
\begin{align}
 F_0&=\frac{2zh}{5}+\frac{\nu}{2}\oint_B y\,\dd x,
 \label{eq:TR-F0}\\
 F_1&=-\frac1{12}\log\!\left(\omega_A^6\Delta_x\right),
 \quad \omega_A=\oint_A\frac{\dd x}{y},\quad
 \Delta_x=\prod_{i<j}(e_i-e_j)^2,
 \label{eq:TR-F1}\\
 F_g&=\frac1{2-2g}\sum_{j=1}^3\Res_{p=r_j}\Phi(p)\omega_{g,1}(p),
 \quad \dd\Phi=y\,\dd x,\quad g\geq2.
 \label{eq:TR-Fg}
\end{align}
Here $e_j=x(r_j)$. Branches of
logarithms are fixed locally. In particular,
\begin{equation}
 \partial_zF_0 =h,\qquad
 \partial_\nu F_0 =\oint_B y\,\dd x.
 \label{eq:TR-special-geometry}
\end{equation}

Define the partition function and its Zak transform by
\begin{align}
 Z(z,\nu;g_s)&=\exp\!\left(\sum_{g\geq0}g_s^{2g-2}F_g(z,\nu)\right),
 \label{eq:IM-partition}\\
 \tau(z,\nu,\rho;g_s)&=\sum_{k\in\mathbb Z}
 e^{2\pi i k\rho/g_s}Z(z,\nu+kg_s;g_s).
 \label{eq:IM-Zak}
\end{align}
The second parameter $\rho$ is the 
Fourier parameter, not the classical $B$-period.
The construction of \cite{Iwaki2019}, transferred by
\eqref{eq:rescaling-Iwaki}, gives the formal identities
\begin{equation}
 u=-2g_s^2\partial_z^2\log\tau,\qquad
 u^2-\frac{g_s^2}{3}\partial_z^2u=z,\qquad
 \mathcal H=g_s^2\partial_z\log\tau.
 \label{eq:TR-tau-PI}
\end{equation}
The Fourier series is understood in the formal sense of that
construction; an analytic interpretation requires additional summability
and convergence assumptions. Its overall parameter-independent
normalization is immaterial here.

The same construction also provides solutions of the isomonodromy system
\eqref{eq:PI-isomonodromy-system} as nonperturbative wave functions.
They are obtained by applying the Zak transform to the product of the
TR partition function and each of the two TR wave functions, using the
same shifts of $\nu$ and Fourier weights as in \eqref{eq:IM-Zak}, and
normalizing by the tau-function. This construction, and the method of
computing the Stokes multipliers from the WKB connection formulas for
the TR wave functions through this Fourier sum, were developed in
\cite{Iwaki2019}. The computation reviewed below is based on this method.

\begin{rem}
Let $(t,q,p,\hbar)$ denote the variables used in
\cite{Iwaki2019}, and let $(z,u,v,g_s)$ denote those used
in the present paper. The two normalizations are related by
\begin{equation}\label{eq:rescaling-Iwaki}
  t=az,\qquad
  q=\frac{u}{2a^2},\qquad
  p=2a^2v,\qquad
  \hbar=g_s,
\end{equation}
where $a$ is a fixed fifth root of $-3/2$.
When a real value of $a$ is required, we take
$a=-\left(3/2\right)^{1/5}$.
Indeed, under \eqref{eq:rescaling-Iwaki}, the Painlev\'e I equation
\begin{equation}
  \hbar^2\frac{\dd^2q}{\dd t^2}=6q^2+t
\end{equation}
is transformed into \eqref{eq:PI-gs}.

For the associated isomonodromy system and its spectral curve,
the corresponding change from the present normalization to
that of \cite{Iwaki2019} is
\begin{equation}
  (x,y)\longmapsto
  \left(\frac{x}{2a^2},\,2a^2y\right).
\end{equation}
This change preserves the initial differential $\omega_{0,1}$ of topological
recursion, and consequently, 
the filling fraction \eqref{eq:A-period-condition} is also preserved.
We may therefore identify the parameter $\nu$ used in the present
paper with that of \cite{Iwaki2019}, once the marking and
orientation of the $A$-cycle have been fixed.
\end{rem}

\begin{rem}
\label{rem:unit-coupling-parameters}
For comparison with the unit-coupling convention of
Section~\ref{sec:bssv}, the rescaling of equation
\eqref{eq:PI-gs}, leading to rescaled variable $\widehat z$ and rescaled solution $\widehat u(\widehat z)$, is accompanied by the parameter rescaling
\[
 \widehat\nu=\frac{\nu}{g_s},
 \qquad
 \widehat\rho=\frac{\rho}{g_s}.
\]
In the Zak representation \eqref{eq:IM-Zak}, these variables
give unit shifts $\widehat\nu\mapsto\widehat\nu+k$ and
Fourier weights $e^{2\pi i k\widehat\rho}$, as can be seen for example in equation \eqref{eq:pinched-Fourier}. Note that in Sections~\ref{sec:bssv} and~\ref{sec:resurgence-modularity},
we use the rescaled variables and omit the hats from
$(\widehat z,\widehat u,\widehat\nu,\widehat\rho)$.
In the present subsection and
Section~\ref{sec:modular-transformation}, we retain explicit
$g_s$ dependence.

\end{rem}

\begin{rem} 
The TR construction above is formal. Its interpretation through analytic
solutions and linear Stokes multipliers uses the Borel summability,
Fourier convergence, and WKB connection assumptions of
\cite[Section~2.2]{IwakiMarino2024}. We adopt these assumptions wherever
Borel-resummed TR wave functions or tau-functions are used, including
Section~\ref{sec:modular-transformation}; the formal construction does
not by itself establish them.
\end{rem}

\subsubsection{The jump associated with a $B$-cycle saddle connection}
\label{sec:linear-Stokes-preliminaries}
The Stokes graph associated with
\eqref{eq:PI-isomonodromy-system} is described by the quadratic
differential 
\begin{equation}
\left( \frac{x^3}{18}-\frac{zx}{6}-\frac{h(z,\nu)}3 \right) \, dx^2.
\end{equation}
Its edges emanate from the simple turning
points $e_j$ and satisfy, on a locally chosen sheet,
\begin{equation}\label{eq:linear-Stokes-trajectories}
 \operatorname{Im}\!\left(\int_{e_j}^x y\,\dd x\right)=0.
\end{equation}
A saddle connection joins two turning
points; a graph without such connections is called saddle-free.
A change in the topology when a saddle connection appears and resolves
is called a mutation.

We call this graph the \emph{linear Stokes graph}, and the Stokes
multipliers of the associated isomonodromy linear system the
\emph{linear Stokes multipliers}. This terminology distinguishes them
from the nonlinear Stokes phenomenon of Painlev\'e~I, which changes the
trans-series description of a Painlev\'e solution. Isomonodromy preserves
the linear Stokes data, even though their expressions in the
parameters used before and after a mutation may differ. Comparing these
expressions is the method used below.

For the numerical computation of the linear Stokes graphs displayed in
this paper, we set $\nu=1$ and approximate $h(z,\nu)$ by truncating its
large-$z$ asymptotic expansion after the first six terms:
\begin{equation}\label{eq:H0-truncated}
  \begin{aligned}
  h(z,\nu) \simeq{}&
  -\frac13 z^{3/2}
  - \sqrt{6}\,\nu z^{1/4}
  - \frac{5\nu^2}{8}z^{-1}
  + \frac{235\nu^3}{192\sqrt{6}}z^{-9/4}\\
  &-\frac{38585\nu^4}{55296}z^{-7/2}
  + \frac{2663129\nu^5}{884736\sqrt{6}}z^{-19/4}.
  \end{aligned}
\end{equation}
This expansion is determined by imposing the period condition
\eqref{eq:A-period-condition}.
It can be derived by the same term-wise residue method as that employed in~\cite{IILZ2025}. 

The choice $\nu=1$ is only for visualization: the multiplier formulas below
retain general $\nu$ and $\rho$, as long as the Stokes graph has the same
topological configuration and the normalization is transported consistently.

\begin{rem} \label{rem:leading-branch}
Note that the resulting expansion is
different from the one presented there, because we choose a different
degenerating $A$-cycle, and hence a different degeneration branch of
the spectral curve. Our choice is characterized by
\begin{equation}
  u(z)\sim+\sqrt{z}
  \qquad (z\longrightarrow\infty),
\end{equation}
or, equivalently, by the degeneration of
\eqref{eq:PI-spectral-curve} to a curve with a double point at
$x=+\sqrt{z}$. The choice made in~\cite{IILZ2025}, on the other hand,
corresponds to
\begin{equation}
  u(z)\sim-\sqrt{z}
  \qquad (z\longrightarrow\infty).
\end{equation}
\end{rem}

\begin{figure}[htbp]
 \centering
 \begin{minipage}[t]{0.31\textwidth}
  \centering
  \paperfigure[width=\linewidth]{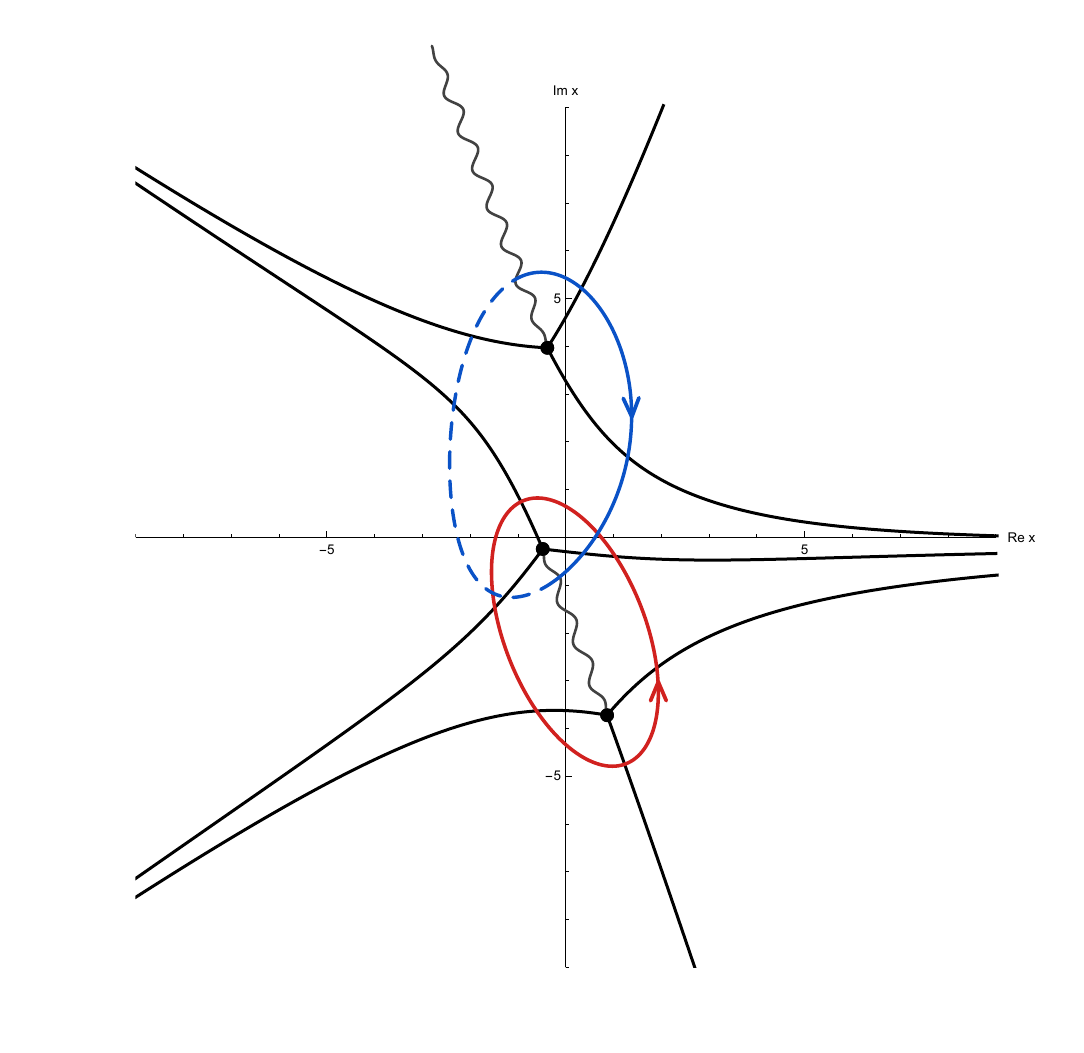}\\[2mm]
  $z_-=5e^{-9\pi i/10}$
 \end{minipage}\hfill
 \begin{minipage}[t]{0.31\textwidth}
  \centering
  \paperfigure[width=\linewidth]{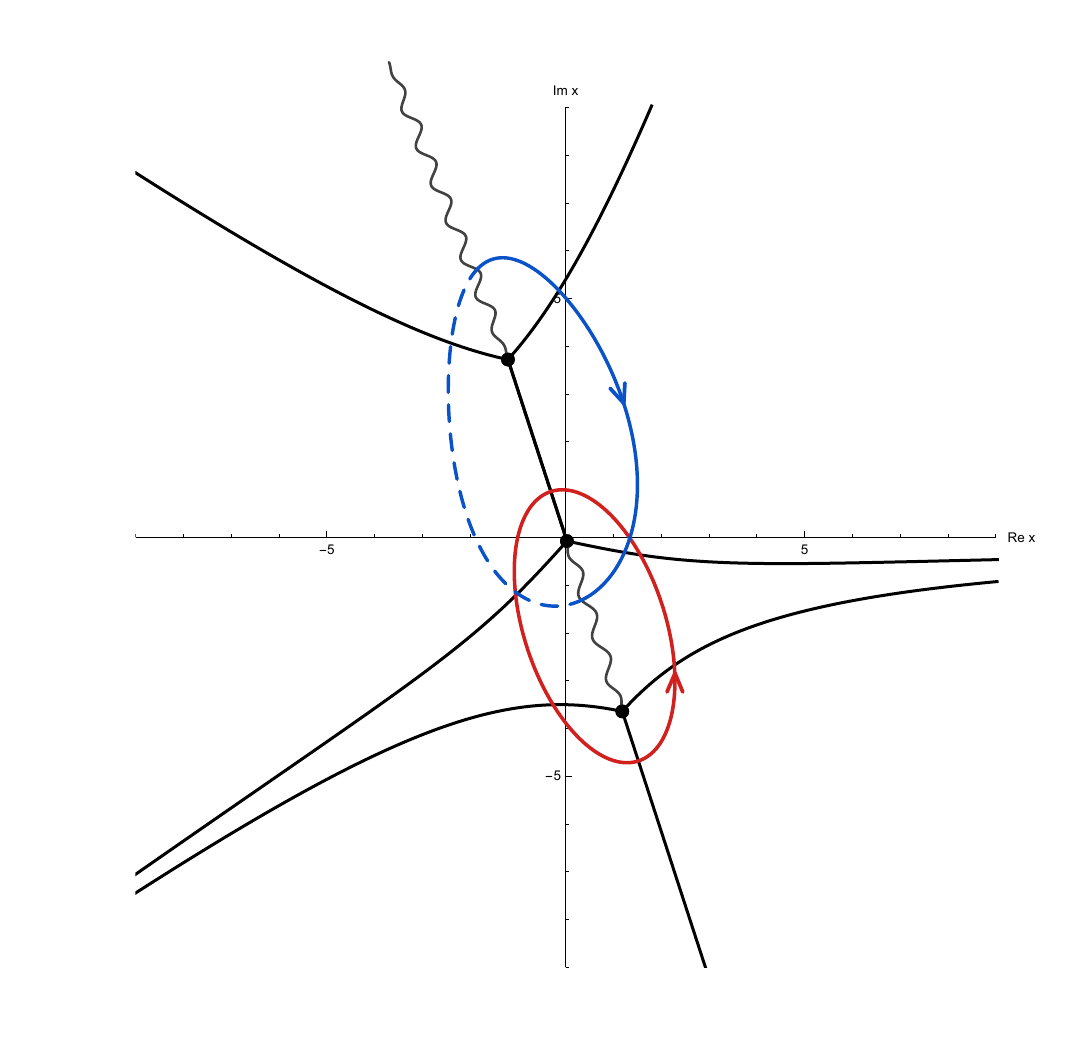}\\[2mm]
  $z=5e^{-4\pi i/5}$
 \end{minipage}\hfill
 \begin{minipage}[t]{0.31\textwidth}
  \centering
  \paperfigure[width=\linewidth]{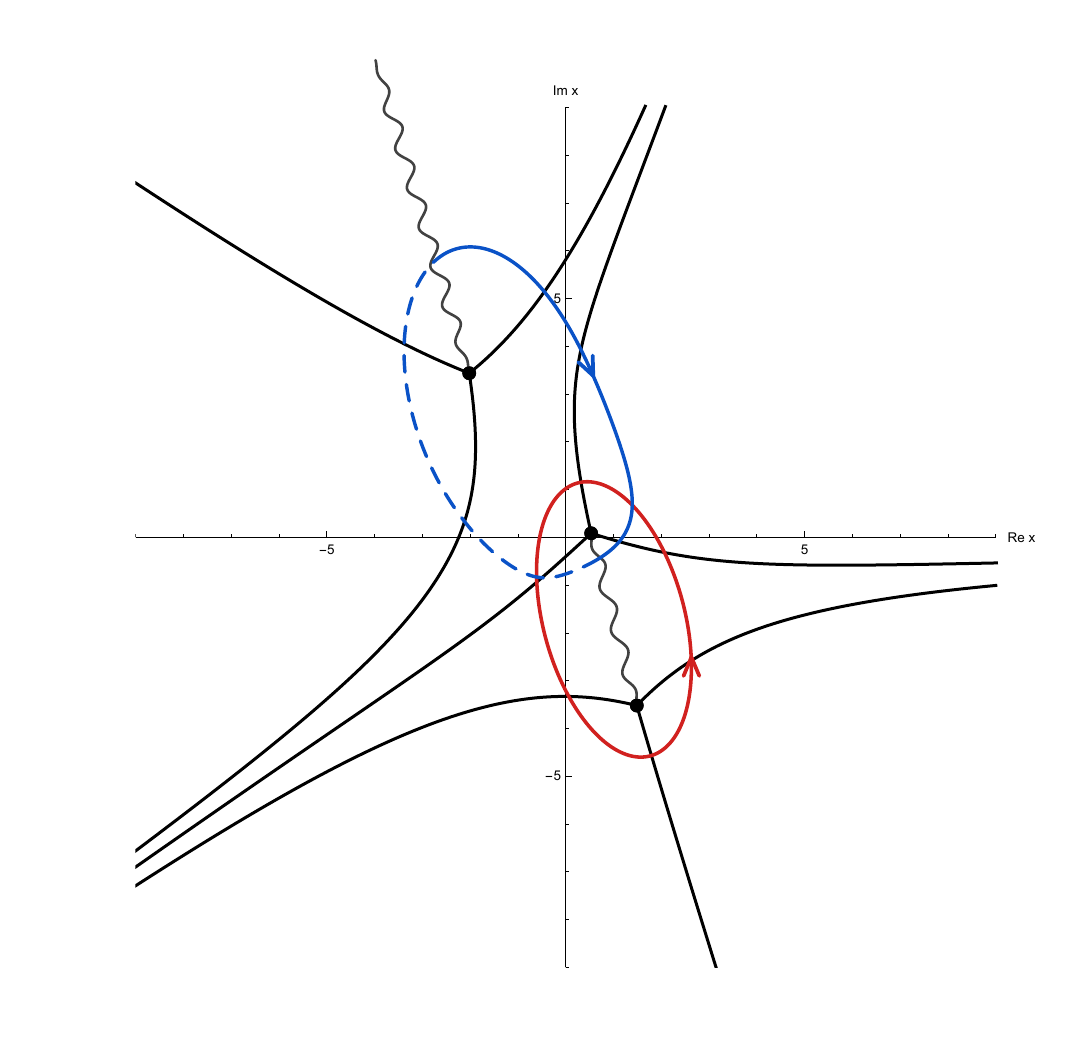}\\[2mm]
  $z_+=5e^{-7\pi i/10}$
 \end{minipage}
 \caption{Mutation of the linear Stokes graph across a $B$-cycle saddle
 connection. 
 } 
 \label{fig:IM-B-mutation}
\end{figure}

Consider a local deformation of \eqref{eq:PI-spectral-curve} across a wall
where a single saddle connection joins two simple turning points.
Its lift to the double cover represents the $B$-cycle, with orientation
chosen so that $\oint_B y\,\dd x$ is positive real on the wall.
The two nearby configurations are saddle-free and differ by a mutation.
Figure~\ref{fig:IM-B-mutation} depicts the linear Stokes graphs on either
side of the Stokes direction $\arg z=-4\pi/5$ in the nonlinear sense.
More precisely, it displays the linear Stokes graphs at
$z_\pm=5e^{i(-4\pi/5\pm\pi/10)}$, together with the Stokes graph at
$z=5e^{-4\pi i/5}$, which contains a saddle connection.
The wavy lines indicate branch cuts, while the red and blue curves
represent the $A$-cycle and $B$-cycle, respectively.
The solid and dashed portions of these cycles lie on the first and second
sheets of the spectral curve, respectively.

We use the sector convention $\arg x=2\pi k/5$ and the red $A$-cycle
and blue $B$-cycle shown in Figure~\ref{fig:IM-B-mutation}, with the
orientations and sheet conventions indicated there.
We use the same convention as in \cite{Iwaki2019}. Writing the ordered
canonical pair in the reference sectors as
$\boldsymbol\Psi^{(k)}=(\psi_+^{(k)},\psi_-^{(k)})$,
we define the Stokes matrices by\footnote{
The direction of the relation in equations~(2.16)--(2.17) of the published version
of \cite{IwakiMarino2024} is a typographical error and should be reversed
as in \eqref{eq:canonical-connection-convention}.
}
\begin{equation}\label{eq:canonical-connection-convention}
 \boldsymbol\Psi^{(k)}=\boldsymbol\Psi^{(k+1)}S_k,\qquad
 S_k=\begin{cases}
 \begin{pmatrix}1&0\\s_k&1\end{pmatrix},&k=0,\pm2,\\[+1.5em]
 \begin{pmatrix}1&s_k\\0&1\end{pmatrix},&k=\pm1.
 \end{cases}
\end{equation}
The sectors are ordered counterclockwise, and the matrices act on the right. 

The exact WKB connection rules therefore give the following multipliers.
The parameters
$(\nu_-,\rho_-)$ and $(\nu_+,\rho_+)$ describe the solutions at $z_-$
and $z_+$, respectively. We denote the corresponding linear Stokes
multipliers by $s^-_k(\nu_-,\rho_-)$ and $s^+_k(\nu_+,\rho_+)$:
\begin{equation}
\renewcommand{\arraystretch}{1.7}
\setlength{\arraycolsep}{5pt}
\begin{array}{c|c|c}
 k&s^-_k(\nu_-,\rho_-)\quad(z=z_-)&s^+_k(\nu_+,\rho_+)\quad(z=z_+)\\ \hline
 -2&i\,\bigl(e^{-2\pi i\rho_-/g_s}-e^{-2\pi i(\nu_-+\rho_-)/g_s}\bigr)
 &i\,\bigl(e^{-2\pi i\nu_+/g_s}-e^{-2\pi i(\nu_++\rho_+)/g_s}+e^{-2\pi i\rho_+/g_s}\bigr)\\
 -1&i\,e^{2\pi i\rho_-/g_s}&i\,e^{2\pi i\rho_+/g_s}\\
 0&i\,\bigl(e^{2\pi i\nu_-/g_s}-e^{2\pi i(\nu_--\rho_-)/g_s}+e^{-2\pi i\rho_-/g_s}\bigr)
 &i\,e^{-2\pi i\rho_+/g_s}(1-e^{2\pi i\nu_+/g_s})\\
 1&i\,e^{-2\pi i\nu_-/g_s}
 &i\,e^{-2\pi i\nu_+/g_s}(1-e^{2\pi i\rho_+/g_s})\\
 2&i\,e^{2\pi i\nu_-/g_s}(1-e^{2\pi i\rho_-/g_s})
 &i\,e^{2\pi i\nu_+/g_s}
\end{array}
 \label{eq:IM-multiplier-table}
\end{equation}
Both columns satisfy the cyclic consistency relations
\begin{equation}\label{eq:Stokes-cyclic-consistency}
 s^\pm_{k+5}=s^\pm_k,\qquad
 1+s^\pm_{k-1}s^\pm_k+i\,s^\pm_{k+2}=0.
\end{equation}

For the descriptions before and after the mutation to represent the same Painlev\'e
solution, isomonodromy requires equality of their linear Stokes multipliers.
Comparing the entries for $k=-1$ and $k=2$ first gives
\begin{equation}
 e^{2\pi i\nu_+/g_s}
 =e^{2\pi i\nu_-/g_s}(1-e^{2\pi i\rho_-/g_s}),
 \qquad e^{2\pi i\rho_+/g_s}=e^{2\pi i\rho_-/g_s}.
 \label{eq:IM-DDP}
\end{equation}
Consequently, away from $e^{2\pi i\rho_-/g_s}=1$ and with a consistent logarithm,
we have
\begin{equation} \label{eq:IM-matching-solution}
u_+(z, \nu_+, \rho_+) 
= u_-(z, \nu_-, \rho_-) 
\end{equation}
with 
\begin{equation}
 \nu_+=\nu_- +\frac{g_s}{2\pi i}\log(1-e^{2\pi i\rho_-/g_s}),
 \qquad 
 \rho_+=\rho_-.
 \label{eq:IM-matching-parameters}
\end{equation}
Here \(u_\pm\) denote the solutions obtained, as in
\cite{IwakiMarino2024}, by Borel resumming the formal series in \(g_s\)
near \(z_\pm\), respectively.
Relation \eqref{eq:IM-matching-solution} describes how their analytic continuations are related across the Stokes direction \(\arg z=-4\pi/5\). We expect that, under suitable conditions, Borel resummation in \(g_s\) of the formal solutions constructed by topological recursion and Borel resummation of the formal expansions in \(z\) appearing in the transseries solutions of the preceding section yield the same exact solutions.
Indeed, setting \(g_s=1\) in \eqref{eq:IM-matching-parameters}, the map $(\nu_-, \rho_-) \mapsto (\nu_+, \rho_+)$ reproduces the formula for $S_\pi^{-1}$ in \eqref{eq:resurgent-Spi} of the preceding section. In what follows, we assume this identification.

The multipliers determine these parameters modulo $g_s\mathbb Z$. 
This ambiguity affects the representation of the tau-function, 
but it does not affect the Painlev\'e transcendents obtained 
by taking its logarithmic derivatives. 
See Remark \ref{rem:monodromy-parameter-branches} below for further details.

Thus the mutation changes the coordinates used to describe the same monodromy data.
In particular, equality of the linear Stokes data determines the
parameter transformation, but does not by itself determine the
parameter-dependent normalization of the tau-function. That additional
normalization supplies the dilogarithmic factor in the jump formula.

We retain the convention of \cite{IwakiMarino2024} for $\mathfrak S_B$.
For the present $z_\pm$, this operator describes the crossing from the
$+$ side to the $-$ side; the crossing from $z_-$ to $z_+$ uses its inverse.
The jump formula reads directly in terms of $\nu$ and $\rho$:
\begin{equation}
    \begin{aligned}
 \mathfrak S_B\tau(z,\nu,\rho;g_s)
 &=\exp\!\left(\frac{\operatorname{Li}_2(e^{2\pi i\rho/g_s})}{2\pi i}\right)\\
 &\quad\times\tau\!\left(z,\nu-\frac{g_s}{2\pi i}
 \log(1-e^{2\pi i\rho/g_s}),\rho;g_s\right).
 \end{aligned}
 \label{eq:IM-tau-jump}
\end{equation}

For the crossing from $z_-$ to $z_+$, the inverse operator is
\begin{equation}
 \begin{aligned}
 \mathfrak S_B^{-1}\tau(z,\nu,\rho;g_s)
 &=\exp\!\left(-\frac{\operatorname{Li}_2(e^{2\pi i\rho/g_s})}{2\pi i}\right)\\
 &\quad\times\tau\!\left(z,\nu+\frac{g_s}{2\pi i}
 \log(1-e^{2\pi i\rho/g_s}),\rho;g_s\right).
 \end{aligned}
 \label{eq:IM-inverse-jump}
\end{equation}

The dilogarithmic multiplier uses the extended tau-function normalization
of \cite{IwakiMarino2024}, characterized here by
\begin{equation}
 \dd\!\left(-\frac{\operatorname{Li}_2(e^{2\pi i\rho_-/g_s})}{2\pi i}\right)
 =-\frac{2\pi i}{g_s^2}
 (\nu_-\,\dd\rho_- -\nu_+\,\dd\rho_+),
 \label{eq:IM-normalization}
\end{equation}
where $\dd$ differentiates monodromy parameters at fixed $g_s$.
Together with the Hamiltonian definition of the tau-function, this relation
determines \eqref{eq:IM-tau-jump} or \eqref{eq:IM-inverse-jump} up to a multiplicative constant
independent of $z,\nu,\rho$ (with $g_s$ held fixed).
Since we are comparing the two lateral realizations of the same normalized
formal tau-function across a Stokes direction, 
they must therefore have the same leading asymptotic expansion, 
including its multiplicative normalization. 
Matching the common normalized leading sector fixes the remaining constant to $1$, 
as used in \eqref{eq:IM-tau-jump} and \eqref{eq:IM-inverse-jump}.
This uses the common lateral Borel-sum normalization, rather than merely
the equality of the associated Painlev\'e solutions.
The dilogarithmic multiplier itself disappears under the $z$-derivatives
defining $u$.
The analytic interpretation assumes the Borel summability and Fourier
convergence hypotheses stated in \cite[Section~2.2]{IwakiMarino2024}.

\begin{rem}%
\label{rem:monodromy-parameter-branches}
Throughout Sections~\ref{sec:stokes-review} and
\ref{sec:modular-transformation}, equality of the multiplier formulas first
identifies the exponentials of $2\pi i\nu/g_s$ and $2\pi i\rho/g_s$.
The additive parameters are therefore determined modulo $g_s\mathbb Z$.
For the Fourier normalization \eqref{eq:IM-Zak}, reindexing the sum gives
\begin{equation}\label{eq:tau-lattice-shifts}
 \tau(z,\nu+m g_s,\rho+n g_s;g_s)
 =e^{-2\pi i m\rho/g_s}\tau(z,\nu,\rho;g_s),\qquad m,n\in\mathbb Z,
\end{equation}
whenever reindexing is legitimate, formally or under the stated convergence assumptions. 
This factor is independent of $z$ and does not change $u$ or $\mathcal H$. 
\end{rem}

\section{Modular Transformation and Rotational Symmetry}
\label{sec:modular-transformation}

In what follows, following the proposal of \cite{Iwaki2019}, we compute
the Stokes multipliers of the isomonodromy linear equation for various
values of \(z\).  In the latter part, we use these results to investigate
the modular properties of the Painlev\'e function.

We first recall the rotational symmetry in
Section~\ref{sec:Z5-symmetry-Stokes}, and then discuss the \(T\)-transformation.  We continuously rotate the
Painlev\'e variable \(z\) and follow the Stokes graph and 
the Torelli marking along this deformation.  
We then consider the \(S\)-transformation, for which a different rotation of \(z\), 
together with the corresponding rotation of the Stokes sectors in the \(x\)-plane, 
realizes the \(S\)-action on the tau-function parameters $(\nu,\rho)$.

\subsection{The ${\mathbb Z}_5$-symmetry and Stokes multipliers}
\label{sec:Z5-symmetry-Stokes}

We also recall here the rotational ${\mathbb Z}_5$-symmetry 
that will play an important role in the discussion below. 
The same idea has already been used in previous works
including \cite{Kapaev2004, LisovyyRoussillon2017, vanSpaendonckVonk2022}. 

Let
\[
  \zeta_n:=e^{2\pi i n/5},
  \qquad n\in\mathbb Z.
\]
If \(u(z)\) is a solution of Painlev\'e I, then
\begin{equation}
  \widetilde{u}(z)
  :=
  \zeta_n^2 u(\zeta_n z)
  \label{eq:Z5-action-u}
\end{equation}
is again a solution of Painlev\'e I.  
In the Hamiltonian formulation, the transformation law 
for the other canonical variable \(v\) is given by
\begin{equation}
  \widetilde{v}(z)
  :=
  \zeta_n^3 v(\zeta_n z)
\label{eq:Z5-action-v}
\end{equation}
Thus Painlev\'e~I admits a natural \(\mathbb Z_5\)-rotational symmetry.

This symmetry lifts to the associated linear isomonodromy system \eqref{eq:PI-isomonodromy-system}.  
More precisely, let $ \psi(x,z;u(z),v(z))$ 
denote a solution of the isomonodromy linear equations \eqref{eq:PI-isomonodromy-system}
associated with the solution $(u(z),v(z))$ of the Hamiltonian system \eqref{eq:PI-Hamiltonian-system} 
which is equivalent to Painlev\'e I.  
Then 
\begin{equation}
  \widetilde{\psi}\bigl(x,z;\widetilde{u}(z), \widetilde{v}(z)\bigr)
  :=
  \psi\bigl(\zeta_n^{-2} x,\zeta_n z;u(\zeta_n z), v(\zeta_n z)\bigr)
  \label{eq:Z5-wavefunction}
\end{equation}
is a solution of the isomonodromy system \eqref{eq:PI-isomonodromy-system} 
associated with the solution $(\widetilde{u}(z), \widetilde{v}(z))$ 
given in \eqref{eq:Z5-action-u} and \eqref{eq:Z5-action-v}.

An immediate consequence is that the linear monodromy data are preserved
up to a cyclic relabeling of the Stokes sectors.
Let \(s_k\) and \(\widetilde{s}_k\) denote the Stokes multipliers of
\[
  \psi(x,z;u(z),v(z))
  \qquad\text{and}\qquad
  \widetilde{\psi}(x,z;\widetilde{u}(z),\widetilde{v}(z)),
\]
respectively, associated with the direction
\[
  \arg x=\frac{2\pi k}{5},
  \qquad
  k\in\mathbb Z/5\mathbb Z.
\]
Since the rotation of the \(x\)-plane induces a cyclic shift of the
Stokes sectors by \(2n\) units, we obtain
\begin{equation}
  \boxed{
    s_k=\widetilde{s}_{k+2n},
    \qquad
    k\in\mathbb Z/5\mathbb Z.
  }
  \label{eq:Z5-Stokes-multipliers}
\end{equation}
In other words, the \(\mathbb Z_5\)-symmetry of Painlev\'e~I acts on the
linear monodromy data by a cyclic relabeling of the five Stokes
multipliers.

This observation will be used in the next two subsections.  To compare Painlev\'e
solutions associated with two points in the \(z\)-plane related by a
\(\mathbb Z_5\)-rotation, we first use
\eqref{eq:Z5-Stokes-multipliers} to identify the corresponding Stokes
sectors and then compare the associated Stokes multipliers.  Together
with the change of Torelli marking obtained by continuously tracking the
cycles, this comparison determines the corresponding transformation of
the parameters \((\nu,\rho)\).


\subsection{The \(T\)-transformation}
\label{sec:T-transformation-Stokes}

\subsubsection{Rotation of linear Stokes graphs and Torelli markings}

We now apply the framework of Section~\ref{sec:stokes-review} to
rotations of $z$. We use the curve \eqref{eq:PI-spectral-curve} with
its marking transported continuously, and take $|z|$ sufficiently large.

\begin{figure}[htbp]
  \centering

  \begin{minipage}{0.19\textwidth}
    \centering
    \paperfigure[width=\linewidth]{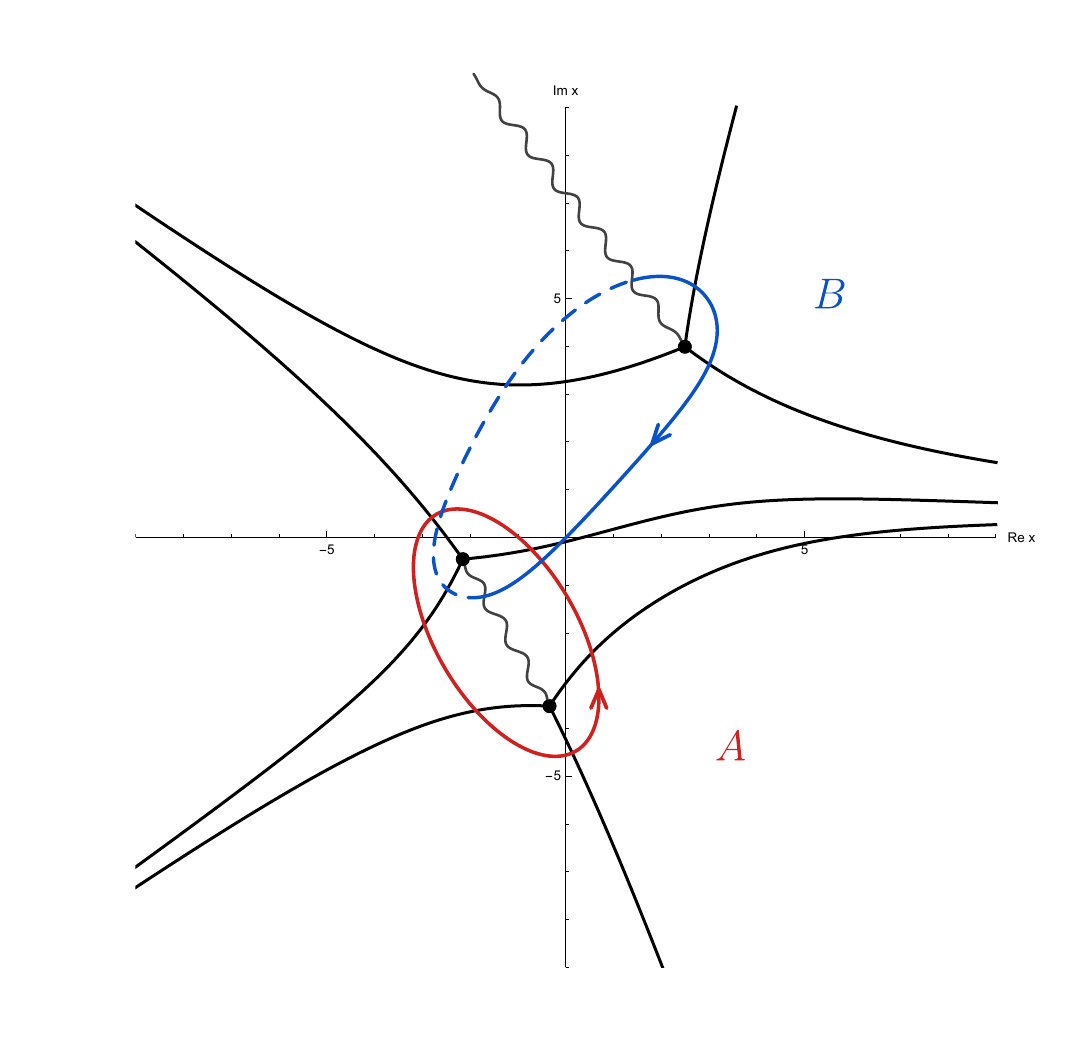} \\ {\tiny $\arg z = -13\pi/10$}
  \end{minipage}
  \begin{minipage}{0.19\textwidth}
    \centering
    \paperfigure[width=\linewidth]{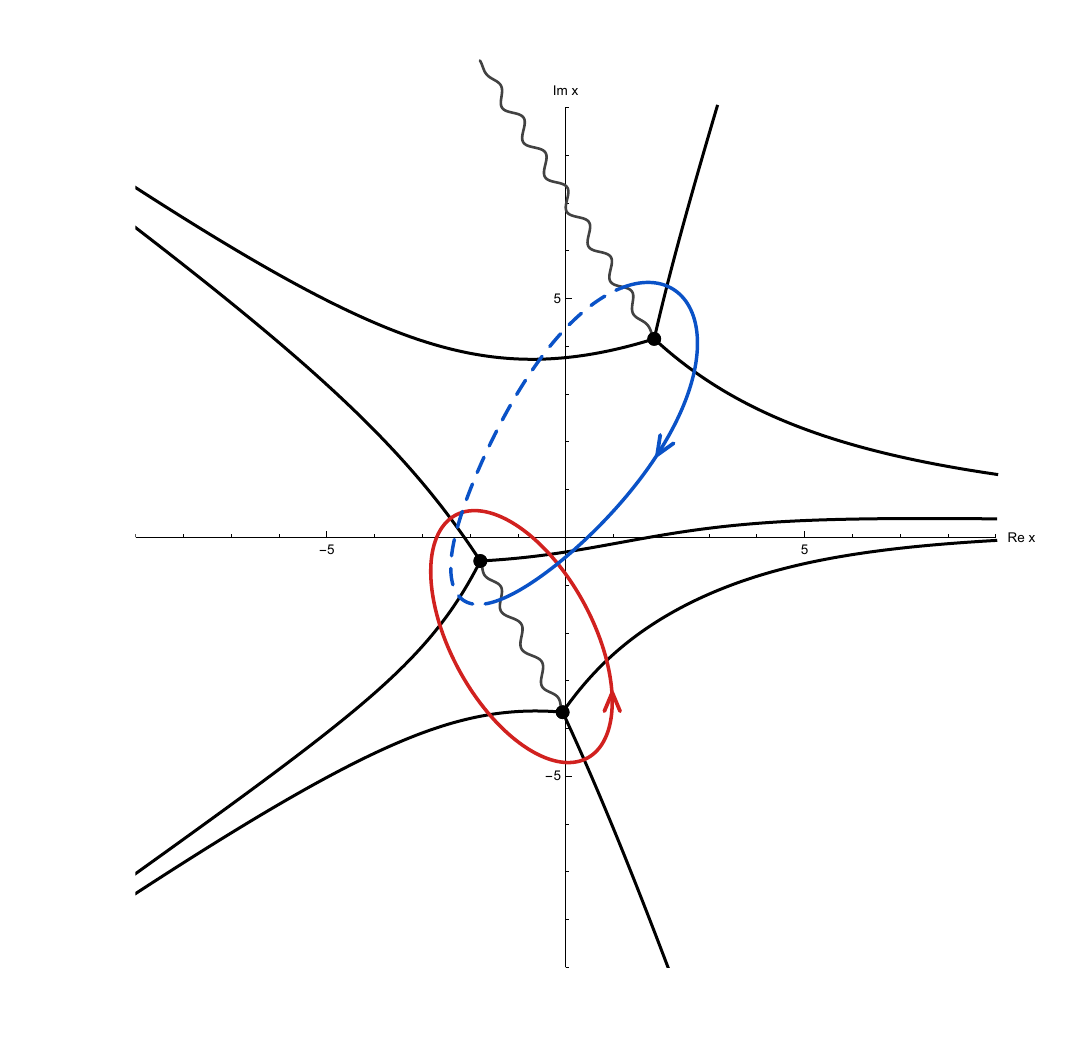} \\ {\tiny $\arg z = -12\pi/10$}
  \end{minipage}
  \begin{minipage}{0.19\textwidth}
    \centering
    \paperfigure[width=\linewidth]{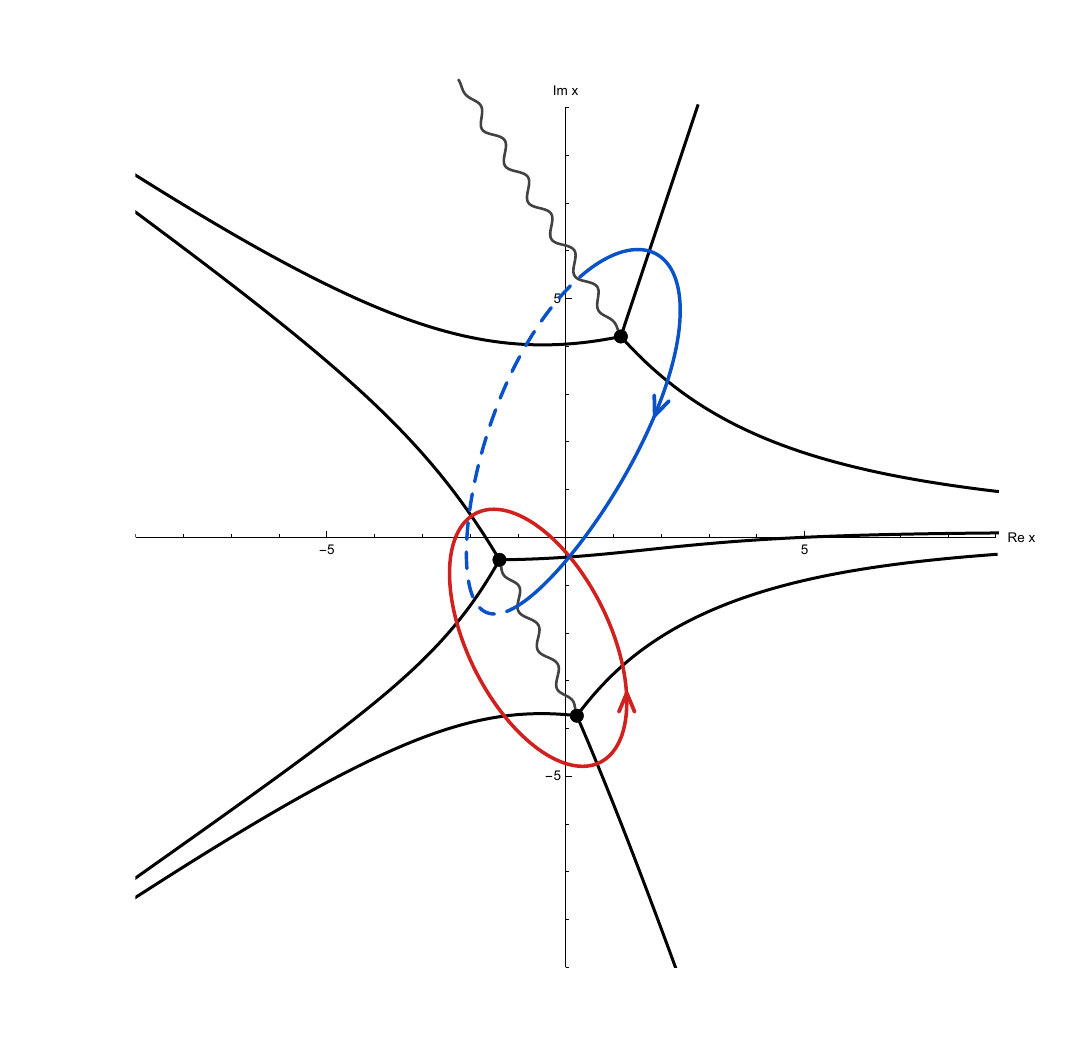} \\ {\tiny $\arg z = -11\pi/10$}
  \end{minipage}
  \begin{minipage}{0.19\textwidth}
    \centering
    \paperfigure[width=\linewidth]{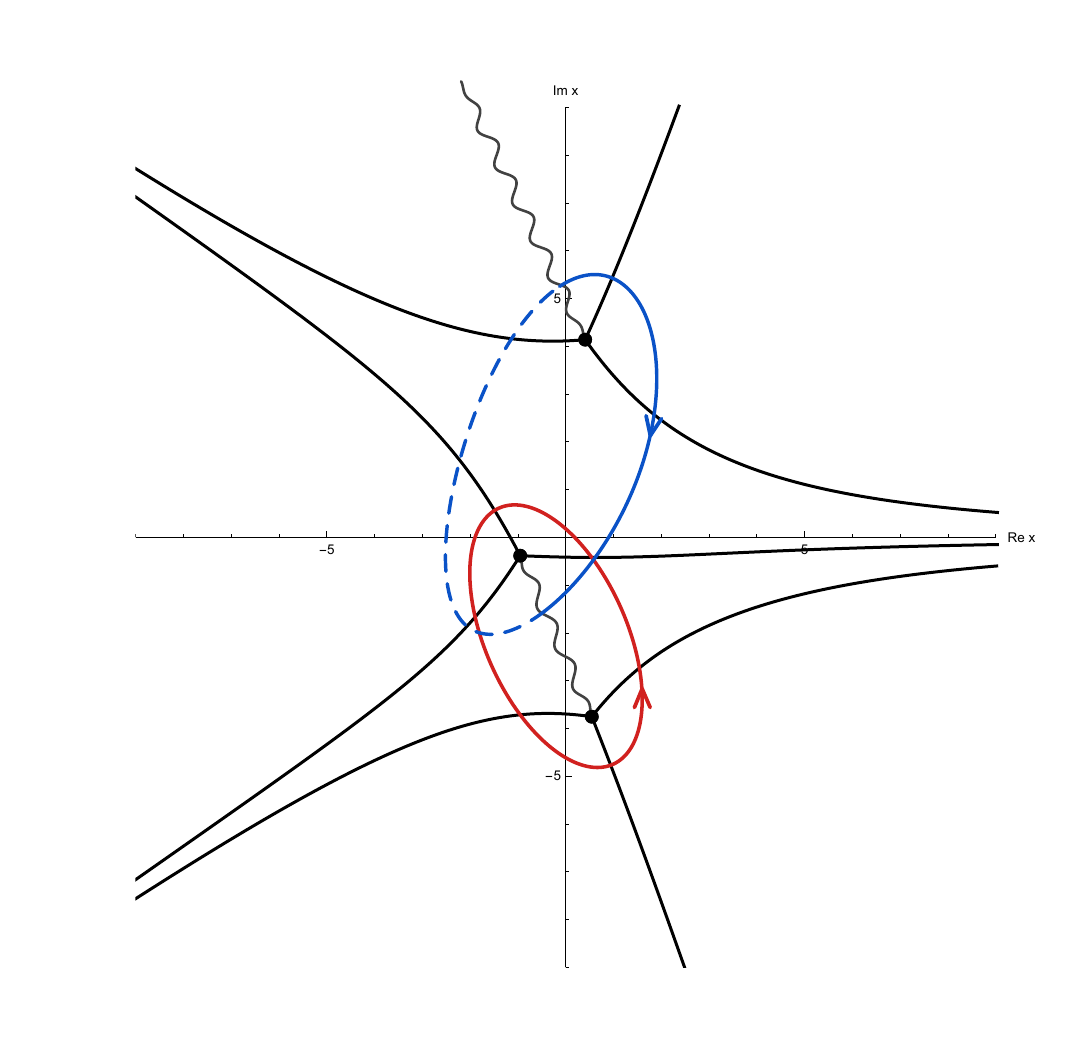} \\ {\tiny $\arg z = -10\pi/10$}
  \end{minipage}
  \begin{minipage}{0.19\textwidth}
    \centering
    \paperfigure[width=\linewidth]{SG3} \\ {\tiny $\arg z = -9\pi/10$}
  \end{minipage}

  \vspace{3mm}

  \begin{minipage}{0.19\textwidth}
    \centering
    \paperfigure[width=\linewidth]{SG4-saddle} \\ {\tiny $\arg z = -8\pi/10$}
  \end{minipage}
  \begin{minipage}{0.19\textwidth}
    \centering
    \paperfigure[width=\linewidth]{SG5} \\ {\tiny $\arg z = -7\pi/10$}
  \end{minipage}
  \begin{minipage}{0.19\textwidth}
    \centering
    \paperfigure[width=\linewidth]{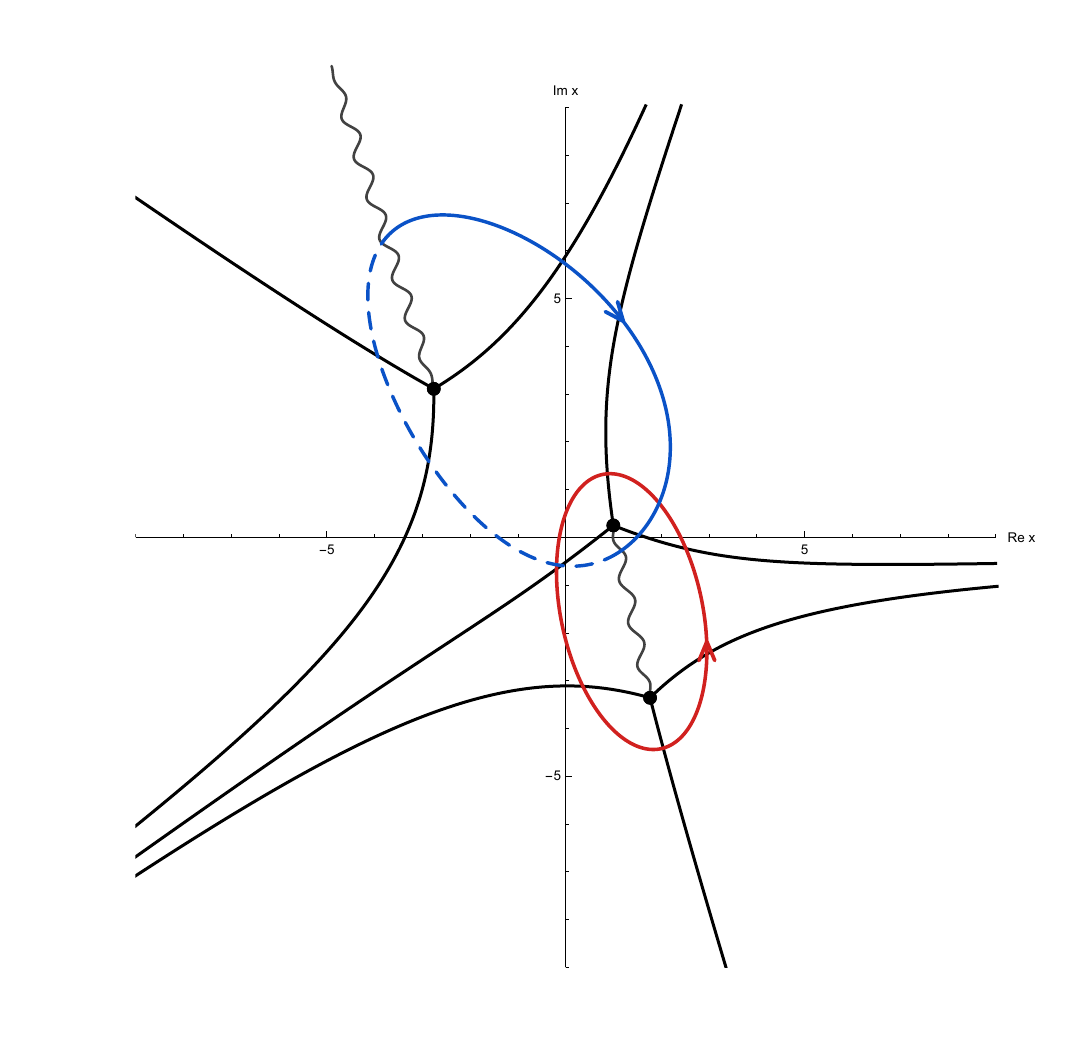} \\ {\tiny $\arg z = -6\pi/10$}
  \end{minipage}
  \begin{minipage}{0.19\textwidth}
    \centering
    \paperfigure[width=\linewidth]{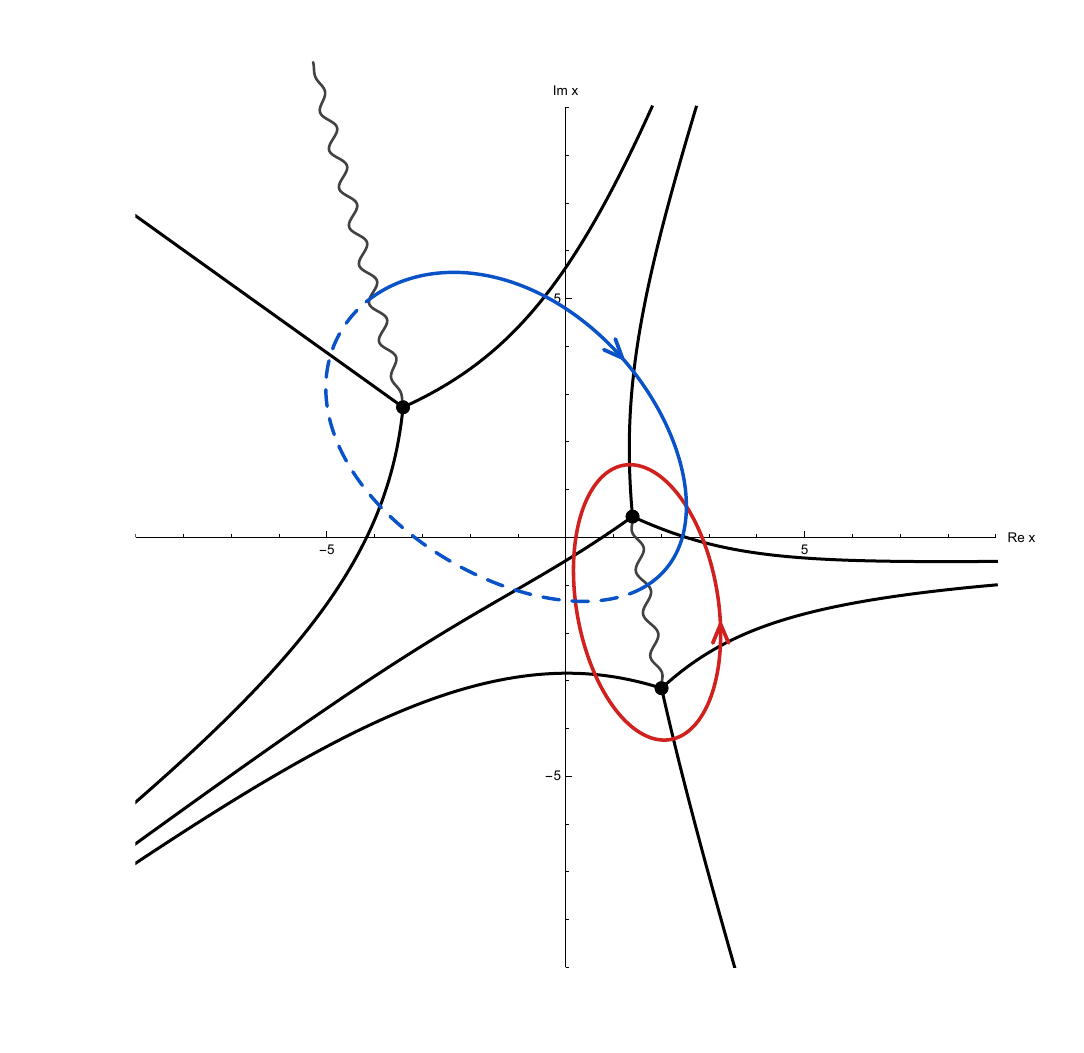} \\ {\tiny $\arg z = -5\pi/10$}
  \end{minipage}
  \begin{minipage}{0.19\textwidth}
    \centering
    \paperfigure[width=\linewidth]{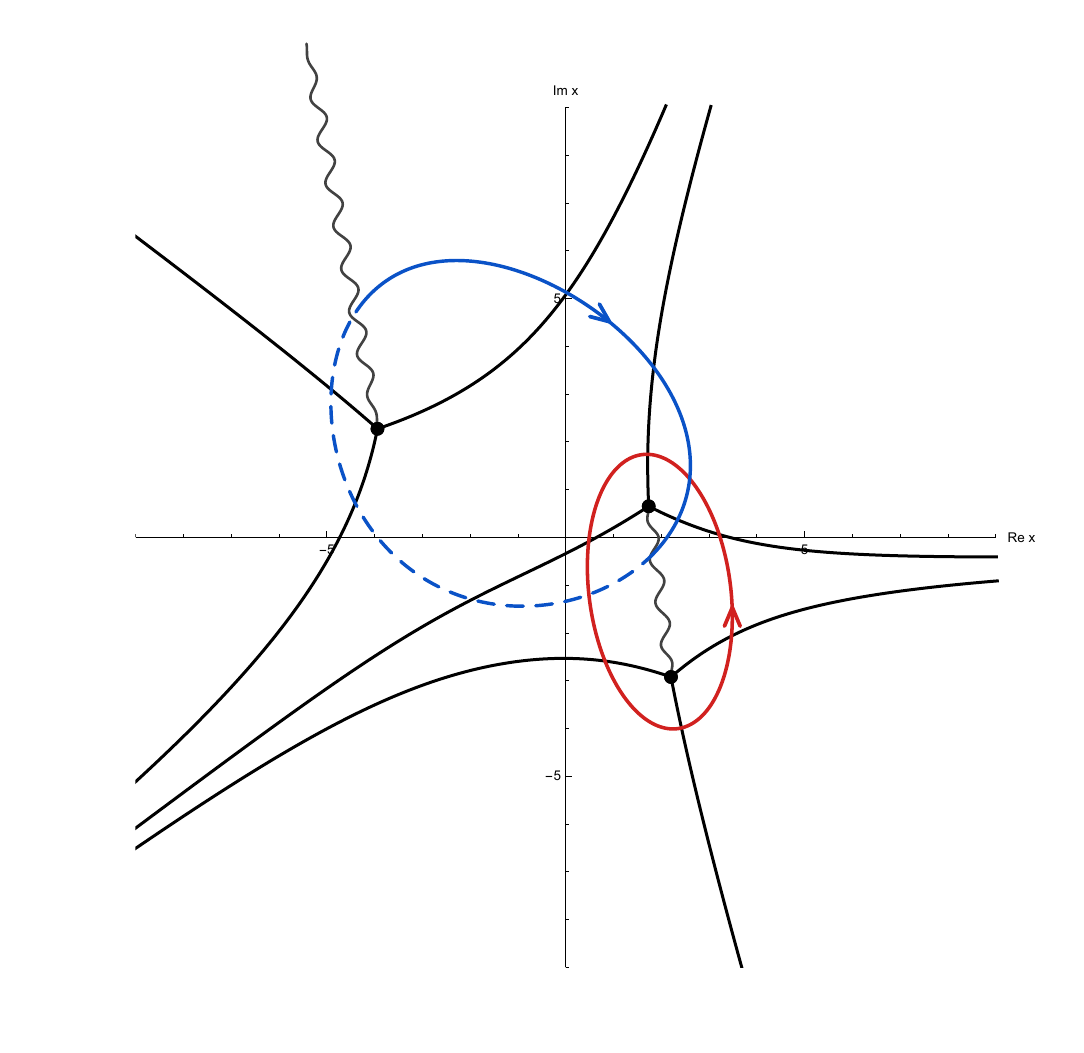} \\ {\tiny $\arg z = -4\pi/10$}
  \end{minipage}

  \vspace{3mm}

  \begin{minipage}{0.19\textwidth}
    \centering
    \paperfigure[width=\linewidth]{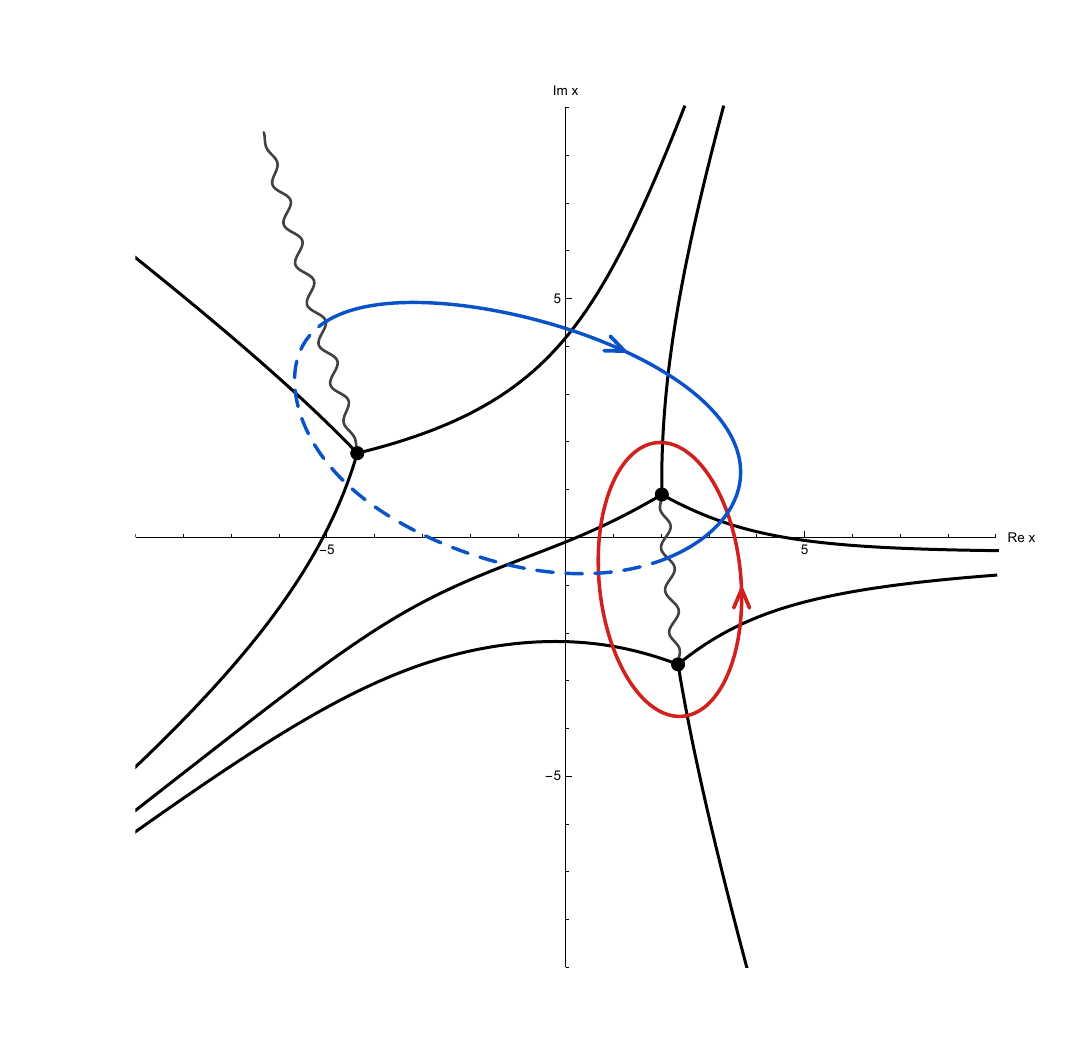} \\ {\tiny $\arg z = -3\pi/10$}
  \end{minipage}
  \begin{minipage}{0.19\textwidth}
    \centering
    \paperfigure[width=\linewidth]{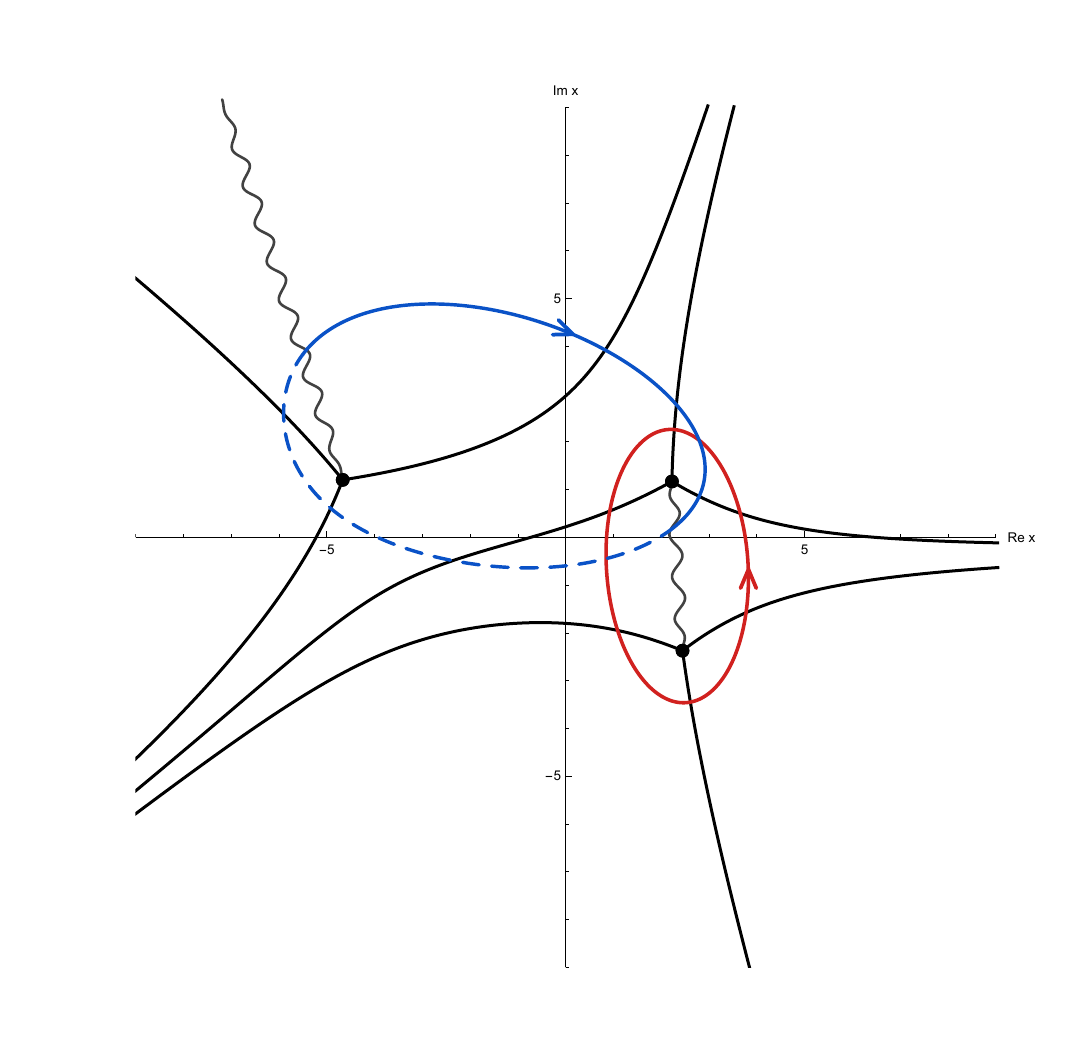} \\ {\tiny $\arg z = -2\pi/10$}
  \end{minipage}
  \begin{minipage}{0.19\textwidth}
    \centering
    \paperfigure[width=\linewidth]{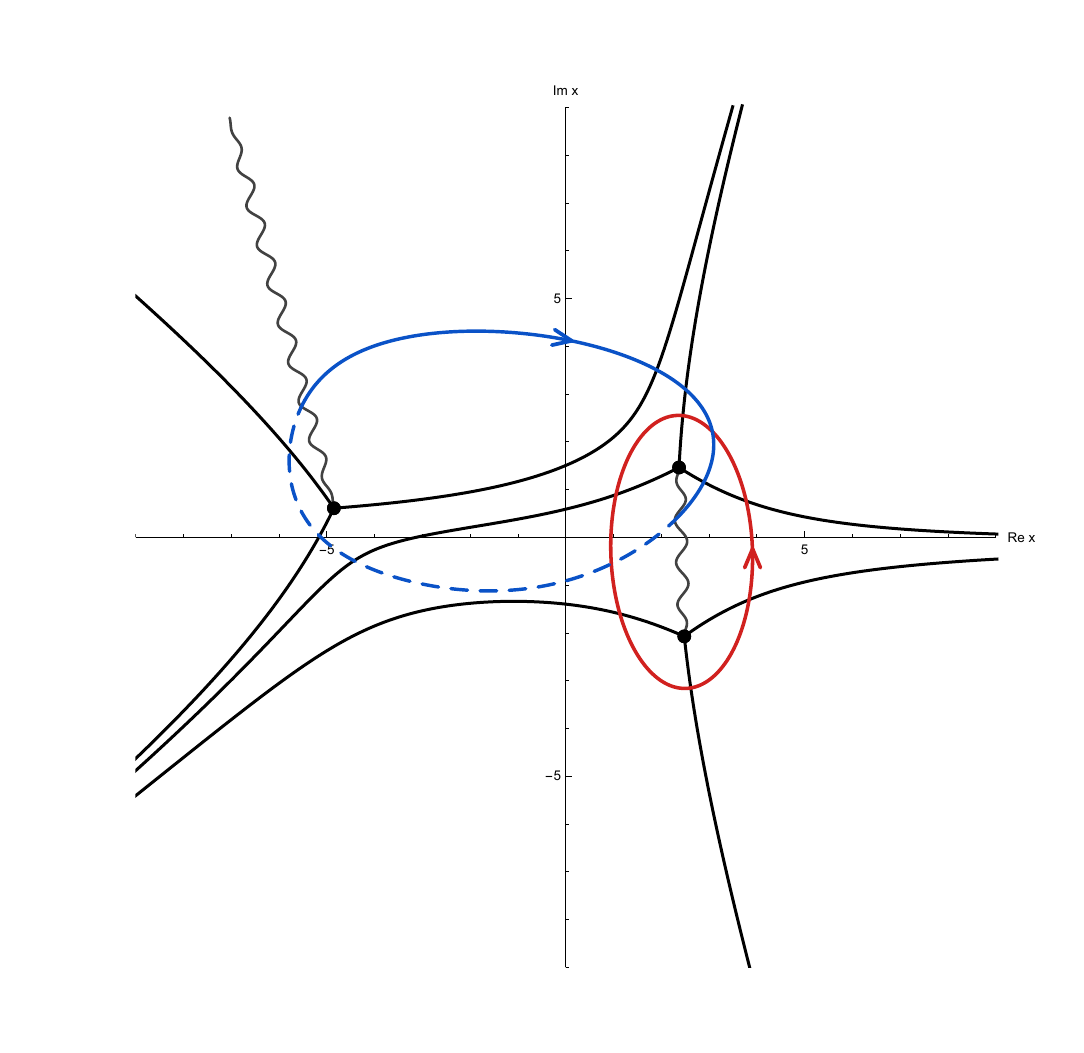} \\ {\tiny $\arg z = -\pi/10$}
  \end{minipage}
  \begin{minipage}{0.19\textwidth}
    \centering
    \paperfigure[width=\linewidth]{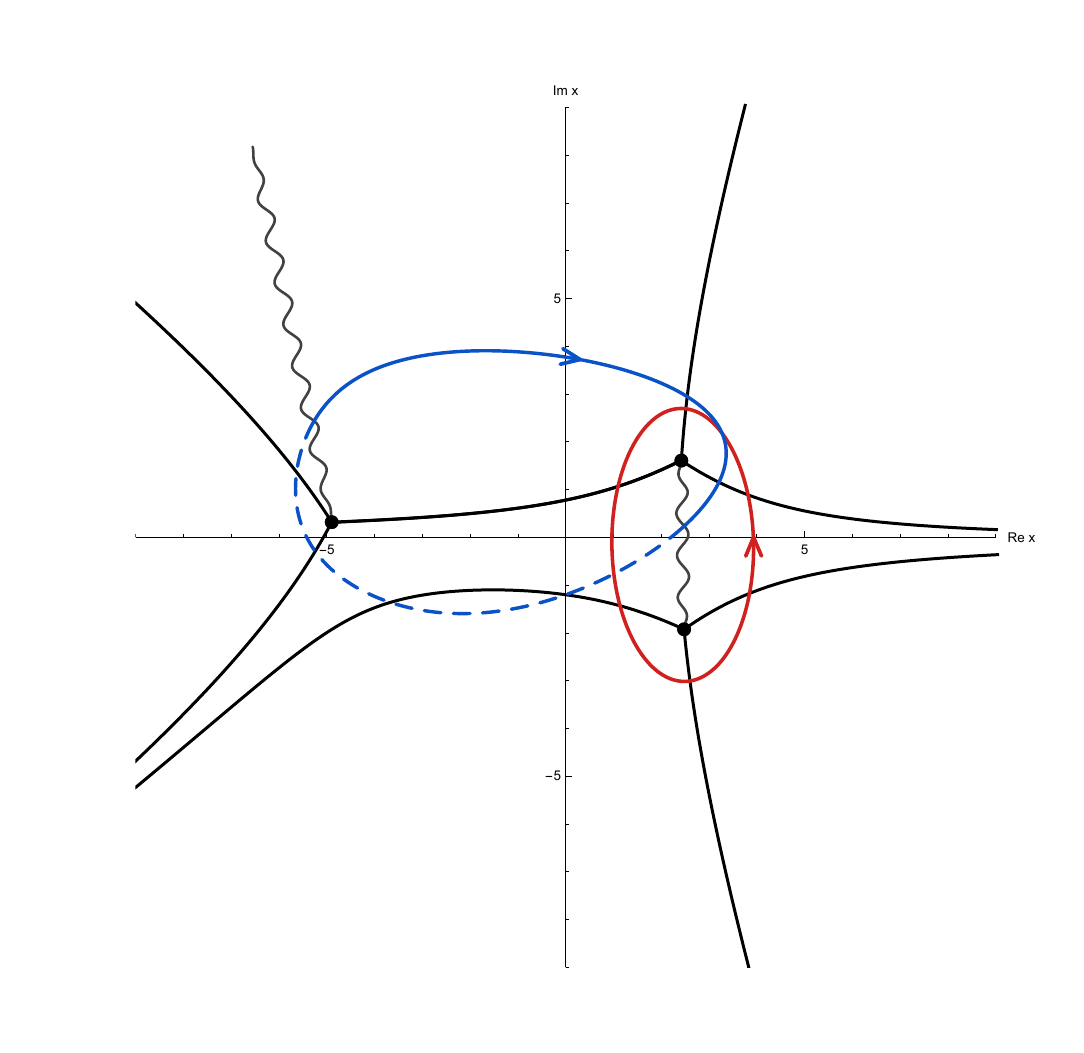} \\ {\tiny $\arg z = -0.052\pi$}
  \end{minipage}
  \begin{minipage}{0.19\textwidth}
    \centering
    \paperfigure[width=\linewidth]{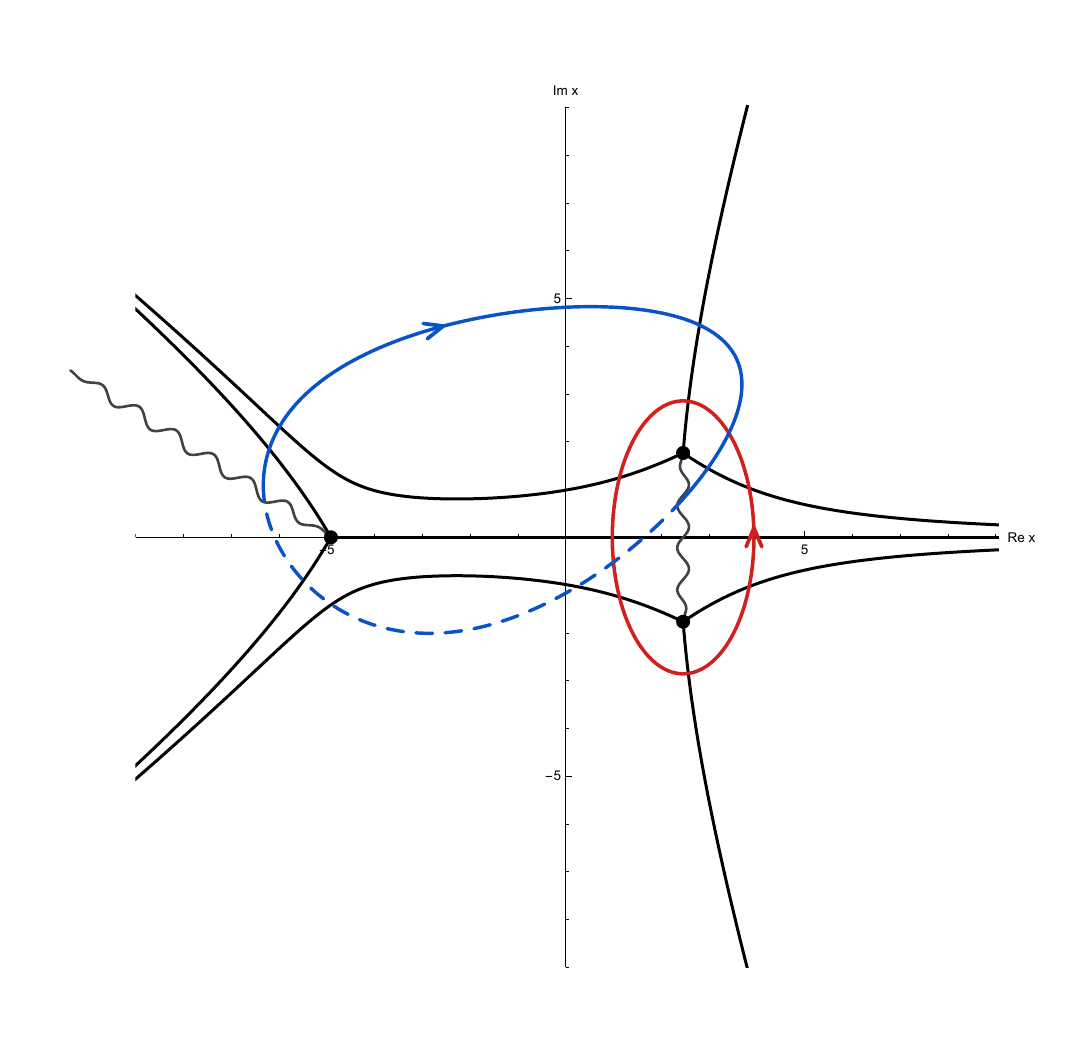} \\ {\tiny $\arg z = 0$}
  \end{minipage}

  \vspace{3mm}

  \begin{minipage}{0.19\textwidth}
    \centering
    \paperfigure[width=\linewidth]{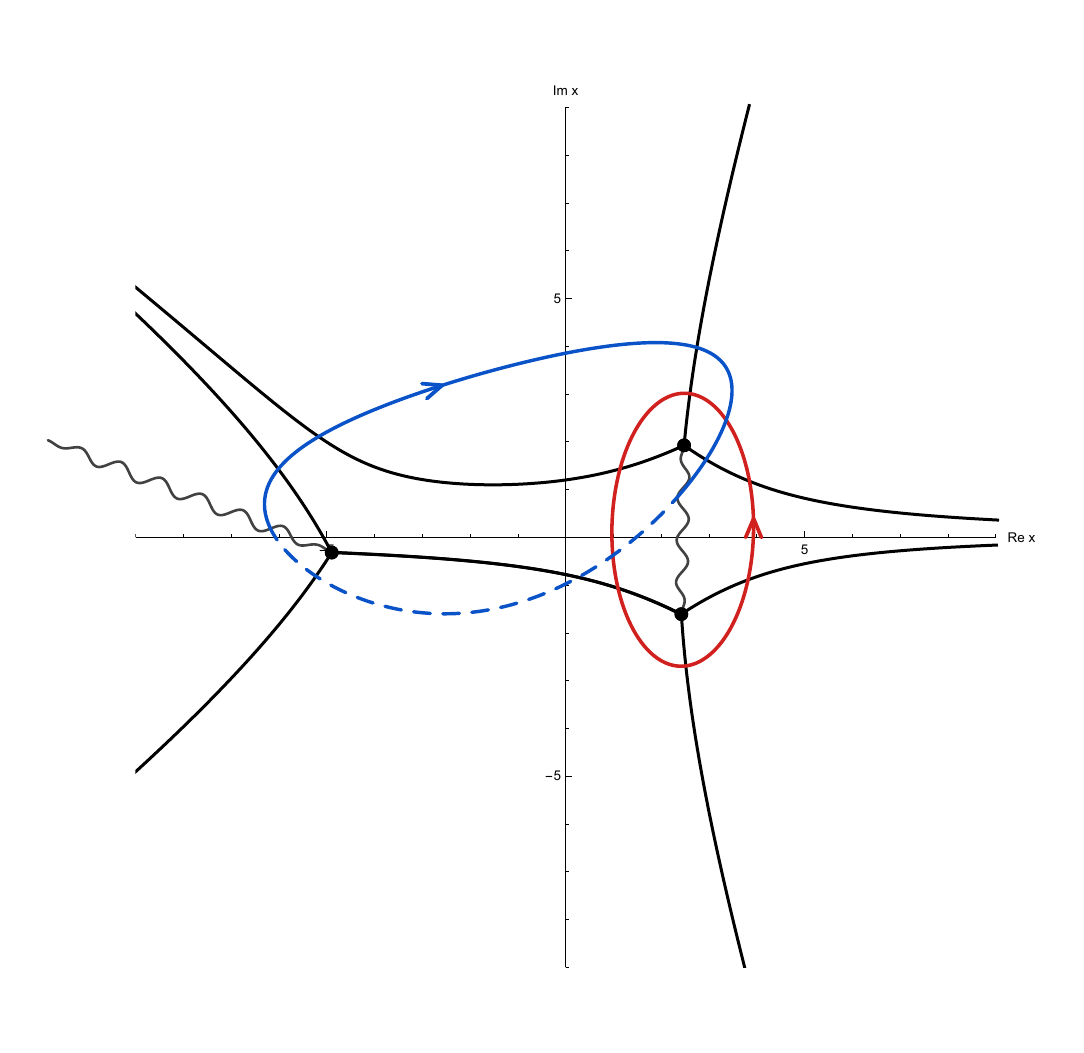} \\ {\tiny $\arg z = 0.052\pi$}
  \end{minipage}
  \begin{minipage}{0.19\textwidth}
    \centering
    \paperfigure[width=\linewidth]{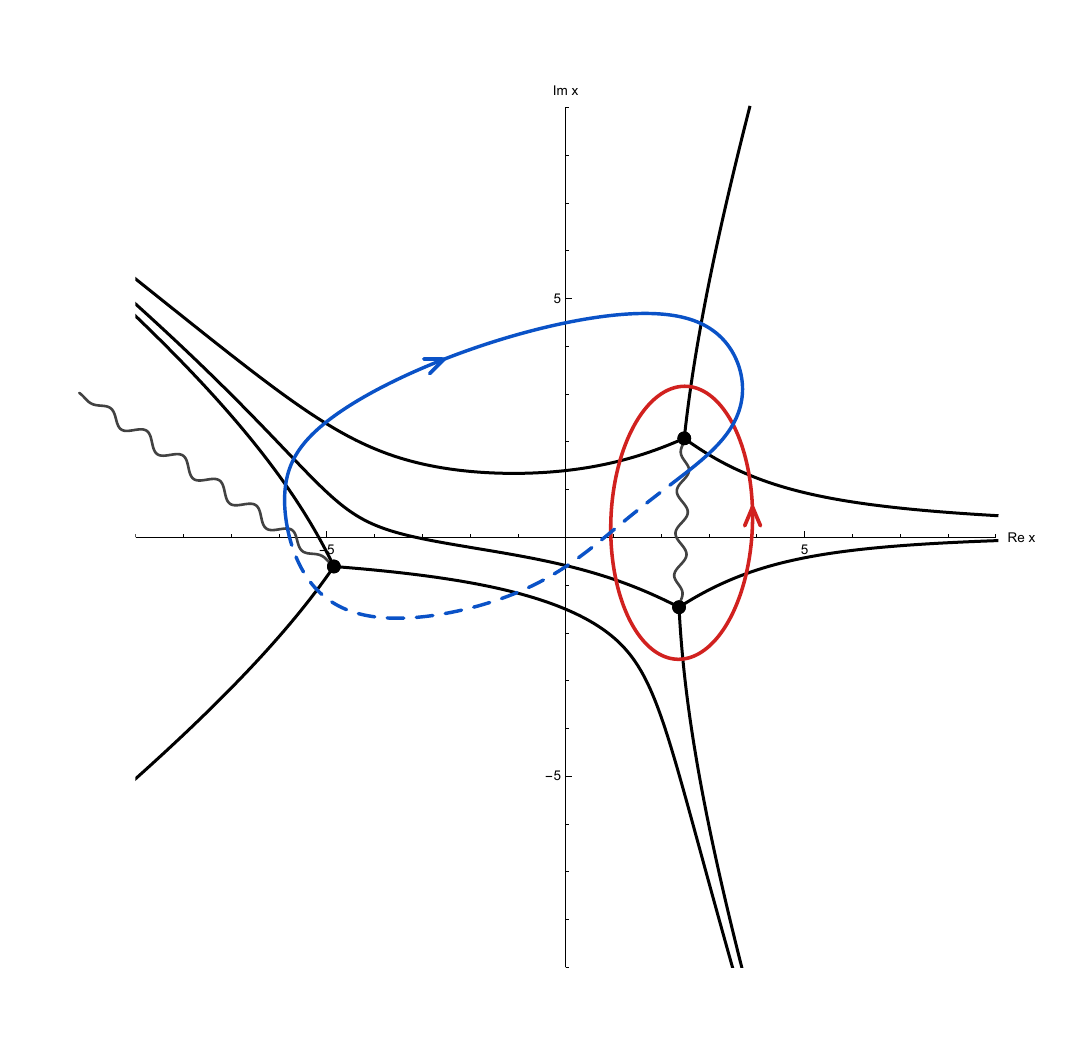}  \\ {\tiny $\arg z = \pi/10$}
  \end{minipage}
  \begin{minipage}{0.19\textwidth}
    \centering
    \paperfigure[width=\linewidth]{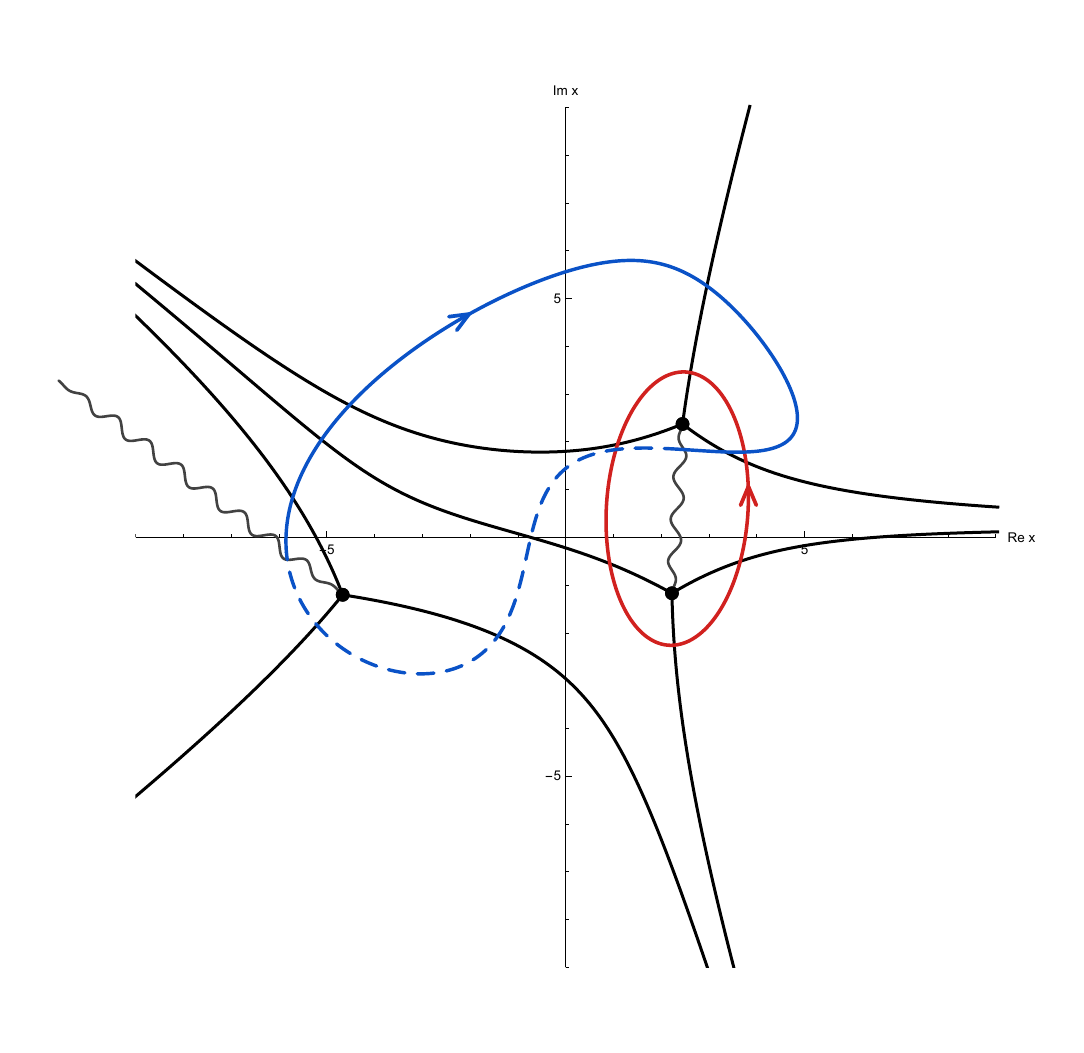} \\ {\tiny $\arg z = 2\pi/10$}
  \end{minipage}
  \begin{minipage}{0.19\textwidth}
    \centering
    \paperfigure[width=\linewidth]{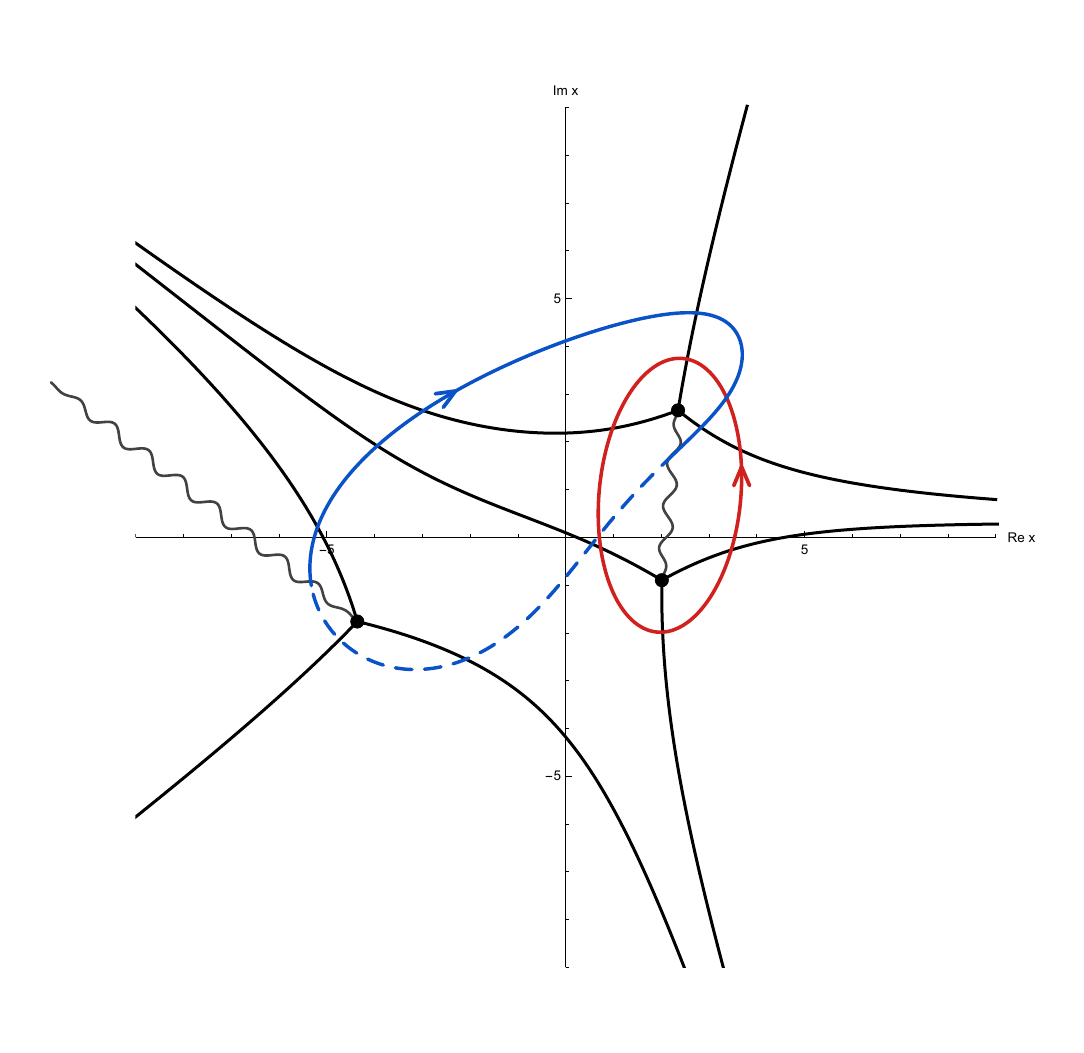} \\ {\tiny $\arg z = 3\pi/10$}
  \end{minipage}
  \begin{minipage}{0.19\textwidth}
    \centering
    \paperfigure[width=\linewidth]{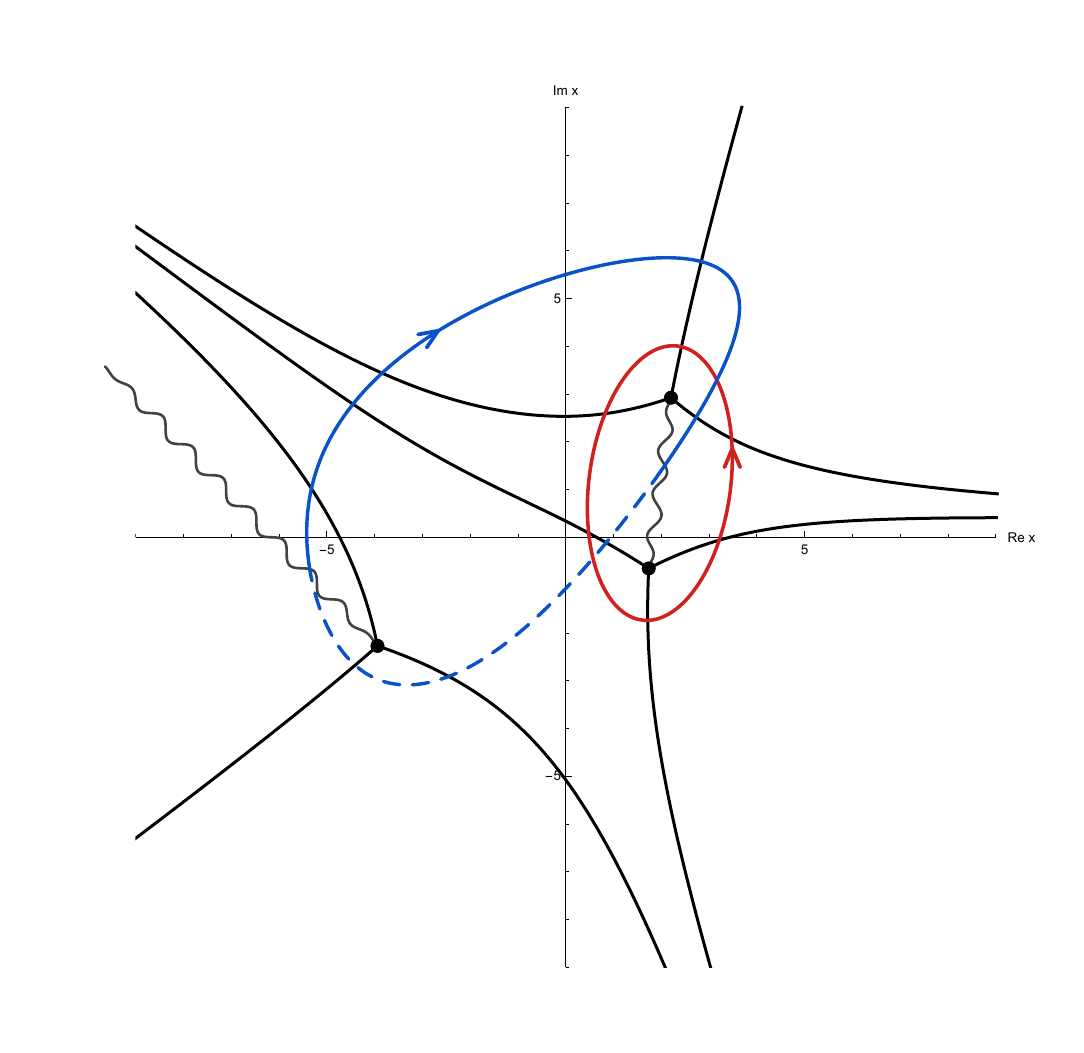} \\ {\tiny $\arg z = 4\pi/10$}
  \end{minipage}

  \vspace{3mm}

  \begin{minipage}{0.19\textwidth}
    \centering
    \paperfigure[width=\linewidth]{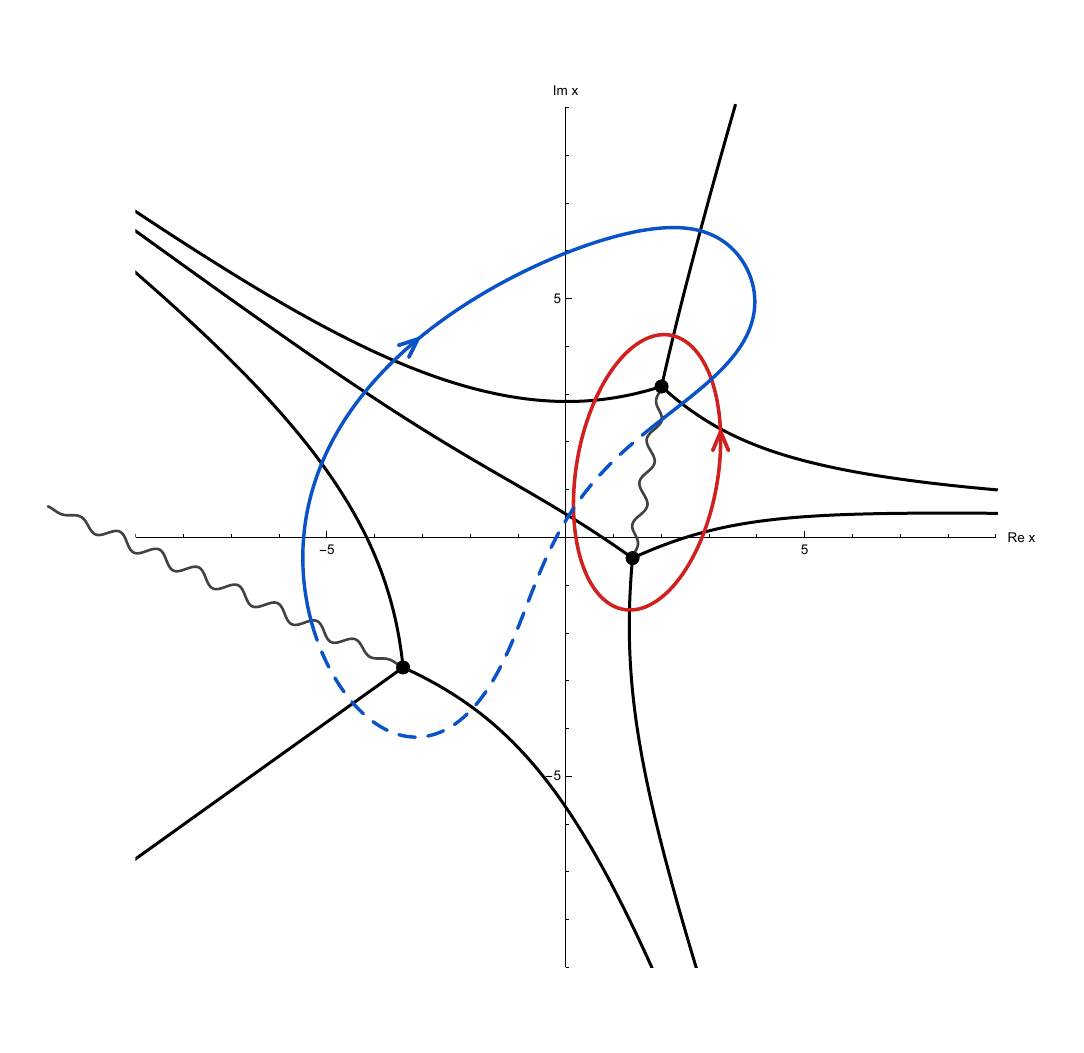} \\ {\tiny $\arg z = 5\pi/10$}
  \end{minipage}
  \begin{minipage}{0.19\textwidth}
    \centering
    \paperfigure[width=\linewidth]{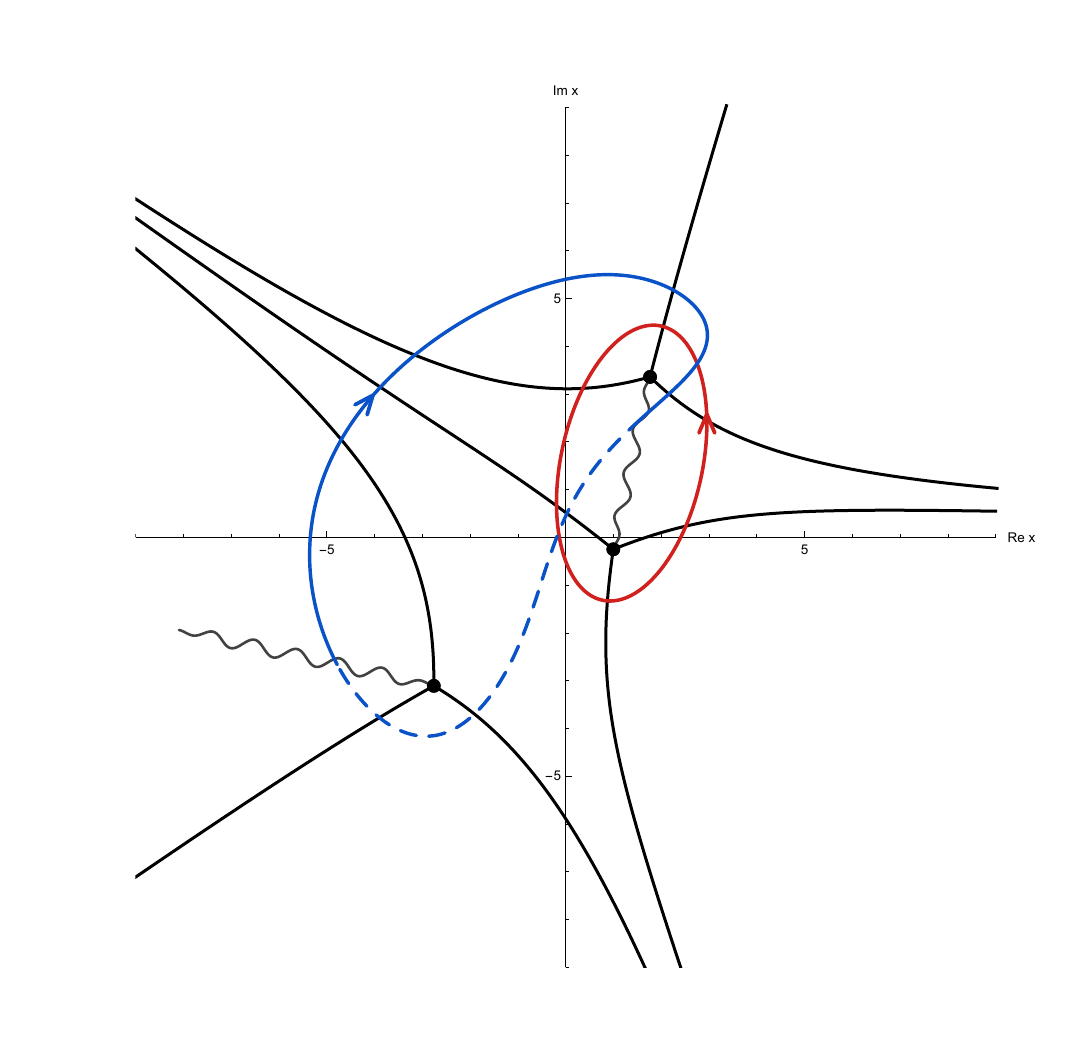} \\ {\tiny $\arg z = 6\pi/10$}
  \end{minipage}
  \begin{minipage}{0.19\textwidth}
    \centering
    \paperfigure[width=\linewidth]{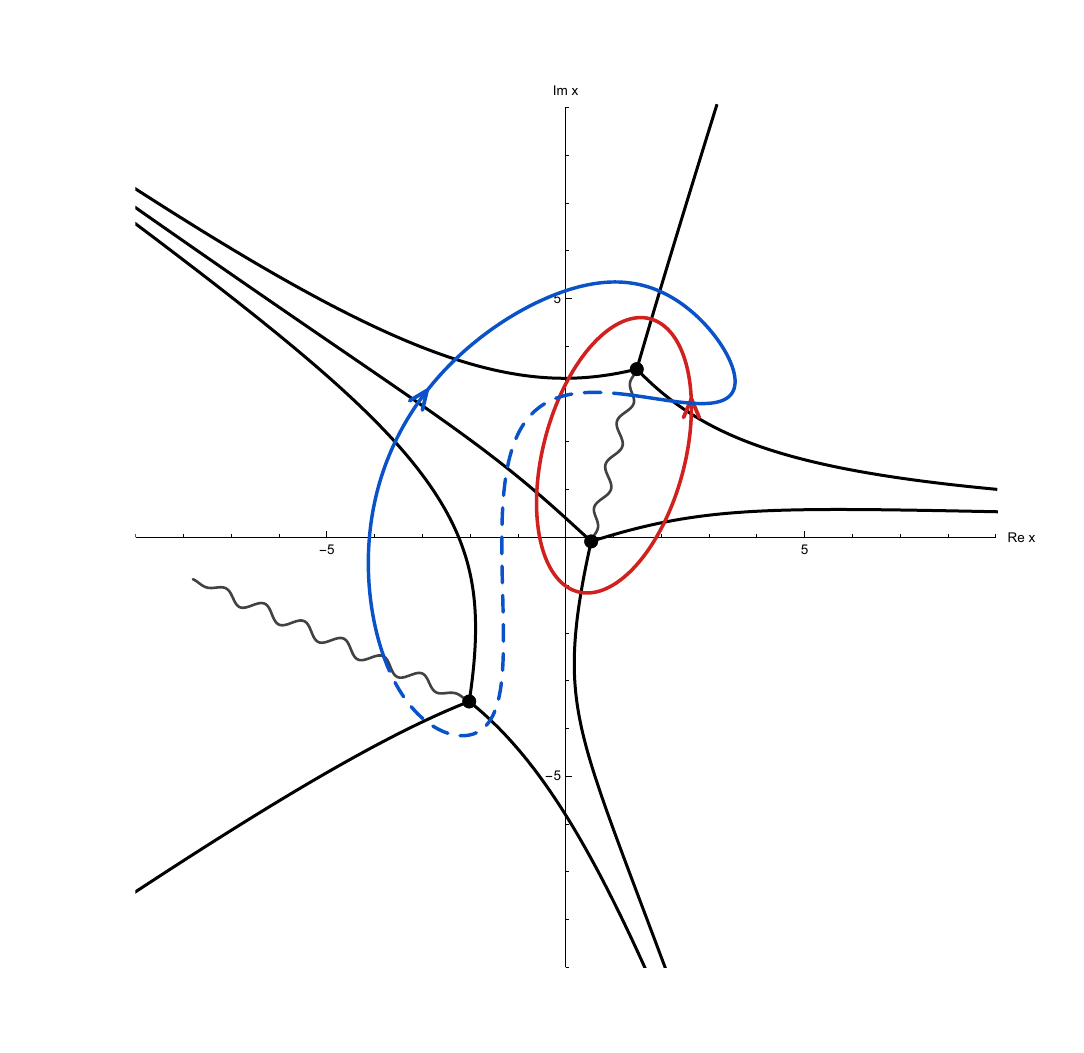} \\ {\tiny $\arg z = 7\pi/10$}
  \end{minipage}
  \begin{minipage}{0.19\textwidth}
    \centering
    \paperfigure[width=\linewidth]{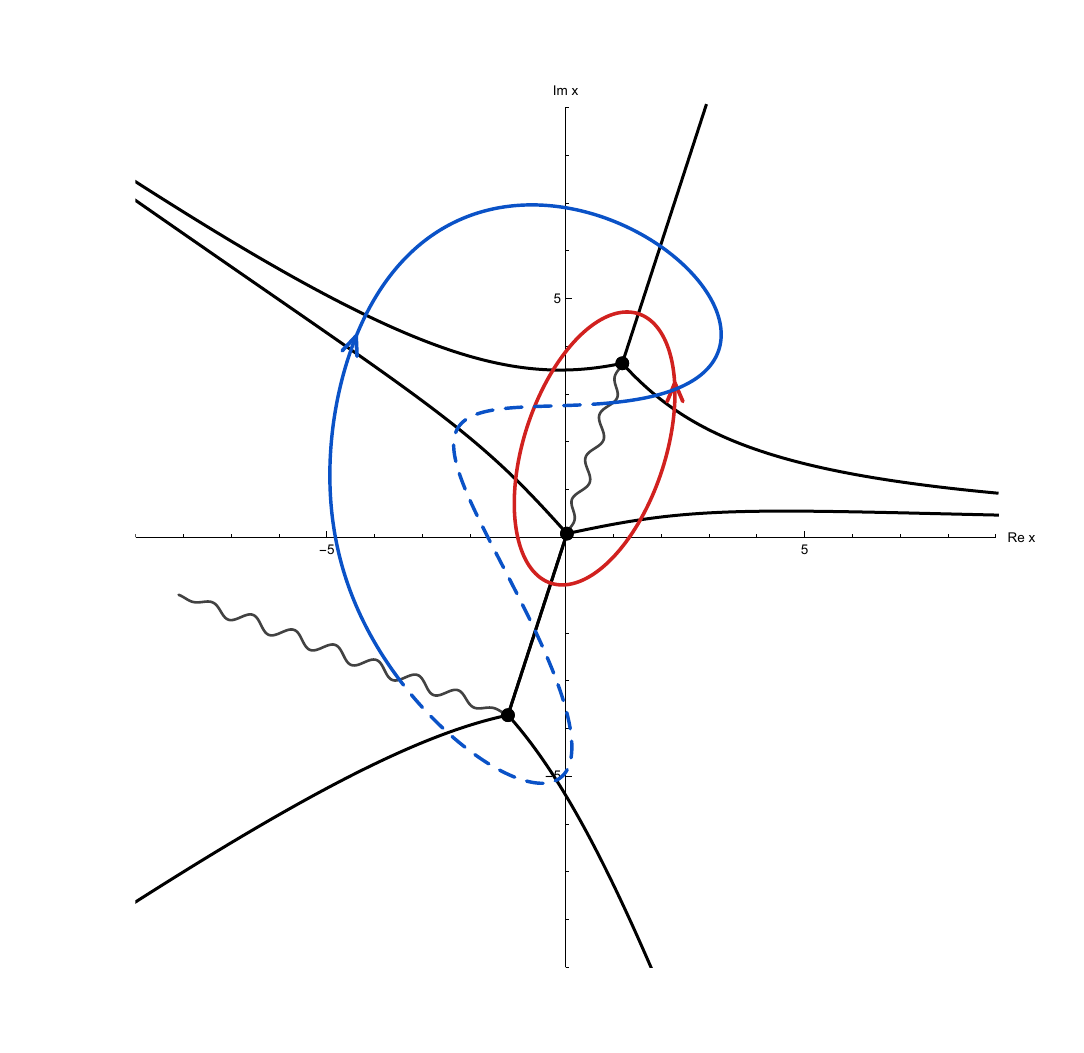} \\ {\tiny $\arg z = 8\pi/10$}
  \end{minipage}
  \begin{minipage}{0.19\textwidth}
    \centering
    \paperfigure[width=\linewidth]{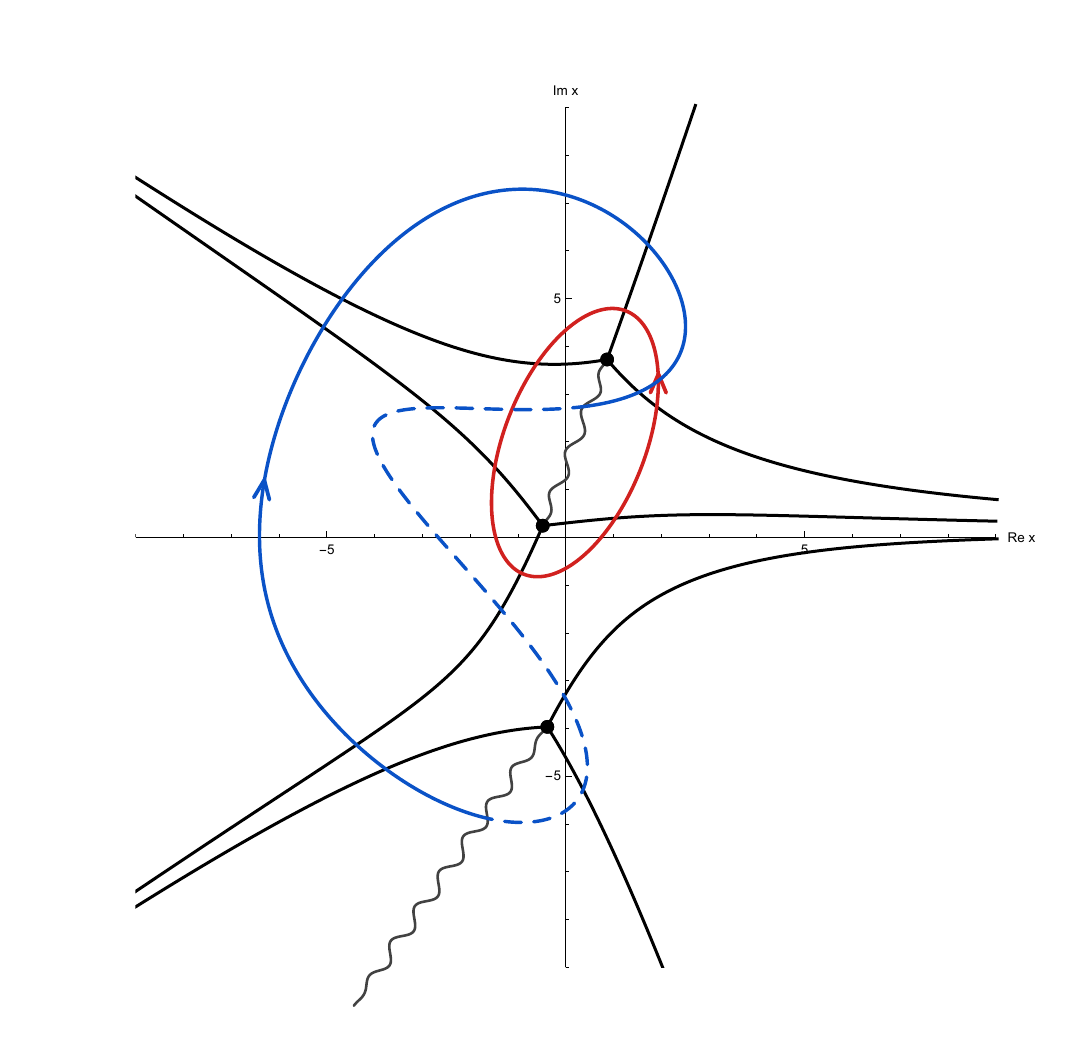} \\ {\tiny $\arg z = 9\pi/10$}
  \end{minipage}

\smallskip
  \caption{Linear Stokes graphs as \(\arg z\) varies from \(-13\pi/10\) to \(9\pi/10\). 
  The Torelli marking is transported continuously along this deformation: 
  the \(A\)-cycle is shown in red and the \(B\)-cycle in blue.}
  \label{fig:SGs}
\end{figure}

Figure \ref{fig:SGs} shows the linear Stokes graphs for
\[
  |z|=5,
  \qquad
  \nu=1,
\]
obtained by varying \(\arg z\) once around the \(z\)-plane, starting from
\[
  \arg z=-\frac{6\pi}{5} - \frac{\pi}{10} = - \frac{13\pi}{10}.
\]
As in Section~\ref{sec:stokes-review}, these Stokes graphs are drawn
using the approximation to $h$ given by \eqref{eq:H0-truncated}, with
the degeneration branch specified in Remark~\ref{rem:leading-branch}.

The direction $\arg z=-{6\pi}/{5}$ is an anti-Stokes direction of Painlev\'e~I.  
We therefore start from a slightly smaller value of \(\arg z\) and 
rotate \(z\) in the \(z\)-plane through an angle slightly larger than \(2\pi\).
In the figures below, \(\arg z\) is varied essentially in increments of
\(\pi/10\).  To illustrate more clearly the mutation of the linear Stokes graph
near \(\arg z=0\), we have inserted several additional plots, including
the critical configuration in which a saddle connection appears.

The directions
$\arg z={4\pi}/{5}$ and $\arg z=-{6\pi}/{5}$
represent the same ray in the \(z\)-plane.  
However, because of the multi-valuedness of \(h\), the analytic continuation
across the branch cut changes the former into a Stokes direction of Painlev\'e I.
This distinction is reflected in the corresponding linear Stokes graphs:
the graph at \(\arg z=-6\pi/5\) is saddle-free, whereas the graph at
\(\arg z=4\pi/5\) contains a saddle connection.

In these figures, we track not only the deformation of the Stokes graph
but also the corresponding evolution of the Torelli marking.  
At $\arg z=-{13\pi}/{10}$, we choose the \(A\)-cycle and 
the \(B\)-cycle as the initial Torelli marking, 
and then follow their continuous deformation as
\(\arg z\) is varied in the subsequent figures.  
Thus Figure \ref{fig:SGs} displays not only the changes in the topology of the linear Stokes graphs,  
but also the corresponding evolution of the chosen homology basis. 
This tracking of the cycles will play an important role later 
when we discuss the modular transformation laws.

\subsubsection{Comparison of Torelli markings}

We first discuss the \(T\)-transformation by continuously rotating the
Painlev\'e variable \(z\).  
The
\(T\)-transformation corresponds to a rotation through an angle
\(8\pi/5\) in the \(z\)-plane. 
Here, we choose a reference point \(z_{p}\) satisfying
\[
  |z_p| = 5, \qquad \arg z_{p}={- \frac{11 \pi}{10}},
\]
and trace the changes in the topology of the linear Stokes graph
and Torelli marking as \(z\) is rotated from \(z_{p}\) to
\begin{equation}
  z_p'=e^{8\pi i/5}z_p .
  \label{eq:T-z-rotation}
\end{equation}
This analysis allows us to derive the transformation formulas relating
the corresponding solutions of Painlev\'e~I and their tau-functions.

\begin{figure}[htbp]
  \centering

  \begin{minipage}[t]{0.45\textwidth}
    \centering
    \paperfigure[width=\textwidth]{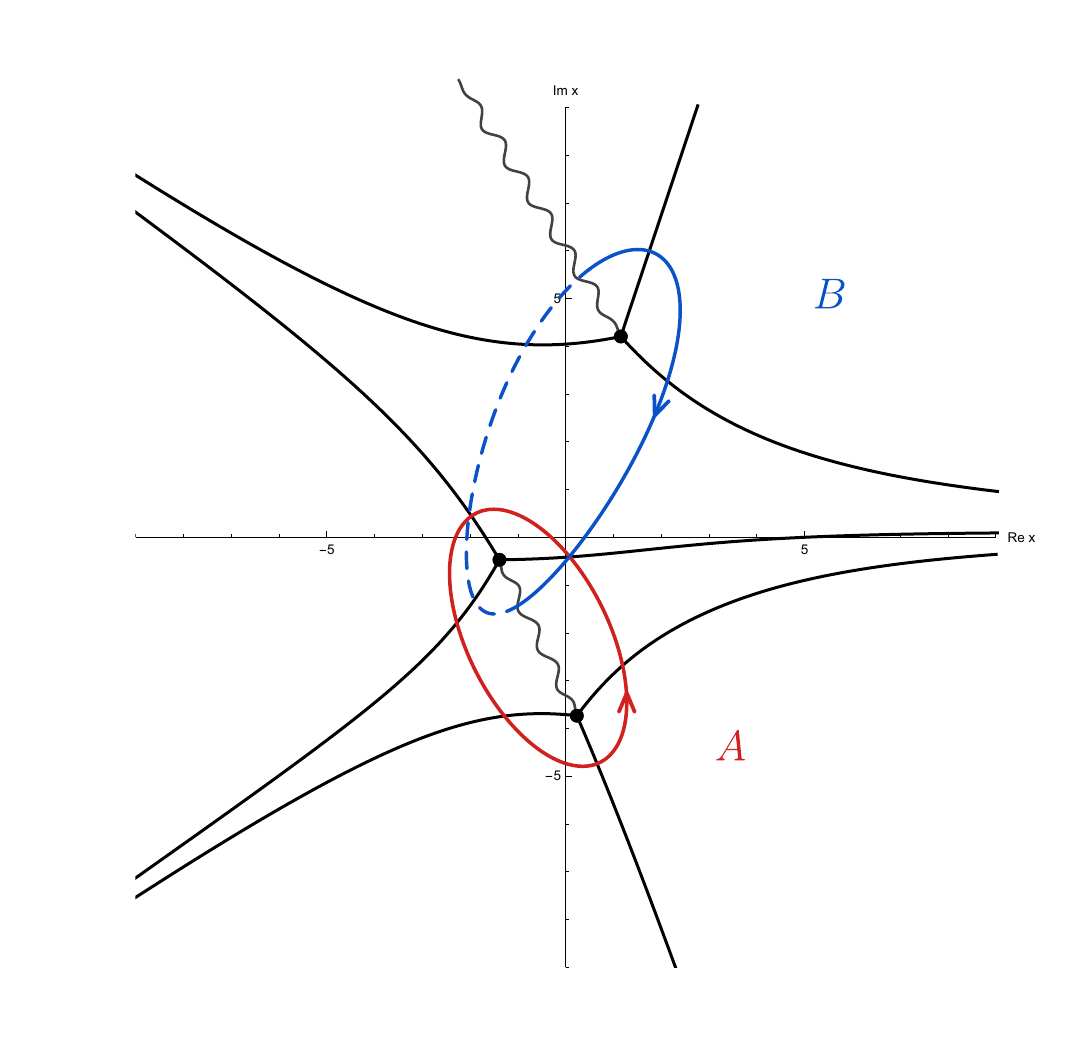}
    
    \vspace{2mm}
    
    \({\rm (i)} ~~ \arg z_p = - 11 \pi/10\)
  \end{minipage}
  \hfill
  \begin{minipage}[t]{0.45\textwidth}
    \centering
    \paperfigure[width=\textwidth]{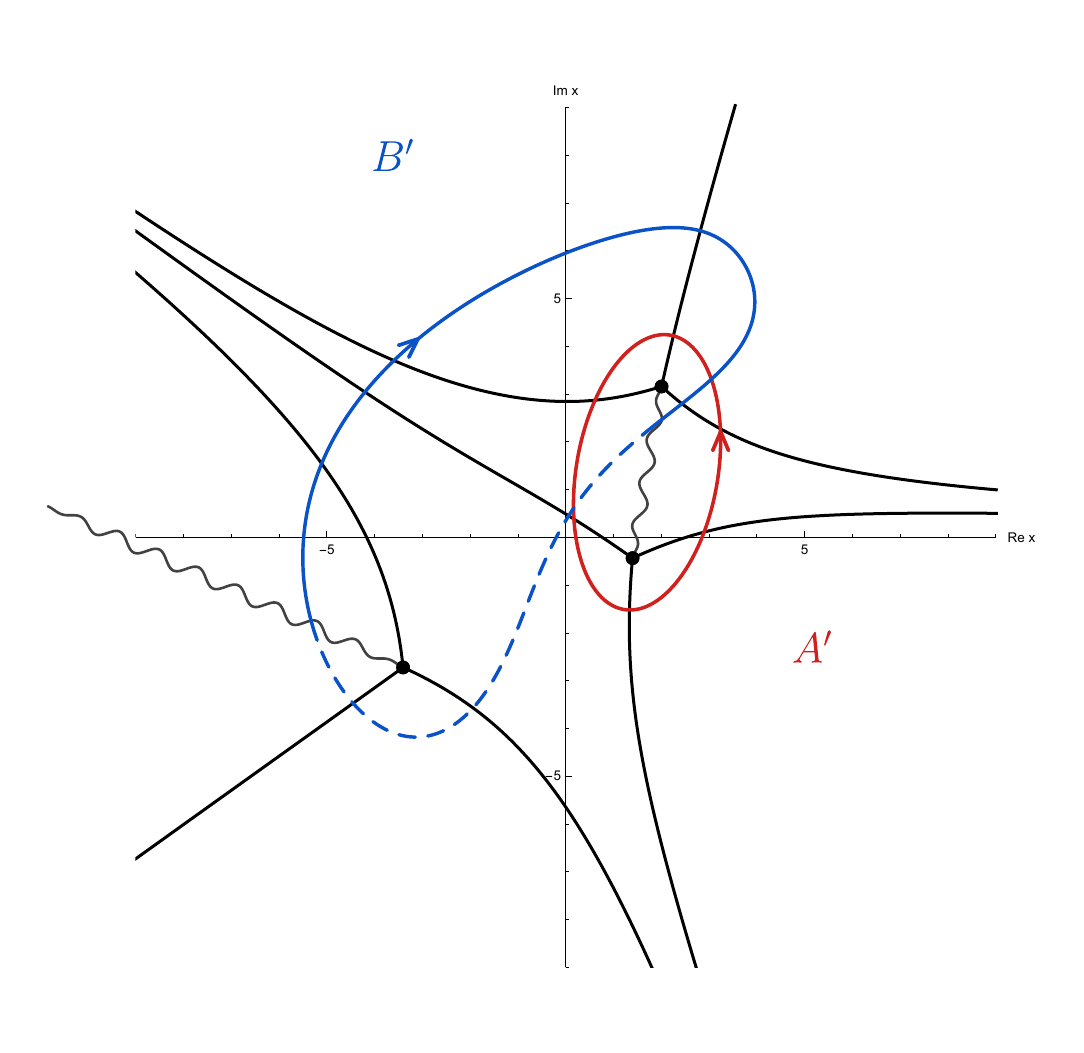}
    
    \vspace{2mm}
    
    \( {\rm (ii)} ~~ \arg z_{p}' = 5 \pi/10\)
  \end{minipage}

  \caption{Linear Stokes graphs at \(z_p\) and \(z_p'\). 
  As mentioned in the main text, 
  equation \eqref{eq:T-transform-of-cycles} 
  describes the relation between the Torelli marking \((A,B)\) 
  of the Stokes graph obtained by rotating the graph in (i) 
  by \(+4\pi/5\) and the Torelli marking \((A',B')\) 
  of the Stokes graph in (ii).}
   \label{fig:Stokes-T}
\end{figure}

Figure \ref{fig:Stokes-T} extracts from Figure~\ref{fig:SGs} 
the two linear Stokes graphs corresponding to \(z_p\) and \(z_p'\).
As the parameter varies from $z_{p}$ to $z_{p}'$,
the Stokes graph undergoes several mutations and hence, the exact solutions obtained by Borel resummation exhibit Stokes jumps. 
Our purpose here, however, is not to describe this analytic continuation in detail. 
Instead, by comparing the topological types of the Stokes graphs 
at the initial and final reference points, we derive the
correspondence between the Painlev\'e I solution associated with
$z_{p}$ and that associated with $z_{p}'$.

At \(z=z_p\), three linear Stokes curves are asymptotic to 
the positive real direction in the \(x\)-plane, 
whereas at \(z=z_p'\) three linear Stokes curves are asymptotic to $\arg x={4\pi}/{5}$.
Accordingly, after rotating the linear Stokes graph in 
Figure~\ref{fig:Stokes-T} (i) by
$-{16\pi}/{5}\equiv + {4\pi}/{5}\pmod{2\pi}$ in the \(x\)-plane, 
it can be topologically identified with the linear Stokes graph in Figure~\ref{fig:Stokes-T} (ii).

Moreover, along the path from $z_{p}$  to $z_{p}'$, 
the turning points and the Torelli marking can be followed continuously throughout the deformation.  
By continuously rotating \(z\) through an angle \(8\pi/5\), 
we obtain at \(z_p'\) a linear Stokes graph equipped with a Torelli marking \((A',B')\).  
On the other hand, 
rotating the linear Stokes graph at \(z_p\) through an angle \(4\pi/5\) 
in the \(x\)-plane naturally induces another Torelli marking. 
We denote this rotated marking by the same symbols \((A,B)\) as the original Torelli marking at \(z_p\).
Then we observe that these two Torelli markings are related by the following transformation:
\begin{equation}  \label{eq:T-transform-of-cycles}
\boxed{(A,B) = (A', B'+ A') = (A', B') \, T} 
\quad {\rm equivalently} \quad
\boxed{(A', B') = (A, B-A) = (A,B) \, T^{-1}}, 
\end{equation}
where 
\begin{equation}
T = \begin{pmatrix} 1 & 1 \\ 0 & 1 \end{pmatrix}.
\end{equation}


\subsubsection{Comparison of Stokes multipliers and transformation laws}

We now determine the transformation of the parameters directly from the
linear Stokes multipliers.
Let \(s_k = s_k(\nu,\rho) \) and \(s_k' = s_k'(\nu',\rho')\) 
denote the Stokes multipliers at \(z_p\) and
\(z_p'\), respectively.
Here, \((\nu,\rho)\) and \((\nu',\rho')\) denote the parameters entering
the Painlev\'e tau-functions 
\[
\tau(z, \nu, \rho; g_s)
\quad \text{and} \quad
\tau'(z', \nu', \rho'; g_s)
\]
via the topological recursion construction
followed by the Zak transform in \cite{Iwaki2019}, 
for the Torelli markings \((A,B)\) and \((A',B')\), respectively.
Here, \(z\) and \(z'\) are variables ranging in neighborhoods of \(z_p\)
and \(z_p'\), respectively.  
Later, we will impose the relation $z' = e^{8 \pi i /5} z$ 
and discuss the relation between the corresponding solutions
\begin{equation}
u(z, \nu, \rho; g_s) = - 2g_s^2 \frac{d^2}{dz^2} \log \tau (z, \nu, \rho; g_s), 
\quad
u'(z', \nu', \rho'; g_s) = - 2g_s^2 \frac{d^2}{d{z'}^2} \log \tau' (z', \nu', \rho'; g_s)
\end{equation}
induced by the rotational symmetry described in
Section~\ref{sec:Z5-symmetry-Stokes}.

Since neither of the endpoint Stokes graphs used in the comparison
contains a saddle connection, the exact WKB formulas of \cite{Iwaki2019} can be applied. 
The five linear Stokes multipliers at the two endpoints are listed
side by side below, using the same sector index $k$ in both columns.
The cyclic shift needed to compare the rotated solutions is imposed
in \eqref{eq:T-Stokes-correspondence}.
\begin{equation}
\renewcommand{\arraystretch}{1.8}
\setlength{\arraycolsep}{5pt}
\begin{array}{c|c|c}
 k&s_k(\nu,\rho)\quad(z=z_p)&s'_k(\nu',\rho')\quad(z'=z'_p)\\ \hline
-2&i\,\bigl(e^{-2 \pi i \rho/g_s} - e^{- 2 \pi i (\nu+\rho)/g_s} \bigr)&i\,e^{- 2\pi i \nu'/g_s}\\
-1&i\,e^{2 \pi i \rho/g_s}&i \, \bigl( e^{2\pi i \nu'/g_s} + e^{2\pi i \rho'/g_s} \bigr)\\
0&i\,\bigl(e^{2\pi i\nu/g_s}-e^{2\pi i(\nu-\rho)/g_s}+e^{-2\pi i\rho/g_s}\bigr)&i \, \bigl( e^{-2\pi i \rho'/g_s} - e^{2\pi i (\nu'-\rho')/g_s} \bigr)\\
1&i\,e^{-2 \pi i \nu/g_s}&-i\,e^{2\pi i (\rho'-\nu')/g_s}\\
2&i\,\bigl( e^{2 \pi i \nu/g_s} - e^{2 \pi i (\nu+\rho)/g_s} \bigr)&i\,\bigl(e^{2\pi i\nu'/g_s}-e^{2\pi i(\nu'-\rho')/g_s}+e^{2\pi i(2\nu'-\rho')/g_s}\bigr)
\end{array}
\label{eq:T-Stokes-zp}
\end{equation}

If \(u\) and \(u'\) are related by the rotational symmetry
\begin{equation}
u'(z',\nu',\rho') = e^{-16\pi i/5} u(e^{-8 \pi i /5}z'; \nu, \rho),
\end{equation}
then, according to \eqref{eq:Z5-Stokes-multipliers},  
the corresponding linear Stokes multipliers satisfy
\begin{equation}
  s_k(\nu,\rho)
  =
  s'_{k-8}(\nu',\rho')
  =
  s'_{k+2}(\nu',\rho'),
  \qquad
  k\in\mathbb Z/5\mathbb Z.
  \label{eq:T-Stokes-correspondence}
\end{equation}
By comparing the two columns of \eqref{eq:T-Stokes-zp}
with the cyclic shift \eqref{eq:T-Stokes-correspondence}, 
we have the following relation between the parameters:
\begin{equation}
  \boxed{
    \nu=\nu',
    \quad
    \rho=\rho'-\nu'+\frac{g_s}{2}
  }
  \quad {\rm equivalently} \quad 
    \boxed{
    \nu'=\nu,
    \quad
    \rho'=\rho+\nu-\frac{g_s}{2}
  }
  \label{eq:T-affine-periods}
\end{equation}
These relations hold modulo $g_s\mathbb Z$ in each additive parameter;
we use the logarithmic branches and local representatives fixed in
Remark~\ref{rem:monodromy-parameter-branches}.

In particular, by the uniqueness of the solution to the associated
Riemann--Hilbert problem, the two Painlev\'e~I solutions \(u\) and \(u'\)
are related by
\begin{equation}
\boxed{ 
 u'\left(
    z',\nu',\rho'; g_s
  \right)
  =
  e^{- 16\pi i/5}
  u\left(
    e^{- 8\pi i/5} z', \nu',\rho'-\nu'+\frac{g_s}{2} ; g_s
  \right).
}
  \label{eq:T-PI-rotation}
\end{equation}
Equivalently, 
\begin{equation}
\boxed{ 
 u\left(
    z,\nu,\rho; g_s
  \right)
  =
  e^{16\pi i/5}
  u'\left(
    e^{8\pi i/5} z, \nu,\rho+\nu - \frac{g_s}{2} ; g_s
  \right).
}
  \label{eq:T-PI-rotation-inverse}
\end{equation}

With the extended tau-function normalization discussed in
Remark~\ref{rem:modular-tau-normalization}, the corresponding relation takes the form
\begin{equation}
  \boxed{
  \tau(z, \nu,\rho;g_s)
  =
  C_T \, e^{\pi i \nu^2 / g_s^2} \, 
  \tau' \left( e^{8\pi i/5}  z, \nu,\rho+\nu - \frac{g_s}{2} ;g_s \right)
  \label{eq:T-tau}.
  }
\end{equation}

\subsection{The \(S\)-transformation}
\label{sec:S-transformation-Stokes}

\subsubsection{Rotation of the Stokes graph and change of Torelli marking}

We next consider the \(S\)-transformation.  
In this case we again rotate the Painlev\'e variable \(z\), 
but the identification of the two endpoint Stokes graphs also 
involves a rotation of the Stokes sectors in the \(x\)-plane.

We take the same reference point $z_p$ as in the previous subsection, and 
we move \(z\) continuously from \(z_p\) to 
\[
z_{p}'' = e^{2 \pi i/5} z_p. 
\]
We also simultaneously track the turning points and the Torelli marking.

\begin{figure}[htbp]
  \centering

  \begin{minipage}[t]{0.48\textwidth}
    \centering
    \paperfigure[width=\textwidth]{SG1-labeled}
    
    \vspace{2mm}
    
     \( {\rm (i)} ~~ \arg z_p = - 11 \pi/10\)
  \end{minipage}
  \hfill
  \begin{minipage}[t]{0.48\textwidth}
    \centering
    \paperfigure[width=\textwidth]{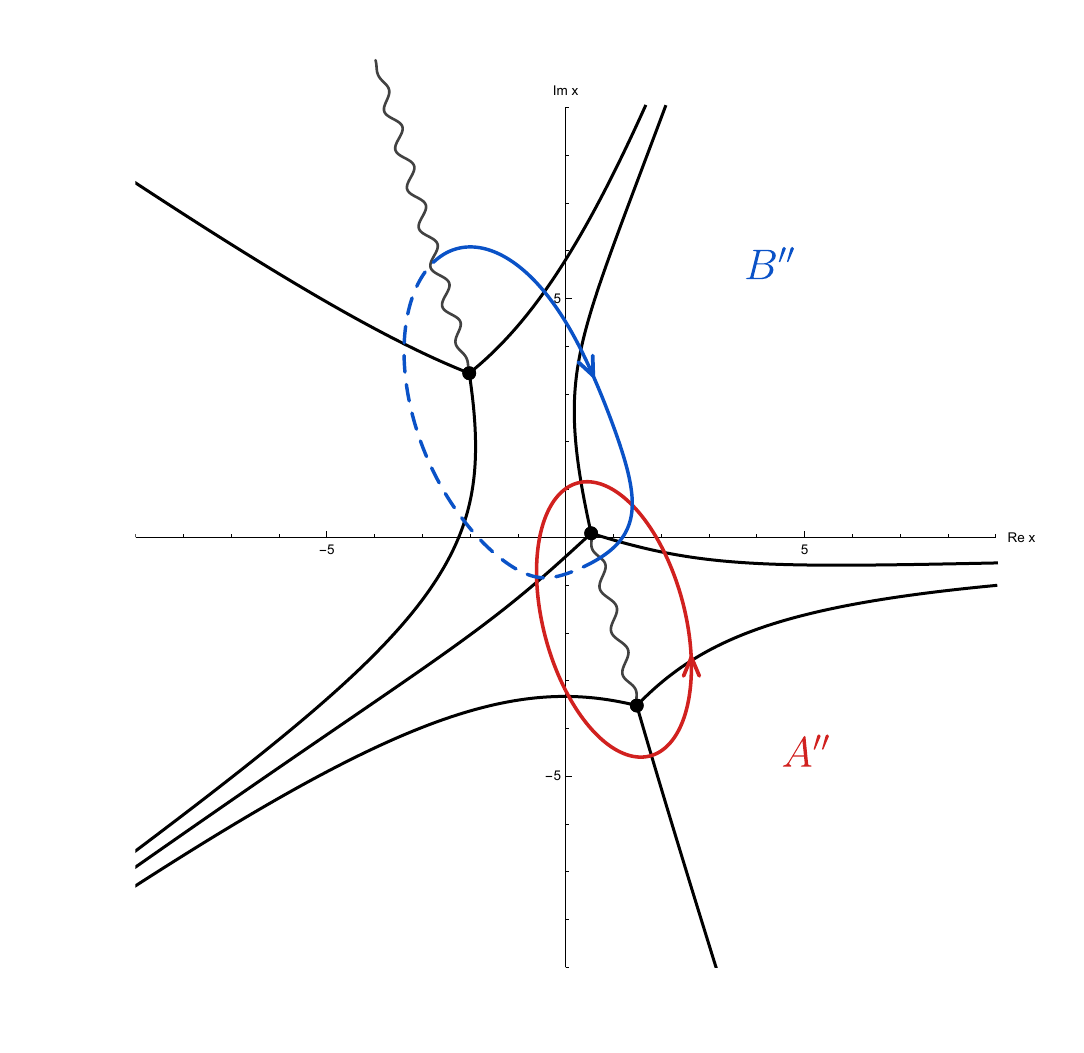}
    
    \vspace{2mm}
    
     \({\rm (ii)} ~~ \arg z_{p}'' = -7 \pi/10\)
  \end{minipage}

  \caption{Linear Stokes graphs at \(z_p\) and \(z_{p}''\).}
  \label{fig:Stokes-S}
\end{figure}

In view of Figure~\ref{fig:SGs}, 
the path from \(z_p\) to \(z_p''\) crosses a Stokes curve of the first
Painlev\'e equation, and hence the Stokes graph undergoes a mutation.
Figure~\ref{fig:Stokes-S} displays only the linear Stokes graphs at
\(z_p\) and \(z_p''\).  Rotating the linear Stokes graph in
Figure~\ref{fig:Stokes-S} (i) by \(-4\pi/5\) in the \(x\)-plane, 
we find that it is topologically identified with 
the linear Stokes graph in Figure~\ref{fig:Stokes-S} (ii).  
Furthermore, as seen from the tracking of the cycles, 
their Torelli markings are related by the \(S\)-transformation: 
\begin{equation} \label{eq:S-transform-of-cycles} 
\boxed{(A,B) = (B'', -A'') = (A'', B'') \, S} 
\quad {\rm equivalently} \quad
\boxed{(A'', B'') = (-B, A) = (A,B) \, S^{-1}}, 
\end{equation}
where 
\begin{equation}
S = \begin{pmatrix}
    0 & -1\\
    1 & 0
  \end{pmatrix}.
\end{equation}
Here, similarly to the $T$-transformation 
\eqref{eq:T-transform-of-cycles}, 
equation \eqref{eq:S-transform-of-cycles} 
is obtained by rotating the Stokes graph in (i) by \(-4\pi/5\) 
and comparing the corresponding Torelli markings.

\subsubsection{Comparison of Stokes multipliers and transformation laws}

We now compare the linear Stokes multipliers, following the same
procedure as for the \(T\)-transformation.
Let
\[
  \tau(z,\nu,\rho;g_s)
  \qquad\text{and}\qquad
  \tau''(z'',\nu'',\rho'';g_s)
\]
be the tau-functions associated with the Torelli markings \((A,B)\) and
\((A'',B'')\), respectively, and let
\[
  u(z;\nu,\rho)
  \qquad\text{and}\qquad
  u''(z'';\nu'',\rho'')
\]
be the corresponding Painlev\'e~I solutions.  Here, \(z\) and \(z''\)
are regarded as variables ranging in neighborhoods of \(z_p\) and
\(z_p''\), respectively.
The simple $S$-action derived below applies to these endpoint
sectors, which straddle the nonlinear Stokes direction
$\arg z=-4\pi/5$.
If the endpoints instead straddle an anti-Stokes direction,
the parameter relation is different; the two descriptions
are related by Stokes transformations, consistently with
the global $\mathbb Z_5$ symmetry.
This chamber dependence will be explained in more detail
in an upcoming paper.


Let
$s_k(\nu,\rho)$ and $s_k''(\nu'',\rho'')$
denote the linear Stokes multipliers of the corresponding isomonodromy
systems in the direction $\arg x={2\pi k}/{5}$ with $k\in\mathbb Z/5\mathbb Z$.
For ease of comparison, we repeat the multipliers at $z_p$ alongside
those at $z_p''$. Both columns use the same sector index $k$; the
cyclic shift for the rotated solutions is specified in
\eqref{eq:S-Stokes-comparison}.
\begin{equation}
\renewcommand{\arraystretch}{1.8}
\setlength{\arraycolsep}{5pt}
\begin{array}{c|c|c}
 k&s_k(\nu,\rho)\quad(z=z_p)&s''_k(\nu'',\rho'')\quad(z''=z''_p)\\ \hline
 -2&i\,\bigl(e^{-2\pi i\rho/g_s}-e^{-2\pi i(\nu+\rho)/g_s}\bigr)
 &i\,\bigl(e^{-2\pi i\nu''/g_s}-e^{-2\pi i(\nu''+\rho'')/g_s}+e^{-2\pi i\rho''/g_s}\bigr)\\
 -1&i\,e^{2\pi i\rho/g_s}&i\,e^{2\pi i\rho''/g_s}\\
 0&i\,\bigl(e^{2\pi i\nu/g_s}-e^{2\pi i(\nu-\rho)/g_s}+e^{-2\pi i\rho/g_s}\bigr)
 &i\,\bigl(e^{-2\pi i\rho''/g_s}-e^{2\pi i(\nu''-\rho'')/g_s}\bigr)\\
 1&i\,e^{-2\pi i\nu/g_s}
 &i\,\bigl(e^{-2\pi i\nu''/g_s}-e^{2\pi i(\rho''-\nu'')/g_s}\bigr)\\
 2&i\,\bigl(e^{2\pi i\nu/g_s}-e^{2\pi i(\nu+\rho)/g_s}\bigr)
 &i\,e^{2\pi i\nu''/g_s}
\end{array} 
\label{eq:S-Stokes-zpp}
\end{equation}

According to \eqref{eq:Z5-Stokes-multipliers}, 
in order for the two solutions to be related by the rotational symmetry
\begin{equation}
u''(z'',\nu'',\rho'') = e^{-4\pi i/5} u(e^{-2 \pi i /5}z''; \nu, \rho),
\end{equation}
their linear Stokes multipliers must satisfy
\begin{equation}
  \boxed{
    s_k(\nu,\rho)
    =
    s_{k-2}''(\nu'',\rho''),
    \qquad
    k\in\mathbb Z/5\mathbb Z.
  }
  \label{eq:S-Stokes-comparison}
\end{equation}
By comparing the two columns of \eqref{eq:S-Stokes-zpp}
with the cyclic shift \eqref{eq:S-Stokes-comparison}, 
we have the following relation between the parameters:
\begin{equation}
  \boxed{
    \nu = - \rho'',
    \quad
    \rho = \nu''
  }
  \quad {\rm equivalently} \quad 
    \boxed{
    \nu''=\rho,
    \quad
    \rho''=-\nu
   }
  \label{eq:S-periods}
\end{equation}
These relations hold modulo integer multiples of the string coupling constant in each additive parameter; 
the choice of local representatives is understood as in Remark \ref{rem:monodromy-parameter-branches}.

As in the case of the $T$-transformation,
we obtain the following relation between the two Painlev\'e~I solutions \(u\) and \(u''\):
\begin{equation}
\boxed{ 
 u''\left(
    z'';\nu'',\rho''
  \right)
  =
  e^{- 4\pi i/5}
  u\left(
    e^{- 2\pi i/5} z'' ; -\rho'',\nu''
  \right).
}
  \label{eq:S-PI-rotation}
\end{equation}
Equivalently, 
\begin{equation}
\boxed{ 
 u\left(
    z;\nu,\rho
  \right)
  =
  e^{4\pi i/5}
  u''\left(
    e^{2\pi i/5} z; \rho, - \nu 
  \right).
}
  \label{eq:S-PI-rotation-inverse}
\end{equation}

With the extended tau-function normalization discussed in
Remark~\ref{rem:modular-tau-normalization}, the corresponding relation takes the form\footnote{ 
The parameter exchange, parameter-dependent exponential
factor, as well as the numerical factor $C_S=e^{\pi i/20}$ in \eqref{eq:S-tau}, also arise formally from Poisson
resummation of the Zak representation \eqref{eq:IM-Zak},
with the partition-function block replaced by its Fourier
transform.
Obtaining the rotation formula by this route additionally
requires identifying the transformed block with that for
the rotated spectral curve and Torelli marking, with
compatible integration contours and normalization. This route is more involved and will be addressed in an upcoming paper.
}
\begin{equation}
  \boxed{
  \tau(z,\nu,\rho;g_s)
  =
  C_S \, e^{- 2 \pi i \nu \rho / g_s^2} \, 
  \tau'' \left( e^{2\pi i/5}  z, \rho, - \nu ;g_s \right)
  \label{eq:S-tau}.
  }
\end{equation}

\begin{rem}%
\label{rem:modular-tau-normalization}
The comparison of Stokes multipliers determines the parameter maps, but
not the normalization of the tau-function. The Hamiltonian definition
$g_s^2\partial_z\log\tau=\mathcal H$, together with its rotational
transformation, determines the ratio of the corresponding tau-functions
only up to a factor independent of $z$; that factor may initially depend
on $\nu,\rho$ and $g_s$. Equality of $u$ alone would even allow a factor
$\exp(c_1z+c_0)$, so the Hamiltonian normalization is needed here.

The Gaussian factors in \eqref{eq:T-tau} and \eqref{eq:S-tau} use the extended normalization discussed in \cite{IwakiMarino2024}, with monodromy symplectic potential $-2\pi i\nu\,\dd\rho/g_s^2$. At fixed $g_s$, the differences of this potential under the parameter maps are exact:
\begin{align}
 -\frac{2\pi i}{g_s^2}
 (\nu\,\dd\rho-\nu'\,\dd\rho')
 &=\dd\!\left(\frac{\pi i\nu^2}{g_s^2}\right),
 &&\text{for }(\nu',\rho')=(\nu,\rho+\nu-g_s/2),\notag\\
 -\frac{2\pi i}{g_s^2}
 (\nu\,\dd\rho-\nu''\,\dd\rho'')
 &=\dd\!\left(-\frac{2\pi i\nu\rho}{g_s^2}\right),
 &&\text{for }(\nu'',\rho'')=(\rho,-\nu).
 \label{eq:modular-generating-factors}
\end{align}
If the tau-functions in all three markings are normalized compatibly with this prescription, the remaining $C_T$ and $C_S$ are independent of $z,\nu,\rho$ on the chosen local domain. In Section~\ref{sec:RM-tau-constants} we fix the normalization by the large-$z$ Fourier expansion \eqref{eq:pinched-Fourier}--\eqref{eq:pinched-block},
and obtain $C_T=e^{\pi i/30}$ and $C_S=e^{\pi i/20}$ in the
$g_s=1$ variables of \eqref{eq:resurgent-gs-scaling}.
These constants refer to that specified Fourier normalization; the comparison of multiplier tables alone does not fix the additive constants in the TR free energies.
\end{rem}

\section{Resurgence and Modularity}
\label{sec:resurgence-modularity}

We now combine the Stokes transformations of Section~\ref{sec:stokes-review}
with the modular transformations of Section~\ref{sec:modular-transformation}.
We will first demonstrate the monodromy closure for the specific heat,
where there are no tau-function prefactors to keep track of. We then
repeat the calculation for the tau function, and show how the
additional prefactors cancel by Abel's dilogarithm identity.

Let us first make precise the resummation convention used below.
We set $g_s=1$ as in \eqref{eq:PI-unit-coupling} and write
$u(z;\nu,\rho)$ for the transseries of Section~\ref{sec:bssv}, resummed
in the Borel direction
\begin{equation}
 \theta(z)=\arg(z^{-5/4})=-\frac54\arg z.
 \label{eq:RM-resummation-direction}
\end{equation}
Both arguments are followed continuously, rather than reduced modulo
$2\pi$; on a singular direction the appropriate lateral prescription
is understood. This allows us to use the same symbol $u$ in different
sectors. In particular, the primes in
\eqref{eq:T-PI-rotation-inverse} and
\eqref{eq:S-PI-rotation-inverse} keep track of sectorial descriptions
which are included in this convention. We do not identify Borel sums
in different directions at fixed parameters. To analytically continue
the same solution across a Stokes line, we still have to transform its
parameters by the corresponding Stokes map. We use the same convention
for $\tau$, with the Fourier normalization
\eqref{eq:pinched-Fourier}--\eqref{eq:pinched-block} and the analytic
assumptions stated in Section~\ref{sec:iwaki-marino}.

In the convention of the theta-function transformations in Appendix~A,
the parameter maps appearing in \eqref{eq:S-periods} and
\eqref{eq:T-affine-periods} are
\begin{equation}
    \boxed{\mathsf S^{-1}(\nu,\rho)=(\rho,-\nu)},
    \qquad
    \boxed{\mathsf T^{-1}(\nu,\rho)
    =\left(\nu,\rho+\nu-\frac12\right)}.
    \label{eq:RM-parameter-maps}
\end{equation}
Thus one also has
\begin{equation}
    \boxed{\mathsf S(\nu,\rho)=(-\rho,\nu)},
    \qquad
    \boxed{\mathsf T(\nu,\rho)
    =\left(\nu,\rho-\nu+\frac12\right)}.
    \label{eq:RM-forward-parameter-maps}
\end{equation}
The inverse maps in \eqref{eq:RM-parameter-maps} correspond to the endpoint markings $(A,B)S^{-1}$ and $(A,B)T^{-1}$ in Section~\ref{sec:modular-transformation}.

\subsection{One-loop monodromy of the specific heat}
\label{sec:RM-specific-heat}

Having seen the Stokes structure on the complex $z$-plane, we can now
ask what happens to a transseries solution after a full counterclockwise
turn. Choose
\begin{equation}
 \arg z_{\rm p}=-\frac{6\pi}{5}+\epsilon,
 \qquad 0<\epsilon<\frac\pi5,
 \qquad z_1=e^{2\pi i/5}z_{\rm p}.
 \label{eq:RM-base-point}
\end{equation}
The reference point of Section~\ref{sec:modular-transformation}
corresponds to $\epsilon=\pi/10$. The point $z_{\rm p}$ lies on sheet
$-1$, while $z_1$ has argument $-4\pi/5+\epsilon$ on the principal
sheet. During the full turn the continuously followed Borel angle
crosses $\pi$, $0$, and $-\pi$, in that order.

\begin{figure}[htbp]
    \centering

    \begin{subfigure}[t]{0.48\textwidth}
        \centering
        \resizebox{\linewidth}{!}{%
        \begin{tikzpicture}[scale=0.75]

        \node[black] at (6.2,6.4) {\small sheet $-1$};
        
            \draw[gray, line width=1pt, ->] (-7,0) -- (7,0);
            \draw[gray, line width=1pt, ->] (0,-7) -- (0,7);

            \draw[black, line width=1pt, decorate,
                  decoration={zigzag, segment length=7pt, amplitude=1.6pt}]
                  (-6.8,0) -- (-0.25,0);
                  
            \node[gray] at (-5.8,6.2) {\small $z$-plane};

            \draw[red, dotted, line width=2pt] (0,0) -- (7,0);
            \draw[red, dotted,line width=2pt] (0,0) -- ({7*cos(144)},{7*sin(144)});
            \draw[red, dotted, line width=2pt] (0,0) -- ({7*cos(216)},{7*sin(216)});
            
            \draw[blue, line width=3pt] (0,0) -- ({6.5*cos(72)},{6.5*sin(72)});
            \node[blue] at ({6.5*cos(65)}, {6.5*sin(65)}) {$\mathfrak{S}_{2\pi}$};

            \draw[orange, line width=3pt] (0,0) -- ({6.5*cos(-72)},{6.5*sin(-72)});
            \node[orange] at ({6.5*cos(-65)}, {6.5*sin(-65)}) {$\mathfrak{S}_{3\pi}$};

            \fill[black] ({5*cos(162)},{5*sin(162)}) circle (2.4pt);
            \node[black] at (-4.8,2.25) {$z_{\rm p}$};
            \node[black] at (-2.55,1.35) {$(\nu,\rho)$};

            \draw[black, densely dotted, line width=1.4pt, ->]
                ({5*cos(162)},{5*sin(162)})
                arc[start angle=162,end angle=180,radius=5];

            \node[black] at (-5.3,0.65) {\scriptsize cut};
            
        \end{tikzpicture}%
        }
        \caption{Start on sheet $-1$.}
        \label{fig:P1monodromyContinuationSheetMinusOne}
    \end{subfigure}
    \hfill
    \begin{subfigure}[t]{0.48\textwidth}
        \centering
        \resizebox{\linewidth}{!}{%
        \begin{tikzpicture}[scale=0.75]

        \node[black] at (6.2,6.4) {\small sheet $0$};
        
            \draw[gray, line width=1pt, ->] (-7,0) -- (7,0);
            \draw[gray, line width=1pt, ->] (0,-7) -- (0,7);

            \draw[black, line width=1pt, decorate,
                  decoration={zigzag, segment length=7pt, amplitude=1.6pt}]
                  (-6.8,0) -- (-0.25,0);
                  
            \node[gray] at (-5.8,6.2) {\small $z$-plane};

            \draw[blue, line width=3pt] (0,0) -- (7,0);
            \node[blue] at (5.5, 0.5) {$\mathfrak{S}_{0}$};
            \node[black] at (2.3,+0.5) {$\boldsymbol S_{0}^{-1}\circ\boldsymbol S_{\pi}^{-1}(\nu,\rho)$};
            
            \draw[orange, line width=3pt] (0,0) -- ({7*cos(144)},{7*sin(144)});
            \node[orange] at ({7.5*cos(151.2)}, {7.5*sin(151.2)}) {$\mathfrak{S}_{-\pi}$};
            \node[black] at (-2.8,5.6) {$\boldsymbol S_{-\pi}^{-1}\circ\boldsymbol S_{0}^{-1}\circ\boldsymbol S_{\pi}^{-1}(\nu,\rho)$};
            
            \draw[orange, line width=3pt] (0,0) -- ({7*cos(216)},{7*sin(216)});
            \node[orange] at ({7.5*cos(208.8)}, {7.5*sin(208.8)}) {$\mathfrak{S}_{\pi}$};
            \node[black] at (-2.3,-2.55) {$\boldsymbol S_{\pi}^{-1}(\nu,\rho)$};
            
            \draw[red,dotted, line width=2pt] (0,0) -- ({6.5*cos(72)},{6.5*sin(72)});
            \draw[red,dotted, line width=2pt] (0,0) -- ({6.5*cos(-72)},{6.5*sin(-72)});

            \draw[black, densely dotted, line width=1.4pt, ->]
                ({5*cos(180)},{5*sin(180)})
                arc[start angle=180,end angle=522,radius=5];

            \fill[black] ({5*cos(234)},{5*sin(234)}) circle (2.4pt);
            \node[black] at (-2.4,-5.2) {$z_1$};

            \fill[black] ({5*cos(162)},{5*sin(162)}) circle (2.4pt);
            \node[black] at (-4.3,2.25) {$z_{\rm p}e^{2\pi i}$};
            
        \end{tikzpicture}%
        }
        \caption{End on sheet $0$.}
        \label{fig:P1monodromyContinuationSheetZero}
    \end{subfigure}

    \caption{One counterclockwise turn, drawn with $\epsilon=\pi/10$, from $\arg z_{\rm p}=-6\pi/5+\epsilon$. The path starts on sheet $-1$, crosses the branch cut, and continues on sheet $0$. It crosses the Stokes directions $\pi$, $0$, and $-\pi$ in that order. The intermediate point is $z_1=e^{2\pi i/5}z_{\rm p}$, and the endpoint is $z_{\rm p}e^{2\pi i}$.}
    \label{fig:P1monodromyContinuationSheets}
\end{figure}
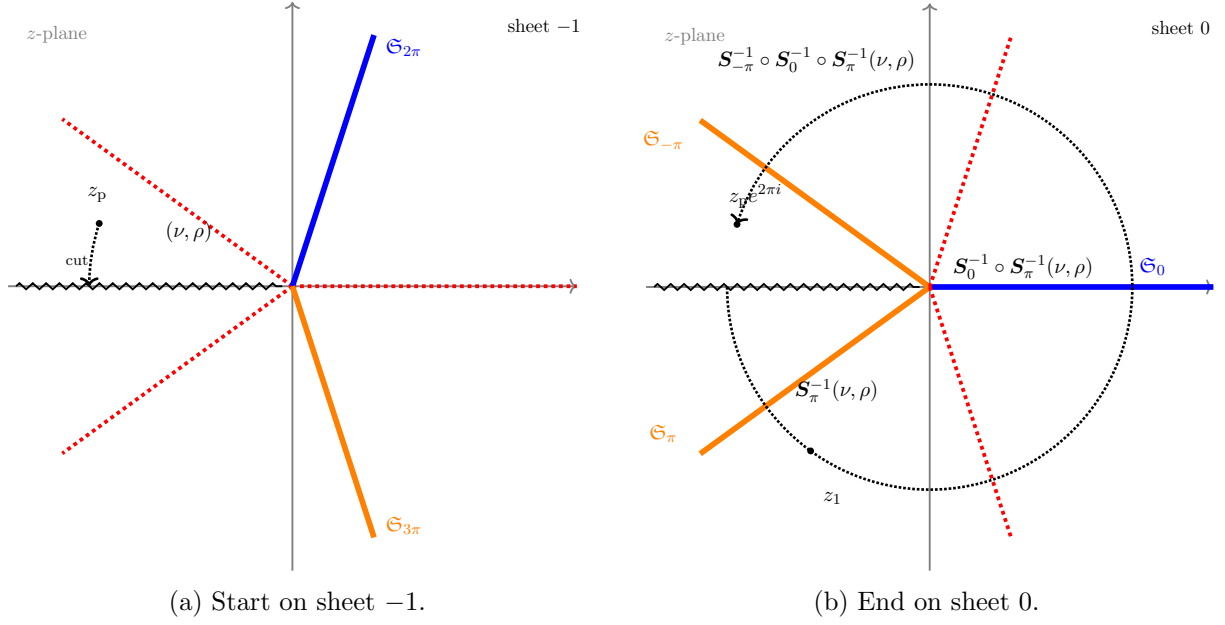

Analytic continuation therefore gives
\begin{equation}
 u(z_{\rm p};\nu,\rho)
 \longrightarrow
 u\left(z_{\rm p}e^{2\pi i};
 \boldsymbol S_{-\pi}^{-1}\circ
 \boldsymbol S_0^{-1}\circ
 \boldsymbol S_\pi^{-1}(\nu,\rho)\right),
 \label{eq:RM-Stokes-loop}
\end{equation}
where the rightmost transformation is applied first. The arrow denotes
analytic continuation, including the changes of transseries parameters.
The monodromy question is whether the expression on the right equals
the initial one.

The useful point is that the initial expression can also be rewritten
at the rotated argument by using modularity. Equations
\eqref{eq:S-PI-rotation-inverse} and
\eqref{eq:T-PI-rotation-inverse}, in the notation
\eqref{eq:RM-parameter-maps}, give
\begin{equation}
\begin{aligned}
 u(z_{\rm p};\nu,\rho)
 &=e^{4\pi i/5}u(z_1;\mathsf S^{-1}(\nu,\rho))\\
 &=e^{4\pi i/5}e^{16\pi i/5}
 u\left(e^{8\pi i/5}z_1;
 \mathsf T^{-1}\circ\mathsf S^{-1}(\nu,\rho)\right)\\
 &=u\left(z_{\rm p}e^{2\pi i};
 \mathsf T^{-1}\circ\mathsf S^{-1}(\nu,\rho)\right).
\end{aligned}
\label{eq:RM-modular-loop}
\end{equation}
The $S$ relation is used in the chamber established in
Section~\ref{sec:S-transformation-Stokes}: $z_{\rm p}$ and $z_1$ have
the Stokes line at $\arg z=-4\pi/5$ between them.

We can now compare the two descriptions of the full turn in
\eqref{eq:RM-Stokes-loop} and \eqref{eq:RM-modular-loop}.
On the modular side, the parameter maps
\eqref{eq:RM-parameter-maps} compose directly. On the resurgent side,
\eqref{eq:resurgent-Spi} and the rotations of the transseries determine
the three Stokes maps; their composition is worked out in
Appendix~\ref{app:parameter-dictionary}. The two calculations give
\begin{equation}
\begin{aligned}
 \mathsf T^{-1}\circ\mathsf S^{-1}(\nu,\rho)
 &=\left(\rho,\rho-\nu-\frac12\right),\\
 \boldsymbol S_{-\pi}^{-1}\circ\boldsymbol S_0^{-1}
 \circ\boldsymbol S_\pi^{-1}(\nu,\rho)
 &=\left(\rho,\rho-\nu-\frac12\right).
\end{aligned}
\label{eq:RM-explicit-loop}
\end{equation}
Thus the changes of parameters agree:
\begin{equation}
 \boxed{\boldsymbol S_{-\pi}^{-1}\circ
 \boldsymbol S_0^{-1}\circ\boldsymbol S_\pi^{-1}
 =\mathsf T^{-1}\circ\mathsf S^{-1}}.
\label{eq:RM-resurgence-modularity-inverse}
\end{equation}
Inverting this relation gives
\begin{equation}
 \boxed{\boldsymbol S_\pi\circ\boldsymbol S_0\circ\boldsymbol S_{-\pi}
 =\mathsf S\circ\mathsf T.}
\label{eq:RM-resurgence-modularity}
\end{equation}
This is the relation between resurgence and modularity: the three
nonlinear Stokes maps of the transseries combine into the affine
modular transformation obtained from the spectral curve. In the
inverse form above, the parameter change accumulated during analytic
continuation is exactly the one required by the modular rewriting of
the full turn. Equations \eqref{eq:RM-Stokes-loop} and
\eqref{eq:RM-modular-loop} therefore return the same specific heat,
closing its monodromy in one loop. For the tau function, it remains
to compare the multiplicative factors accompanying these two
descriptions; this is the next step.

\subsection{The tau-function constants}
\label{sec:RM-tau-constants}

We next determine the constants needed for the tau-function calculation.
Its second logarithmic derivative gives $u$, so a factor independent of
$z$ is invisible to the preceding argument. The Hamiltonian definition
fixes the $z$-dependence, while the extended normalization of
Remark~\ref{rem:modular-tau-normalization} fixes the parameter-dependent
factors. We retain the names $C_T$ and $C_S$ of
\eqref{eq:T-tau} and \eqref{eq:S-tau}; with
\eqref{eq:RM-parameter-maps}, these relations read
\begin{equation}
\begin{aligned}
 \tau(z;\nu,\rho)
 &=C_Te^{\pi i\nu^2}
 \tau(e^{8\pi i/5}z;\mathsf T^{-1}(\nu,\rho)),\\
 \tau(z;\nu,\rho)
 &=C_Se^{-2\pi i\nu\rho}
 \tau(e^{2\pi i/5}z;\mathsf S^{-1}(\nu,\rho)).
\end{aligned}
\label{eq:RM-tau-modular-relations}
\end{equation}

The $T$ constant follows immediately from the Fourier representation.
Under a rotation by $8\pi/5$, both $z^{5/4}$ and $z^{5/2}$ return to
their initial values, and each term $z^{-5k/4}$ is unchanged. The two
remaining powers in \eqref{eq:pinched-block} give
\begin{equation}
 \mathcal Z(e^{8\pi i/5}z,a)
 =e^{-\pi i/30-\pi ia^2}\mathcal Z(z,a).
 \label{eq:RM-block-T}
\end{equation}
Setting $a=\nu+n$ and using $n^2\equiv n\pmod2$,
\begin{equation}
 e^{-\pi i(\nu+n)^2}
 =e^{-\pi i\nu^2}e^{2\pi in(1/2-\nu)}.
\end{equation}
The extra Fourier weight is absorbed by the shift in
\eqref{eq:RM-parameter-maps}. Consequently
\begin{equation}
 \boxed{\tau(e^{8\pi i/5}z;\mathsf T^{-1}(\nu,\rho))
 =e^{-\pi i/30-\pi i\nu^2}\tau(z;\nu,\rho)},
 \qquad \boxed{C_T=e^{\pi i/30}.}
 \label{eq:RM-CT}
\end{equation}
Here the arguments refer to the sectorial prescription
\eqref{eq:RM-resummation-direction}; analytic continuation of a fixed
solution through the intervening Stokes lines also transforms its
parameters.

To determine $C_S$, we compare a single Stokes crossing with a single
rotation at a symmetric solution. The inverse Stokes automorphism
\eqref{eq:IM-inverse-jump} gives the continuation from $z_{\rm p}$ to $z_1$:
\begin{equation}
 \tau(z_{\rm p};\nu,\rho)
 \longrightarrow
 \exp\!\left\{\frac{i}{2\pi}\operatorname{Li}_2(e^{2\pi i\rho})\right\}
 \tau(z_1;\boldsymbol S_\pi^{-1}(\nu,\rho)).
 \label{eq:RM-tau-Stokes}
\end{equation}
We choose parameters for which the Stokes map in this formula gives
exactly the same pair as $\mathsf S^{-1}$ in
\eqref{eq:RM-tau-modular-relations}. Let
\begin{equation}
 y_* =\frac{\sqrt5-1}{2},
 \qquad r=\frac{\log y_*}{2\pi i},
 \qquad (\nu_*,\rho_*)=(-r,r).
 \label{eq:RM-fixed-point}
\end{equation}
Since $1-y_*=y_*^2$, with real logarithms of these positive numbers,
\begin{equation}
 \boldsymbol S_\pi^{-1}(-r,r)
 =\left(-r+\frac{\log(1-y_*)}{2\pi i},r\right)
 =(r,r)=\mathsf S^{-1}(-r,r).
 \label{eq:RM-fixed-point-matching}
\end{equation}
Thus both formulas involve the same tau function at $z_1$.

Substitution of $(-r,r)$ in \eqref{eq:S-Stokes-zpp} gives
$s_k=iy_*$ for all $k$. These are the Stokes data of the regular
symmetric solution used in
\cite[Section~3.3]{LisovyyRoussillon2017}.
For this solution $u(0)=\partial_z u(0)=0$, and $\tau(0)\ne0$.
As explained there, the tau function of this solution is invariant under the rotation:
\begin{equation}
 \tau(e^{2\pi i/5}z)=\tau(z)
 \quad\text{for this symmetric solution.}
 \label{eq:RM-symmetric-tau}
\end{equation}
Here both sides are values of the same analytic solution, continued
from the origin. In particular, the continuation in
\eqref{eq:RM-tau-Stokes} returns its initial value for these parameters.
Comparison with \eqref{eq:RM-tau-modular-relations} therefore gives
\begin{equation}
\begin{aligned}
 \tau(z_{\rm p};-r,r)
 &=\exp\!\left\{\frac{i}{2\pi}\operatorname{Li}_2(y_*)\right\}
   \tau(z_1;r,r)\\
 &=C_Se^{2\pi i r^2}\tau(z_1;r,r).
\end{aligned}
\label{eq:RM-CS-comparison}
\end{equation}
The common factor at $z_1$ cancels, fixing
\begin{equation}
 C_S=\exp\!\left\{
 \frac{i}{2\pi}\operatorname{Li}_2(y_*)-2\pi i r^2\right\}
 =\exp\!\left\{\frac{i}{2\pi}
 \bigl(\operatorname{Li}_2(y_*)+\log^2 y_*\bigr)\right\}.
 \label{eq:RM-CS-evaluation}
\end{equation}
Using the classical value
$\operatorname{Li}_2(y_*)=\pi^2/10-\log^2 y_*$, also recalled in
\cite{LisovyyRoussillon2017}, we obtain
\begin{equation}
 \boxed{C_S=e^{\pi i/20},}
 \qquad
 \boxed{\tau(e^{2\pi i/5}z;\nu,\rho)
 =e^{-\pi i/20-2\pi i\nu\rho}
 \tau(z;-\rho,\nu)}.
 \label{eq:RM-CS}
\end{equation}

\subsection{One-loop monodromy of the tau function}
\label{sec:RM-tau}

With both constants fixed, we can follow the tau function through a
full turn. It is convenient to organize the continuation as five equal
angular steps, using the rotational symmetry after each step to express
the result at the original base point. For the specific heat, a single
step reads
\begin{equation}
 \boxed{u(z_{\rm p};\nu,\rho)
 \longrightarrow u(z_1;\boldsymbol S_\pi^{-1}(\nu,\rho))
 =e^{-4\pi i/5}u(z_{\rm p};\mathrm m(\nu,\rho))},
 \label{eq:RM-cyclic-u}
\end{equation}
where\footnote{
We note that the transformation $m$ can be identified with 
the mutation loop of type $A_2$ studied by 
Ishibashi--Terashima~\cite[Example 2.5]{IT26}, 
after a suitable change of variables.}
\begin{equation}
     \boxed{\mathrm{m}=\mathsf S\circ\boldsymbol S_\pi^{-1}},
    \qquad
    \boxed{\mathrm{m}(\nu,\rho)
    =\left(-\rho,\nu+\frac{\log(1-e^{2\pi i\rho})}{2\pi i}\right)}.
 \label{eq:RM-cyclic-map}
\end{equation}
The arrow is analytic continuation from $z_{\rm p}$ to $z_1$, while
the equality uses the modular relation to rewrite the endpoint value.
Thus the next angular step is obtained by applying the same calculation
to $\mathrm m(\nu,\rho)$. The Stokes crossing is always described in
this rotated frame, as illustrated in
Figure~\ref{fig:P1cyclicMonodromyStep}.

\begin{figure}[htbp]
    \centering

    \resizebox{0.92\textwidth}{!}{%
    \begin{tikzpicture}[scale=0.75]

        \begin{scope}
        
        \node[black] at (6.2,6.8) {\small sheet $-1$};
        
            \draw[gray, line width=1pt, ->] (-7,0) -- (7,0);
            \draw[gray, line width=1pt, ->] (0,-7) -- (0,7);

            \draw[black, line width=1pt, decorate,
                  decoration={zigzag, segment length=7pt, amplitude=1.6pt}]
                  (-6.8,0) -- (-0.25,0);
                  
            \node[gray] at (-5.8,6.2) {\small $z$-plane};

            \draw[red, dotted, line width=2pt] (0,0) -- (7,0);
            \draw[red, dotted,line width=2pt] (0,0) -- ({7*cos(144)},{7*sin(144)});
            \draw[red, dotted, line width=2pt] (0,0) -- ({7*cos(216)},{7*sin(216)});
            
            \draw[blue, line width=3pt] (0,0) -- ({6.5*cos(72)},{6.5*sin(72)});
            \draw[orange, line width=3pt] (0,0) -- ({6.5*cos(-72)},{6.5*sin(-72)});

            \coordinate (startZp) at ({5*cos(162)},{5*sin(162)});
            
            \fill[black] (startZp) circle (2.7pt);
            \node[black] at (-4.8,2.25) {$z_{\rm p}$};
            \node[black] at (-2.55,1.35) {\scriptsize $(\nu_j,\rho_j)$};

            \draw[black, densely dotted, line width=1.5pt, ->]
                ({5*cos(162)},{5*sin(162)})
                arc[start angle=162,end angle=179,radius=5];

            \node[black] at (0,-7.35) {\small (a) Start on sheet $-1$.};

        \end{scope}

        \begin{scope}[xshift=15.5cm]

        \node[black] at (6.2,6.8) {\small sheet $0$};
        
            \draw[gray, line width=1pt, ->] (-7,0) -- (7,0);
            \draw[gray, line width=1pt, ->] (0,-7) -- (0,7);

            \draw[black, line width=1pt, decorate,
                  decoration={zigzag, segment length=7pt, amplitude=1.6pt}]
                  (-6.8,0) -- (-0.25,0);
                  
            \node[gray] at (-5.8,6.2) {\small $z$-plane};

            \draw[blue, line width=3pt] (0,0) -- (7,0);
            \node[blue] at (5.5, 0.5) {$\mathfrak{S}_{0}$};
            
            \draw[orange, line width=3pt] (0,0) -- ({7*cos(144)},{7*sin(144)});
            \node[orange] at ({7.5*cos(151.2)}, {7.5*sin(151.2)}) {$\mathfrak{S}_{-\pi}$};
            
            \draw[orange, line width=3pt] (0,0) -- ({7*cos(216)},{7*sin(216)});
            \node[orange] at ({7.5*cos(208.8)}, {7.5*sin(208.8)}) {$\mathfrak{S}_{\pi}$};
            
            \draw[red,dotted, line width=2pt] (0,0) -- ({6.5*cos(72)},{6.5*sin(72)});
            \draw[red,dotted, line width=2pt] (0,0) -- ({6.5*cos(-72)},{6.5*sin(-72)});

            \draw[black, densely dotted, line width=1.5pt, ->]
                ({5*cos(181)},{5*sin(181)})
                arc[start angle=181,end angle=234,radius=5];

            \coordinate (endZ) at ({5*cos(234)},{5*sin(234)});
            
            \fill[black] (endZ) circle (2.7pt);
            \node[black] at (-2.4,-4.6) {$z_1$};

            \node[black] at (0,-7.35) {\small (b) End on sheet $0$.};

        \end{scope}

    \end{tikzpicture}%
    }

    \vspace{0.12cm}

    \[
    \boxed{
    \begin{gathered}
        u(z_{\rm p};\nu_j,\rho_j)
        \longrightarrow
        u\left(z_1;\boldsymbol S_\pi^{-1}(\nu_j,\rho_j)\right)
        =
        e^{-\frac{4\pi i}{5}}
        u(z_{\rm p};\nu_{j+1},\rho_{j+1}),
        \\[4pt]
        (\nu_{j+1},\rho_{j+1})
        =
        \mathsf S\circ\boldsymbol S_\pi^{-1}(\nu_j,\rho_j).
    \end{gathered}
    }
    \]

    \caption{One cyclic step, drawn with $\epsilon=\pi/10$.
    The black dotted path is analytic continuation from $z_{\rm p}$
    on sheet $-1$ to $z_1=e^{2\pi i/5}z_{\rm p}$ on sheet $0$.
    It crosses the negative-real-axis cut and then the $\pi$ Stokes
    line, changing the parameters by $\boldsymbol S_\pi^{-1}$.
    The equality below the panels uses the modular relation to express
    this endpoint value at $z_{\rm p}$ with parameters
    $(\nu_{j+1},\rho_{j+1})$.
    Solid coloured rays are Stokes lines, and red dotted rays are
    anti-Stokes lines.}
    \label{fig:P1cyclicMonodromyStep}
\end{figure}

Writing $(\nu_{j+1},\rho_{j+1})=\mathrm m(\nu_j,\rho_j)$, two steps give
$e^{-8\pi i/5}u(z_{\rm p};\nu_3,\rho_3)$, and five give
$u(z_{\rm p};\nu_6,\rho_6)$. The phase has returned to one; below we
verify that $\mathrm m^5=\mathrm{id}$ on consistent logarithmic branches.

For the tau function we combine the Stokes factor in
\eqref{eq:RM-tau-Stokes} with the rotation formula \eqref{eq:RM-CS}.
Since $\boldsymbol S_\pi^{-1}(\nu_j,\rho_j)
=(\rho_{j+1},-\nu_{j+1})$, the elementary step is
\begin{equation}
 \boxed{\tau(z_{\rm p};\nu_j,\rho_j)
 \longrightarrow
 P_j\tau(z_{\rm p};\nu_{j+1},\rho_{j+1})},
 \qquad
 \boxed{P_j=\exp\!\left\{-\frac{\pi i}{20}
 +\frac{i}{2\pi}\operatorname{Li}_2(e^{2\pi i\rho_j})
 +2\pi i\nu_{j+1}\rho_{j+1}\right\}}.
 \label{eq:RM-cyclic-tau-factor}
\end{equation}

Introduce $y_j=e^{2\pi i\rho_j}$. The recurrence
\eqref{eq:RM-cyclic-map} gives
\begin{equation}
 \nu_j=-\rho_{j-1},
 \qquad y_{j-1}y_{j+1}=1-y_j.
 \label{eq:RM-five-term-recurrence}
\end{equation}
The five-periodicity can be seen directly. Starting with $y_0=a$ and
$y_1=b$, the next values are
\begin{equation}
 y_2=\frac{1-b}{a},\qquad
 y_3=\frac{a+b-1}{ab},\qquad
 y_4=\frac{1-a}{b},\qquad
 y_5=a,\qquad y_6=b.
 \label{eq:RM-five-term-orbit}
\end{equation}
We first take $0<a,b<1$ and $a+b>1$, so that all five $y_j$ lie in
$(0,1)$ and the logarithms are real. Then
$\rho_j=\log y_j/(2\pi i)$ satisfies the additive recurrence without
integer shifts, and $(\nu_6,\rho_6)=(\nu_1,\rho_1)$.
The identities below extend by analytic continuation with these
logarithms.

After five steps the parameters have returned, and the remaining
monodromy factor is
\begin{equation}
 \boxed{\mathcal P=\prod_{j=1}^5P_j
 =\exp\!\left\{-\frac{\pi i}{4}
 +\frac{i}{2\pi}\sum_{j=1}^5\operatorname{Li}_2(y_j)
 +2\pi i\sum_{j=1}^5\nu_j\rho_j\right\}}.
 \label{eq:RM-total-prefactor}
\end{equation}
The constant $-\pi i/4$ comes from the five factors $C_S^{-1}$.
To evaluate the other terms, recall the Rogers dilogarithm
\begin{equation}
 L(y)=\operatorname{Li}_2(y)+\frac12\log y\log(1-y).
 \label{eq:RM-Rogers}
\end{equation}
Equation \eqref{eq:RM-five-term-recurrence}, with the logarithms just
specified, gives
\begin{equation}
 \log(1-y_j)=2\pi i(\rho_{j-1}+\rho_{j+1}).
\end{equation}
Consequently
\begin{equation}
 \frac{i}{2\pi}\sum_{j=1}^5\operatorname{Li}_2(y_j)
 =\frac{i}{2\pi}\sum_{j=1}^5L(y_j)
 +2\pi i\sum_{j=1}^5\rho_j\rho_{j+1}.
\end{equation}
The second term cancels the quadratic term in
\eqref{eq:RM-total-prefactor}, because
$\sum_j\nu_j\rho_j=-\sum_j\rho_{j-1}\rho_j$.
Thus
\begin{equation}
 \mathcal P=\exp\!\left\{-\frac{\pi i}{4}
 +\frac{i}{2\pi}\sum_{j=1}^5L(y_j)\right\}.
 \label{eq:RM-Rogers-prefactor}
\end{equation}
The five arguments in \eqref{eq:RM-five-term-orbit} are precisely those
of Abel's five-term identity, which in this form reads
\begin{equation}
 \sum_{j=1}^5L(y_j)=\frac{\pi^2}{2}
 \label{eq:RM-Abel}
\end{equation}
\cite{LisovyyRoussillon2017,Tamarin2026Thesis}. Substitution gives
\begin{equation}
 \boxed{\mathcal P=
 \exp\!\left(-\frac{\pi i}{4}+\frac{\pi i}{4}\right)=1.}
 \label{eq:RM-tau-closure}
\end{equation}
The parameter recurrence closes after five steps, and the tau-function
prefactors cancel. The same one-loop continuation that gave
\eqref{eq:RM-resurgence-modularity} for the specific heat therefore
returns the tau function to its initial value as well.

\section{Conclusions and outlook}
\label{sec:conclusions}

The identity \eqref{eq:RM-resurgence-modularity} provides
an explicit relation between resurgence and modularity
for Painlev\'e~I: the product
$\boldsymbol S_\pi\circ\boldsymbol S_0\circ
\boldsymbol S_{-\pi}$ of nonlinear Stokes maps coincides
with the modular composition $\mathsf S\circ\mathsf T$.
Together with the tau-function transformation laws, this
makes the compatibility of the two descriptions of
analytic continuation explicit.
The five-step cyclic parameter map closes after one full
turn around infinity,
and the accumulated tau-function
prefactors cancel by Abel's five-term identity.

A fuller treatment of the chamber dependence of the
$S$-relation and of the Poisson resummation approach will
be presented in forthcoming work.
More broadly, it is natural to ask whether an analogous
Stokes--modular correspondence can be made explicit for
Painlev\'e~II\footnote{ 
For Painlev\'e~II,
\cite[Section~5]{IorgovLisovyyTamarinZhuravlov2026}
derives connection coefficients between rays of the same
type by Stokes continuation and rotational symmetry.
For Painlev\'e~I,
\cite[Section~3.2]{Tamarin2026Thesis}
identifies the cyclic relation for the connection
coefficients of \cite{LisovyyRoussillon2017} with the
dilogarithmic prefactor cancellation in the tau-function
monodromy calculation.
}
and the other Painlev\'e equations, including
the multiplicative factors in their tau-function
transformations.

The higher members of the Painlev\'e~I (KdV) hierarchy
provide another natural setting for this question.
The transseries and monodromy analysis developed in
\cite{Tamarin2026Thesis,KagerEtAl2026Hierarchy} provides
substantial input from resurgence.
It would be interesting to determine whether the cyclic
structures and dilogarithm identities found or conjectured
there admit a corresponding geometric interpretation.
For spectral curves of higher genus, the natural analogue
of the modular transformations considered here would involve
symplectic changes of the larger homology basis.

\appendix

\section{Theta Functions with Characteristics and Modular Transformations}

For $\operatorname{Im}\tau>0$, define
\[
\thc{\epsilon}{\eta}(z\mid\tau)
=\sum_{n\in\mathbb Z}
\exp\!\left(\pi i(n+\epsilon)^2\tau
+2\pi i(n+\epsilon)(z+\eta)\right).
\]
We use $A\cdot B=1$, $\tau=\omega_B/\omega_A$, and
$z=\omega_A^{-1}\int_{p_0}^p\omega$, with a fixed integration path.
The matrices act on the cycle basis from the right:
\[
(A',B')=(A,B)M,\qquad
T=\begin{pmatrix}1&1\\0&1\end{pmatrix},\qquad
S=\begin{pmatrix}0&-1\\1&0\end{pmatrix}.
\]

\[
\begin{array}{c|c|c|c|c}
M&(A',B')&\tau'&z'&(\epsilon',\eta')\\ \hline
T&(A,B+A)&\tau+1&z&(\epsilon,\eta-\epsilon+\tfrac12)\\
T^{-1}&(A,B-A)&\tau-1&z&(\epsilon,\eta+\epsilon-\tfrac12)\\
S&(B,-A)&-1/\tau&z/\tau&(-\eta,\epsilon)\\
S^{-1}&(-B,A)&-1/\tau&-z/\tau&(\eta,-\epsilon)
\end{array}
\]

With these exact representatives of the characteristics,
\begin{align}
\thc{\epsilon}{\eta-\epsilon+\frac12}(z\mid\tau+1)
&=e^{\pi i\epsilon(1-\epsilon)}
\thc{\epsilon}{\eta}(z\mid\tau),\qquad &&(T)\\
\thc{\epsilon}{\eta+\epsilon-\frac12}(z\mid\tau-1)
&=e^{-\pi i\epsilon(1-\epsilon)}
\thc{\epsilon}{\eta}(z\mid\tau), &&(T^{-1})\\
\thc{-\eta}{\epsilon}\!\left(\frac z\tau\Bigm|-\frac1\tau\right)
&=(-i\tau)^{1/2}e^{\pi i z^2/\tau-2\pi i\epsilon\eta}
\thc{\epsilon}{\eta}(z\mid\tau), &&(S)\\
\thc{\eta}{-\epsilon}\!\left(-\frac z\tau\Bigm|-\frac1\tau\right)
&=(-i\tau)^{1/2}e^{\pi i z^2/\tau-2\pi i\epsilon\eta}
\thc{\epsilon}{\eta}(z\mid\tau). &&(S^{-1})
\end{align}
The square root is the holomorphic branch on the upper half-plane
with $(-i\tau)^{1/2}=1$ at $\tau=i$.

For $m,n\in\mathbb Z$, changing representatives gives
\[
\thc{\epsilon+m}{\eta+n}(z\mid\tau)
=e^{2\pi i\epsilon n}\thc{\epsilon}{\eta}(z\mid\tau).
\]
Also,
\[
\thc{-\epsilon}{-\eta}(-z\mid\tau)
=\thc{\epsilon}{\eta}(z\mid\tau).
\]
In particular, $S$ and $S^{-1}$ have the same action on $\tau$
but opposite actions on $z$; indeed $S^2=-I$.
The characteristics are not reduced modulo $\mathbb Z^2$ in the
four transformation identities above, so the displayed phases apply
without additional representative changes.

\section{Transseries normalization and the one-loop Stokes product}
\label{app:parameter-dictionary}

We give two details of the comparison between resurgence and modularity.
First, we check that the Fourier representation of Section~\ref{sec:bssv}
has the same parameter normalization as the transseries for the
Painlev\'e~I transcendent. We then explain the rotations used to relate
the Stokes maps and evaluate their product in
Section~\ref{sec:RM-specific-heat}. Throughout, $g_s=1$ and we use
$\alpha=-2\sqrt{2/3}$ and $\nu=\alpha\sigma_1\sigma_2/2$.

\subsection{Matching the transseries and Fourier normalizations}

The parameters $\sigma_1$ and $\sigma_2$ in
\eqref{eq:resurgent-PI-transseries} were fixed by requiring
$u_0^{(1,0)}=u_0^{(0,1)}=1$. We now verify explicitly that the dictionary
\eqref{eq:resurgent-Fourier-dictionary} and the Fourier block
\eqref{eq:pinched-block} reproduce these leading one-instanton
coefficients of $u$.

Substitute the dictionary into the Fourier sum
\eqref{eq:pinched-Fourier} and expand in $\sigma_1,\sigma_2$.
The terms linear in these parameters come from the modes $n=1$ and
$n=-1$. Using $G_2(w+1)=\Gamma(w)G_2(w)$ to compare each of these modes
with the zero mode gives
\begin{equation}
\begin{aligned}
 \frac{\tau(z;\nu,\rho)}{\mathcal Z(z,\nu)}
 =1-\frac{z^{-5/8}}{12}\Bigl[
 &\sigma_1e^{-A(z)}\bigl(1+O(z^{-5/4})\bigr)\\
 &+\sigma_2e^{A(z)}\bigl(1+O(z^{-5/4})\bigr)\Bigr]
 +O\bigl((\sigma_1,\sigma_2)^2\bigr).
\end{aligned}
\label{eq:app-leading-sector-check}
\end{equation}
Here $A(z)=\widetilde A z^{5/4}$, as in
\eqref{eq:resurgent-background}, and the last term denotes terms of
total degree at least two in the transseries parameters. In particular,
the factor $e^{-2\pi i\rho}$ in the $n=-1$ mode is proportional to
$\sigma_1^{-1}$, but the Barnes ratio contributes
$1/\Gamma(\nu)=\nu+O(\nu^2)$. Their product is therefore proportional
to $\nu/\sigma_1=\alpha\sigma_2/2$, as required.

Taking the logarithm leaves the linear terms unchanged. Moreover,
$\mathcal Z(z,\nu)$ has no linear dependence on $\sigma_1,\sigma_2$,
since $\nu$ is quadratic in them. In $u=-2\partial_z^2\log\tau$,
the leading contribution from each exponential comes from
differentiating it twice. Since $A'(z)^2=6\sqrt z$, the factor
$(-2)(-1/12)A'(z)^2z^{-5/8}$ is $z^{-1/8}$. Consequently,
\begin{equation}
\begin{aligned}
 u(z;\sigma_1,\sigma_2)=u(z;0,0)+z^{-1/8}\Bigl[
 &\sigma_1e^{-A(z)}\bigl(1+O(z^{-5/4})\bigr)\\
 &+\sigma_2e^{A(z)}\bigl(1+O(z^{-5/4})\bigr)\Bigr]
 +O\bigl((\sigma_1,\sigma_2)^2\bigr).
\end{aligned}
\label{eq:app-unit-u-sectors}
\end{equation}
Here $u(z;0,0)$ denotes the perturbative sector in the $\sigma$
parameters. Since $\beta_{1,0}=\beta_{0,1}=1/2$ in
\eqref{eq:resurgent-PI-transseries}, its elementary sectors start with
$\sqrt z\,z^{-5/8}=z^{-1/8}$. Thus
\eqref{eq:app-unit-u-sectors} gives precisely
$u_0^{(1,0)}=u_0^{(0,1)}=1$, checking the normalization used to pass
between the two descriptions.

\subsection{Rotations and the Stokes product}

To evaluate the product of Stokes maps in
\eqref{eq:RM-explicit-loop}, we need the jumps at Borel angles $0$ and
$-\pi$ as well as the jump at $\pi$ given in
\eqref{eq:resurgent-Spi}. They are related by rotations of the
transseries. A rotation $z\mapsto e^{\pm4\pi i/5}z$ changes
$\widetilde x=z^{-5/4}$ to $e^{\mp\pi i}\widetilde x$: it is a
half-turn in the expansion variable $\widetilde x$. This is the
meaning of the \emph{half rotations} used in
\cite{KagerEtAl2026Global} and
\cite[Appendix~A]{Tamarin2026Thesis}.

Under either rotation $A(z)$ changes sign, so the two elementary
exponentials are exchanged. The rotated transseries can therefore be
written at the original argument by exchanging its parameters and
including the phases from the powers of $z$. Explicitly,
\begin{equation}
\begin{aligned}
 u(e^{4\pi i/5}z;\sigma_1,\sigma_2)
 &=e^{2\pi i/5}u(z;R_{-1/2}(\sigma_1,\sigma_2)),\\
 u(e^{-4\pi i/5}z;\sigma_1,\sigma_2)
 &=e^{-2\pi i/5}u(z;R_{1/2}(\sigma_1,\sigma_2)),
\end{aligned}
\label{eq:app-half-rotations-u}
\end{equation}
where
\begin{equation}
\begin{aligned}
 R_{-1/2}(\sigma_1,\sigma_2)
 &=(-i\sigma_2e^{\pi i\nu},-i\sigma_1e^{-\pi i\nu}),\\
 R_{1/2}(\sigma_1,\sigma_2)
 &=(i\sigma_2e^{-\pi i\nu},i\sigma_1e^{\pi i\nu}).
\end{aligned}
\end{equation}
Both maps send $\nu$ to $-\nu$. The factors $\pm i$ come from the
fractional powers of $z$, while $e^{\pm\pi i\nu}$ come from the
power containing $\nu$ in \eqref{eq:resurgent-PI-transseries}.
The relations between the coefficients of the exchanged sectors,
given in \cite[Appendix~A]{Tamarin2026Thesis}, extend these identities
to the full formal transseries.

For the composition below it is convenient to use
$a=e^{2\pi i\nu}$ and $b=e^{2\pi i\rho}$.
Substituting the rotations into
\eqref{eq:resurgent-Fourier-dictionary} and using
$\Gamma(1+\nu)\Gamma(1-\nu)=\pi\nu/\sin\pi\nu$ gives
\begin{equation}
 R_{-1/2}(a,b)=\left(\frac1a,\frac{1-a}{b}\right),
 \qquad
 R_{1/2}(a,b)=\left(\frac1a,\frac{1-a^{-1}}{b}\right).
\label{eq:app-half-rotations-parameters}
\end{equation}
These maps are inverse to one another. Applying each twice gives
\begin{equation}
 R_{-1/2}^2(a,b)=\left(a,-\frac ba\right),
 \qquad
 R_{1/2}^2(a,b)=(a,-ab),
\end{equation}
which are the exponentiated actions of $\mathsf T$ and
$\mathsf T^{-1}$ in \eqref{eq:RM-parameter-maps}.

A half rotation changes the Borel angle by $\pi$. We can therefore
express a jump in a neighbouring direction by rotating to the
$\pi$ description, applying its known jump, and undoing the rotation.
In our conventions this gives
\begin{equation}
\begin{aligned}
 \boldsymbol S_0^{-1}
 &=R_{1/2}\circ\boldsymbol S_\pi^{-1}\circ R_{-1/2},\\
 \boldsymbol S_{-\pi}^{-1}
 &=R_{1/2}^2\circ\boldsymbol S_\pi^{-1}\circ R_{-1/2}^2.
\end{aligned}
\label{eq:app-Stokes-conjugation}
\end{equation}
The inverse rotations cancel between successive jumps, leaving
\begin{equation}
 \boldsymbol S_{-\pi}^{-1}\circ\boldsymbol S_0^{-1}
 \circ\boldsymbol S_\pi^{-1}
 =R_{1/2}^2\circ\boldsymbol S_\pi^{-1}\circ R_{-1/2}
 \circ\boldsymbol S_\pi^{-1}\circ R_{-1/2}
 \circ\boldsymbol S_\pi^{-1}.
\label{eq:app-Stokes-product}
\end{equation}
Now \eqref{eq:resurgent-Spi} reads
$\boldsymbol S_\pi^{-1}(a,b)=(a(1-b),b)$.
The five rightmost maps in \eqref{eq:app-Stokes-product} can thus be
evaluated directly:
\begin{equation}
\begin{aligned}
 (a,b)&\xrightarrow{\ \boldsymbol S_\pi^{-1}\ }
 (a(1-b),b)
 \xrightarrow{\ R_{-1/2}\ }
 \left(\frac1{a(1-b)},\frac{1-a+ab}{b}\right)\\
 &\xrightarrow{\ \boldsymbol S_\pi^{-1}\ }
 \left(\frac{a-1}{ab},\frac{1-a+ab}{b}\right)
 \xrightarrow{\ R_{-1/2}\ }
 \left(\frac{ab}{a-1},\frac1a\right)
 \xrightarrow{\ \boldsymbol S_\pi^{-1}\ }
 \left(b,\frac1a\right).
\end{aligned}
\label{eq:app-five-map-algebra}
\end{equation}
This is the exponentiated action of $\mathsf S^{-1}$. The remaining
$R_{1/2}^2$ acts as $\mathsf T^{-1}$ and sends it to $(b,-b/a)$.
With the representatives used in \eqref{eq:RM-parameter-maps}, the
result is
\begin{equation}
 \boldsymbol S_{-\pi}^{-1}\circ\boldsymbol S_0^{-1}
 \circ\boldsymbol S_\pi^{-1}(\nu,\rho)
 =\left(\rho,\rho-\nu-\frac12\right).
\end{equation}
This gives the Stokes-side calculation in \eqref{eq:RM-explicit-loop}
and completes the comparison with the modular parameter maps.


\begin{thebibliography}{99}

\bibitem{AnicetoSchiappaVonk2012}
I.~Aniceto, R.~Schiappa and M.~Vonk,
\emph{The resurgence of instantons in string theory},
Commun. Number Theory Phys. \textbf{6} (2012), 339--496;
\href{https://arxiv.org/abs/1106.5922}{arXiv:1106.5922}.

\bibitem{BSSV2022}
S.~Baldino, R.~Schiappa, M.~Schwick and R.~Vega,
\emph{Resurgent Stokes Data for Painlev\'e Equations and
Two-Dimensional Quantum (Super) Gravity},
Commun. Number Theory Phys. \textbf{17} (2023), 385--552;
\href{https://doi.org/10.4310/cntp.2023.v17.n2.a5}
{doi:10.4310/cntp.2023.v17.n2.a5};
\href{https://arxiv.org/abs/2203.13726}{arXiv:2203.13726}.

\bibitem{BLMST2016}
G.~Bonelli, O.~Lisovyy, K.~Maruyoshi, A.~Sciarappa and A.~Tanzini,
\emph{On Painlev\'e/gauge theory correspondence},
Lett. Math. Phys. \textbf{107} (2017), 2359--2413;
\href{https://arxiv.org/abs/1612.06235}{arXiv:1612.06235}.

\bibitem{EM11}
B.~Eynard and M.~Mari\~no,
A holomorphic and background independent partition
function for matrix models and topological strings,
J. Geom. Phys. 61 (2011), 1181--1202.
\href{https://doi.org/10.1016/j.geomphys.2010.11.012}
{doi:10.1016/j.geomphys.2010.11.012},
\href{https://arxiv.org/abs/0810.4273}
{arXiv:0810.4273}.

\bibitem{EynardOrantin2007}
B.~Eynard and N.~Orantin,
\emph{Invariants of algebraic curves and topological expansion},
Commun. Number Theory Phys. \textbf{1} (2007), 347--452;
\href{https://arxiv.org/abs/math-ph/0702045}
{arXiv:math-ph/0702045}.

\bibitem{GuMarino2023}
J.~Gu and M.~Mari\~no,
\emph{Exact multi-instantons in topological string theory},
SciPost Phys. \textbf{15} (2023), 179;
\href{https://doi.org/10.21468/SciPostPhys.15.4.179}
{doi:10.21468/SciPostPhys.15.4.179};
\href{https://arxiv.org/abs/2211.01403}{arXiv:2211.01403}.

\bibitem{IILZ2025}
N.~Iorgov, K.~Iwaki, O.~Lisovyy and Yu.~Zhuravlov,
\emph{Many-faced Painlev\'e I: irregular conformal blocks,
topological recursion, and holomorphic anomaly approaches},
\href{https://arxiv.org/abs/2505.16803}{arXiv:2505.16803} (2025).

\bibitem{IorgovLisovyyTamarinZhuravlov2026}
N.~Iorgov, O.~Lisovyy, N.~Tamarin and Yu.~Zhuravlov,
\emph{Painlev\'e II connection problem revisited},
\href{https://arxiv.org/abs/2610.28227}{arXiv:2610.28227} (2026).

\bibitem{GIKM12}
S.~Garoufalidis, A.~Its, A.~Kapaev and M.~Mari\~no,
Asymptotics of the instantons of Painlev\'e I,
Int. Math. Res. Not. IMRN 2012 (2012), no.~3, 561--606.
\href{https://doi.org/10.1093/imrn/rnr029}{doi:10.1093/imrn/rnr029},
\href{https://arxiv.org/abs/1002.3634}{arXiv:1002.3634}.


\bibitem{Iwaki2019}
K.~Iwaki,
\emph{2-parameter $\tau$-function for the first Painlev\'e equation:
Topological recursion and direct monodromy problem via exact WKB analysis},
Commun. Math. Phys. \textbf{377} (2020), 1047--1098;
\href{https://arxiv.org/abs/1902.06439}{arXiv:1902.06439} (2019).

\bibitem{IwakiMarino2024}
K.~Iwaki and M.~Mari\~no,
\emph{Resurgent Structure of the Topological String and the
First Painlev\'e Equation},
SIGMA \textbf{20} (2024), 028, 21~pages;
\href{https://doi.org/10.3842/SIGMA.2024.028}
{doi:10.3842/SIGMA.2024.028};
\href{https://arxiv.org/abs/2307.02080}{arXiv:2307.02080}.

\bibitem{KagerEtAl2026Global}
J.~Kager, J.~Rodrigues, R.~Schiappa, M.~Schwick and N.~Tamarin,
\emph{Exact Solutions to Matrix Models and String Theories:
The Global Construction},
to appear (2026).

\bibitem{KagerEtAl2026Local}
J.~Kager, J.~Rodrigues, R.~Schiappa, M.~Schwick and N.~Tamarin,
\emph{Exact Solutions to Matrix Models and String Theories:
The Local Construction},
\href{https://arxiv.org/abs/2602.15101}{arXiv:2602.15101} (2026).

\bibitem{KagerEtAl2026Hierarchy}
J.~Kager, J.~Rodrigues, R.~Schiappa, M.~Schwick and N.~Tamarin,
\emph{From Resurgence to Monodromy Along the KdV Hierarchy},
to appear (2026).

\bibitem{Kapaev2004}
A.~A.~Kapaev,
\emph{Quasi-linear Stokes phenomenon for the
Painlev\'e first equation},
J. Phys. A: Math. Gen. \textbf{37} (2004), 11149--11167;
\href{https://arxiv.org/abs/nlin/0404026}{arXiv:nlin/0404026}.

\bibitem{KapaevKitaev1993}
A.~A.~Kapaev and A.~V.~Kitaev,
\emph{Connection formulae for the first Painlev\'e
transcendent in the complex domain},
Lett. Math. Phys. \textbf{27} (1993), 243--252;
\href{https://doi.org/10.1007/BF00777371}
{doi:10.1007/BF00777371}.

\bibitem{LisovyyRoussillon2017}
O.~Lisovyy and J.~Roussillon,
\emph{On the connection problem for Painlev\'e I},
J. Phys. A: Math. Theor. \textbf{50} (2017), 255202;
\href{https://arxiv.org/abs/1612.08382}{arXiv:1612.08382}.

\bibitem{MarinoSchiappaSchwick2022}
M.~Mari\~no, R.~Schiappa and M.~Schwick,
\emph{New Instantons for Matrix Models},
\href{https://arxiv.org/abs/2210.13479}{arXiv:2210.13479}.

\bibitem{SchiappaSchwickTamarin2026}
R.~Schiappa, M.~Schwick and N.~Tamarin,
\emph{All the D-Branes of Resurgence},
SciPost Phys. \textbf{20} (2026), 135;
\href{https://arxiv.org/abs/2301.05214}{arXiv:2301.05214}.

\bibitem{Takei1995}
Y.~Takei,
\emph{On the connection formula for the first Painlev\'e
equation---from the viewpoint of the exact WKB analysis},
RIMS K\^oky\^uroku \textbf{931} (1995), 70--99;
\href{https://www.kurims.kyoto-u.ac.jp/~kyodo/kokyuroku/contents/pdf/0931-6.pdf}
{PDF}.

\bibitem{Takei2000}
Y.~Takei,
\emph{An explicit description of the connection formula
for the first Painlev\'e equation},
in \emph{Toward the Exact WKB Analysis of Differential
Equations, Linear or Non-Linear},
Kyoto University Press, Kyoto (2000), 271--296.

\bibitem{Tamarin2026Thesis}
N.~Tamarin,
\emph{Resurgence, monodromy, and cluster algebras along the
Painlev\'e I KdV hierarchy},
PhD thesis, Instituto Superior T\'ecnico,
Universidade de Lisboa (2026).

\bibitem{IT26}
T. Ishibashi and Y. Terashima,
Index for a mutation loop,
arXiv:2610.18940 [math.RT] (2026).
\href{https://arxiv.org/abs/2610.18940}{arXiv:2610.18940}.

\bibitem{vanSpaendonckVonk2022}
A.~van Spaendonck and M.~Vonk,
\emph{Painlev\'e I and exact WKB: Stokes phenomenon for
two-parameter transseries},
J. Phys. A: Math. Theor. \textbf{55} (2022), 454003;
\href{https://arxiv.org/abs/2204.09062}{arXiv:2204.09062}.

\end{thebibliography}
\end{document}